\documentclass[longauth,structabstract]{aa}  
\usepackage{txfonts}
\usepackage{graphicx}
\usepackage{cancel}
\usepackage[authoryear]{natbib}
\usepackage{color}
\usepackage{multirow,bigdelim}

\usepackage{txfonts}
\newcommand{\hel}[2] {He\,{\sc #1}~$\rm{\lambda}$#2}
\newcommand{\Sil}[2] {Si\,{\sc #1}~$\lambda$#2}
\newcommand{\Nl}[2] {N\,{\sc #1}~$\lambda$#2}
\newcommand{\Mgl}[2] {Mg\,{\sc #1}~$\lambda$#2}
\newcommand{\Cl}[2] {C\,{\sc #1}~$\lambda$#2}
\newcommand{\Ol}[2] {O\,{\sc #1}~$\lambda$#2}
\newcommand{\hhel}[1] {He\,{\sc #1}}
\newcommand{\SSil}[1] {Si\,{\sc #1}}
\newcommand{\NNl}[1] {N\,{\sc #1}}

\newcommand{\msun}{\ensuremath{\mathit{M}_{\odot}}}

\newcommand{\vrot}{\ensuremath{\varv_{\rm e}\sin i}}
\newcommand{\veq}{\ensuremath{\varv_{\rm e}}}
\newcommand{\vcrit}{\ensuremath{\varv_{\rm crit}}}

\def\kms{\mbox{${\rm km}\:{\rm s}^{-1}$}}

\def\lesssim{\mathrel{\hbox{\rlap{\hbox{\lower4pt\hbox{$\sim$}}}\hbox{$<$}}}}

\def\gtrsim{\mathrel{\hbox{\rlap{\hbox{\lower4pt\hbox{$\sim$}}}\hbox{$>$}}}}

\defcitealias{evans}{Paper~I}
\defcitealias{dufton}{Paper~X}
\defcitealias{sana}{Paper~VIII}

\titlerunning{Rotational velocities of the single O-type stars in 30\,Dor}
\begin{document}
  \title{The VLT-FLAMES Tarantula Survey\thanks{Based on observations
   collected at the European Southern Observatory under program ID 182.D-0222}}

   \subtitle{XII. Rotational velocities of the single O-type stars}
   \author{
     O.H. Ram\'{i}rez-Agudelo   \inst{1}
     \and
     S. Sim\'on-D\'{i}az   \inst{2,3}
     \and
     H. Sana      \inst{1,4}
     \and
     A. de Koter  \inst{1,5}
     \and
     C. Sab\'{i}n-Sanjul\'{i}an \inst{2,3}
     \and
     S.E. de Mink \inst{6,7}%\footnote{Hubble Fellow}
     \and \\
     P. L. Dufton \inst{8}
    \and
      G. Gr\"afener \inst{9}
       \and 
     C.J. Evans   \inst{10}
     \and
     A. Herrero   \inst{2,3}
     \and
     N. Langer    \inst{11}
	\and	
	D.J. Lennon    \inst{12}
    \and
 	J. Ma\'{i}z Apell\'aniz \inst{13}
 	      \and \\
      N. Markova \inst{14}
      \and    
      F. Najarro \inst{15}     
      \and 
      J. Puls \inst{16}
     \and     
      W.D. Taylor  \inst{10}      
     \and     
      J.S. Vink  \inst{9}      
}
\institute{
     % 1. Oscar, Hugues, Alex
          Astronomical Institute Anton Pannekoek, 
          Amsterdam University,  
          Science Park 904, 1098~XH, 
          Amsterdam, The Netherlands\\
          \email{o.h.ramirezagudelo@uva.nl}
          %\thanks{...}	
\and % 2. Sergio, Carolina, Artemio
           Instituto de Astrof\'{i}sica de Canarias, 
           C/ V\'{i}a L\'{a}ctea s/n, E-38200 La Laguna, Tenerife,
           Spain
\and % 3. Sergio, Carolina, Artemio
           Departamento de Astrof\'{i}sica, 
           Universidad de La Laguna, 
           Avda. Astrof\'{i}sico Francisco S\'{a}nchez s/n, 
           E-38071 La Laguna, Tenerife, Spain
\and %4 Hugues
           Space Telescope Science Institute,
           3700 San Martin Drive,
           Baltimore,
           MD 21218,
           USA 
\and % 5. Alex
           Instituut voor Sterrenkunde, 
           Universiteit Leuven, 
           Celestijnenlaan 200 D, 
           3001, Leuven, Belgium
%\and % 5. Alex
%          Utrecht University,
%           Princetonplein 5, 3584CC,
%           Utrecht, The Netherlands
\and % 6. Selma
			Observatories of the Carnegie Institution for Science, 
			813 Santa Barbara St, 
			Pasadena, 
			CA 91101, 
			USA
\and % 7. Selma
	Cahill Center for Astrophysics, 
	California Institute of Technology, 
	Pasadena, 
	CA 91125, 
	USA
\and % 8. Dufton
	        Astrophysics Research Centre, 
	        School of Mathematics and Physics, 
	        Queen's University of Belfast, 
	        Belfast BT7 1NN, 
	        UK
 \and % 9. Goetz
           Armagh Observatory,
           College Hill,
           Armagh, BT61 9DG,
           Northern Ireland,
           UK 
\and % 10. Chris  and Taylor
           UK Astronomy Technology Centre,
           Royal Observatory Edinburgh,
           Blackford Hill, Edinburgh, EH9 3HJ, UK
\and % 11. Norbert Langer
           Argelander-Institut f\"ur Astronomie, 
           Universit\"at Bonn, 
           Auf dem H\"ugel 71, 
           53121 Bonn, Germany
\and % 12. Danny
           European Space Astronomy Centre (ESAC),
           Camino bajo del Castillo, s/n
           Urbanizacion Villafranca del Castillo,
           Villanueva de la Ca\~nada,
           E-28692 Madrid, Spain
\and % 13. Jesus
           Instituto de Astrof\'{i}sica de Andaluc\'{i}a-CSIC,
           Glorieta de la Astronom\'ia s/n,
           E-18008 Granada, Spain
\and % 14. Markova
			Institute of Astronomy with NAO,
			Bulgarian Academy of Science,
			PO Box 136,
			4700 Smoljan,
			Bulgaria
\and %15 Paco
			Centro de Astrobiolog\'{i}a (CSIC-INTA),
			Ctra. de Torrej\'on a Ajalvir km-4,
			E-28850 Torrej\'on de Ardoz,
			Madrid,
			Spain
\and %16 Puls
			Universit\"atssternwarte,
			Scheinerstrasse 1,
			81679 M\"unchen, 
			Germany
}
             
%   \date{Received September 15, 1996; accepted March 16, 1997}
   \date{Received ....}

% \abstract{}{}{}{}{} 
% 5 {} token are mandatory
 
 \abstract
  % context heading (optional)
  % {} leave it empty if necessary  resultsssss
   {The 30 Doradus (30\,Dor) region of the Large Magellanic Cloud is the nearest starburst region. It contains the richest population of massive stars in the Local Group and it is thus the best possible laboratory to investigate open questions in the formation and evolution of massive stars.}
   %
  % aims heading (mandatory)
   {Using ground based multi-object optical spectroscopy obtained in the  framework of the VLT-FLAMES Tarantula Survey (VFTS), we aim to establish the (projected) rotational velocity distribution for a sample of 216 presumably single O-type stars in 30\,Dor. The size of the sample is large enough to obtain statistically  significant information  and to search for variations among sub-populations -- in terms of spectral type, luminosity class, and spatial location -- in the field of view.}
  % methods heading (mandatory)
   {We measured  projected rotational velocities, \vrot,  by means of a Fourier transform method  and  a profile fitting
      method applied on a set of isolated spectral lines.  We also used an iterative deconvolution procedure to infer the probability density, 
      $\rm{P(\veq)}$, of the equatorial rotational velocity, \veq.}
  % results heading (mandatory) 
   {The distribution of \vrot\  shows a two-component structure: a peak around  80~\kms\ 
   and a high-velocity tail extending up to $\sim$600 \kms. This structure is also present in the inferred distribution $\rm{P(\veq)}$
   with around 80\% of the sample having 0 $<$ \veq\,  $\leq\, 300$~\kms\ and the other 20\%  distributed in the high-velocity 
   region. The presence of the low-velocity peak is consistent with that found in other studies for late O- and early B-type stars. }
  % conclusions heading (optional), leave it empty if necessary 
  {Most of the stars in our sample rotate with a rate less than 20\%\ of their break-up velocity. 
  For the bulk of the sample, mass-loss in a stellar wind and/or envelope expansion is not efficient enough 
  to significantly spin down these stars within the first few Myr of evolution. If massive-star formation 
  results in stars rotating at birth with a large fraction of their break-up velocities, an alternative 
  braking mechanism, possibly magnetic fields, is thus required to explain the present day rotational 
  properties of the O-type stars in 30\,Dor. The presence of a sizeable population of fast rotators is 
  compatible with recent population synthesis computations that investigate the influence of binary evolution 
  on the rotation rate of massive stars. Despite the fact that we have excluded stars that show significant
  radial velocity variations, our sample may have remained contaminated by post-interaction binary products. 
  The fact that the high-velocity tail may be preferentially (and perhaps even exclusively), populated by 
  post-binary interaction products, has important implications for the evolutionary origin of systems 
  that produce gamma-ray bursts.
}


   \maketitle
%
%________________________________________________________________

\section{Introduction}

The distribution of stellar rotation rates at birth is a `fingerprint' of the formation process of a population of stars.  
For massive stars the rotational distribution is especially interesting because so little is known about how these stars form \citep[e.g.][]{zinnecker2007}.  
Considerations of angular momentum conservation during the gravitational collapse of a molecular cloud suggest an 
`angular momentum problem', i.e. it appears difficult for the forming stars -- whether they be of low- or high-mass --
 {\em not} to rotate near critical velocities. However, very few massive O-type stars are known to be extreme rotators.

 If massive stars form through disk accretion, in a similar way to low 
 mass stars, their initial spin rates 
 are likely controlled by gravitational torques \citep{lin}.  Only massive stars that have low accretion rates, long disk lifetimes, 
 weak magnetic coupling with the disk, and/or surface magnetic fields that are significantly stronger than what current 
 observational estimates suggest, may have their initial spin regulated by magnetic torques \citep{rosen}.  Perhaps
 in these cases, intrinsic slow rotators can be formed.
 
The initial rotation rate is also one of the main 
properties affecting the evolution of a massive star.
%quantity of a massive star as it may directly affects its evolution.  
For instance, rotation induces internal mixing and prolongs the 
main-sequence life time \citep[see][]{maederymeynet2000,brott,ekstrom2012}.  A very high initial rotation 
rate may cause rotational distortion and gravity darkening \citep{collins1963,collins1966} and may even lead to homogeneous evolution \citep{maeder1980,brott,brotta}.  In low-metallicity environments this type of 
evolution has been suggested to lead to gamma-ray-bursts \citep{yoon2005,woosley}.

With the above science topics in mind, considerable effort has been invested in establishing the full 
distribution of equatorial rotational velocities (\veq) of massive stars. %\citep{rosen}.  
Spectroscopic studies provide projected rotational velocities (\vrot, where $i$ is the inclination angle of the stellar rotation
axis with respect to the line-of-sight). Large populations are thus preferred in order to confidently deconvolve the observed distribution 
and obtain a true \veq\ distribution \citep{penny,huanggies1,hunter,penny2009,huang2010,dufton}.  
Obviously, the current values of spin rates do 
not necessarily reflect initial values.  Stellar expansion and/or angular momentum loss via stellar winds are proposed
mechanisms that spin down stars as time passes. Both these mechanisms, however, seem rather ineffective for the bulk of the massive
stars.  Internal redistribution of angular momentum effectively prevents the spin down of the surface layers as the
star evolves to larger main-sequence radii and mass loss seems only effective for stars more massive than 
%Prior mass loss in a stellar wind, for instance, may have caused spin down, though this 
%mechanism seems only effective for stars that are more massive than 
40 \msun\ at birth \citep{brott,brotta,vink2010,vink2011}.

%%%%%%%%%%%%%%%%%%%%%%%%%%%%%%%%%%%%%%%%%%%%%%%%%%%%%%%%%%%%%%%
\begin{figure}
\centering
\includegraphics[scale=0.5]{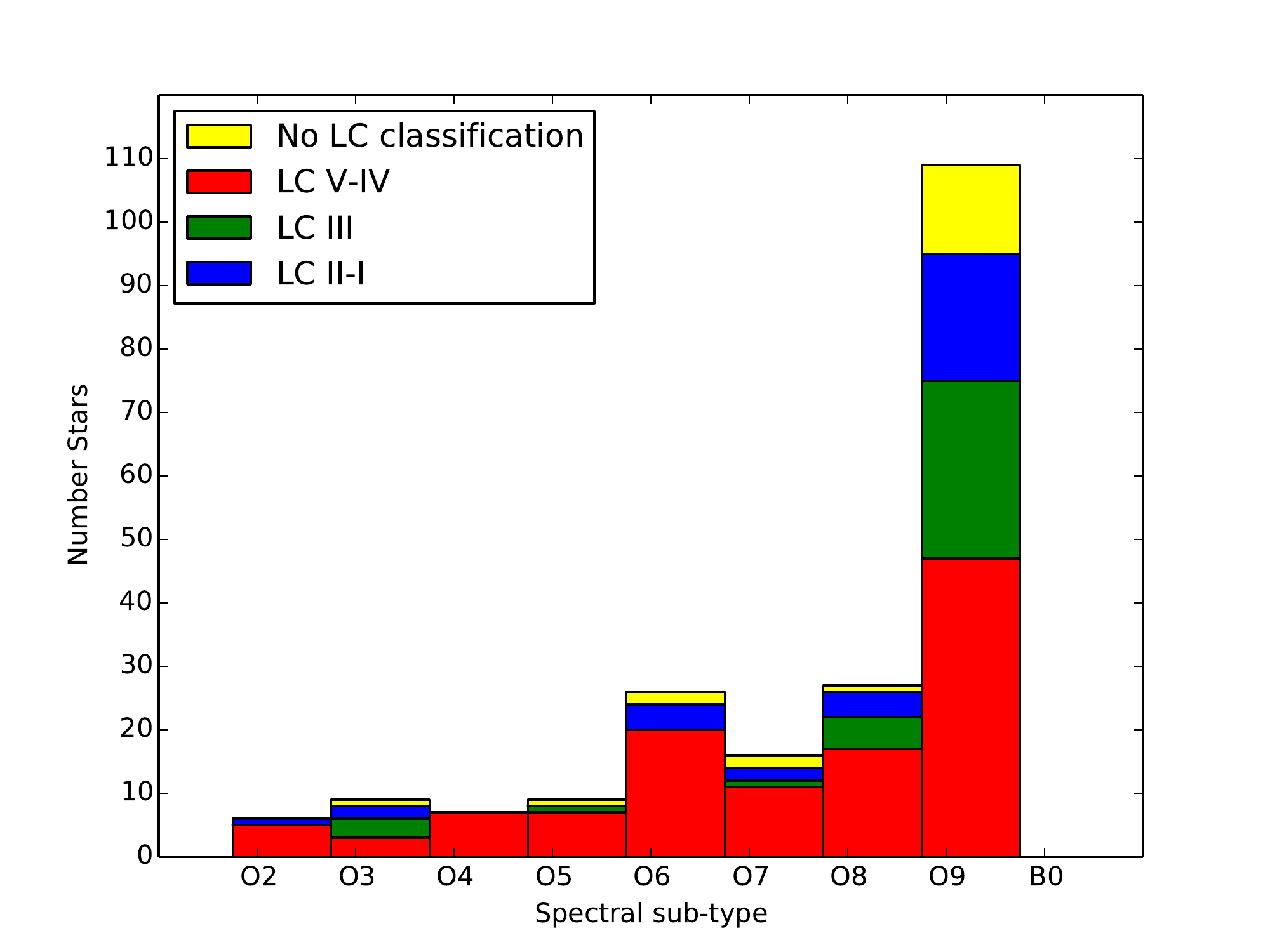}
\caption{Spatial distribution (SpT)  distribution of the O-type stars in our sample,
binned per spectral sub-type. Different colors indicate different Luminosity Class (LC) (see legend).
%[A color version of the figure is available in the electronic version of the journal.]
}
\label{fig:SpTLC}
\end{figure}
%%%%%%%%%%%%%%%%%%%%%%%%%%%%%%%%%%%%%%%%%%%%%%%%%%%%%%%%%%%%%%%

As the intrinsic multiplicity fraction of massive stars at birth seems very high \citep{sanaa}, it is also important to consider 
the effects of binary evolution on the rotational properties.  Interestingly enough, binary interaction quite often
occurs early on in the evolution of these systems, with up to 40\%\ of all stars born as O-type stars being affected before leaving the main sequence \citep{sanaa}.
%Binary evolution may also affect the spin rates, surprisingly enough, sometimes already quite early on in the evolution.  
Using the intrinsic binary properties of 
Galactic O-type stars \citep{sanaa}, \citet{selma} show that binary interaction strongly impacts spin rates, producing a 
`high-velocity tail' in the \vrot\ distribution through angular momentum transfer processes. 
%such as tides, Roche lobe overflow, common envelope evolution and/or stellar mergers.  
Though studies of O-type star populations often try to select samples from 
which the known binaries have been removed, de Mink et al. (in prep.) show that such samples likely remain strongly 
contaminated by (unidentified) binary interaction products.  Whether the results of de Mink et al. are consistent with {\em all} rapidly 
($\veq\, > 200-300$\,\kms) spinning O-type stars being spun-up binary products is an intriguing question.  A positive answer 
may imply that the proposed homogeneous single-star channel to gamma-ray-bursts mentioned above does not occur in nature.
%Whether binarity can also explain very low rotation rates is \textbf{an} interesting question too, but this still needs to be addressed.

The VLT-FLAMES Tarantula Survey (VFTS) is a multi-epoch spectroscopic campaign targeting over 800 massive O and 
and early-B stars across the 30~Doradus (30\,Dor) region in the Large Magellanic Cloud (LMC), including targets in the 
OB clusters NGC\,2070 and NGC\,2060. 
The distance to 30\,Dor is well constrained \citep{gibson2000} 
and its foreground extinction is relatively low \citep[see][hereafter \citetalias{evans}]{evans}.  
The large massive-star population %subset of O-type stars 
is ideal for studying the rotational velocity distribution.
% and even allows to focus in on the high velocity tail.  
The low metal content of the LMC is in this sense even beneficial, as for such an environment the tail of the \vrot\ distribution 
produced by binary effects is expected to be more extended and pronounced \citep{selma}.

The dense core cluster of the Tarantula Nebula, Radcliffe 136 (R136), has a likely age of 1-2 Myr 
\citep{dekoter1998,masseyhunter1998}, but may actually be
a composite of two sub-clusters \citep{sabbi}, with a third population that is a few million years older nearby \citep{selman}. 
The very central regions were excluded from the VFTS because of crowding issues.
% referred to as R136a, are excluded from the VFTS because of crowding issues.
A series of distinct populations, with varying ages reaching up to $\sim$25\,Myr, can be distinguished in the Tarantula field 
\citep{walborn}, suggesting that 30\,Dor has been continuously forming massive stars over the last 25~Myr, although probably with a variable formation rate.

The key questions we want to address in this study are: how are the rotational velocities of O-type stars in 30 Dor distributed?  Which fraction of the stars are slow rotators?  Does the distribution have a high-velocity tail?  If present, is the distribution of rapid
rotators suggestive of a contribution of unidentified binaries?  The layout of the paper is as follows. Section \ref{sec:sample} describes the selection of our sample. The methodology and the results from different diagnostic lines are described in Section \ref{sec:detervsini}. Section \ref{subsec:dist} presents the \vrot\ and \veq\ distributions. The results are discussed in Section \ref{sec:discuss} and our conclusions are summarized (Section \ref{sec:conclusions}). 

%%%%%%%%%%%%%%%%%%%%%%%%%%%%%%%%%%%%%%%%%%%%%%%%%%%%%%%%%%%%%%%
\begin{figure}
\centering
\includegraphics[scale=0.4]{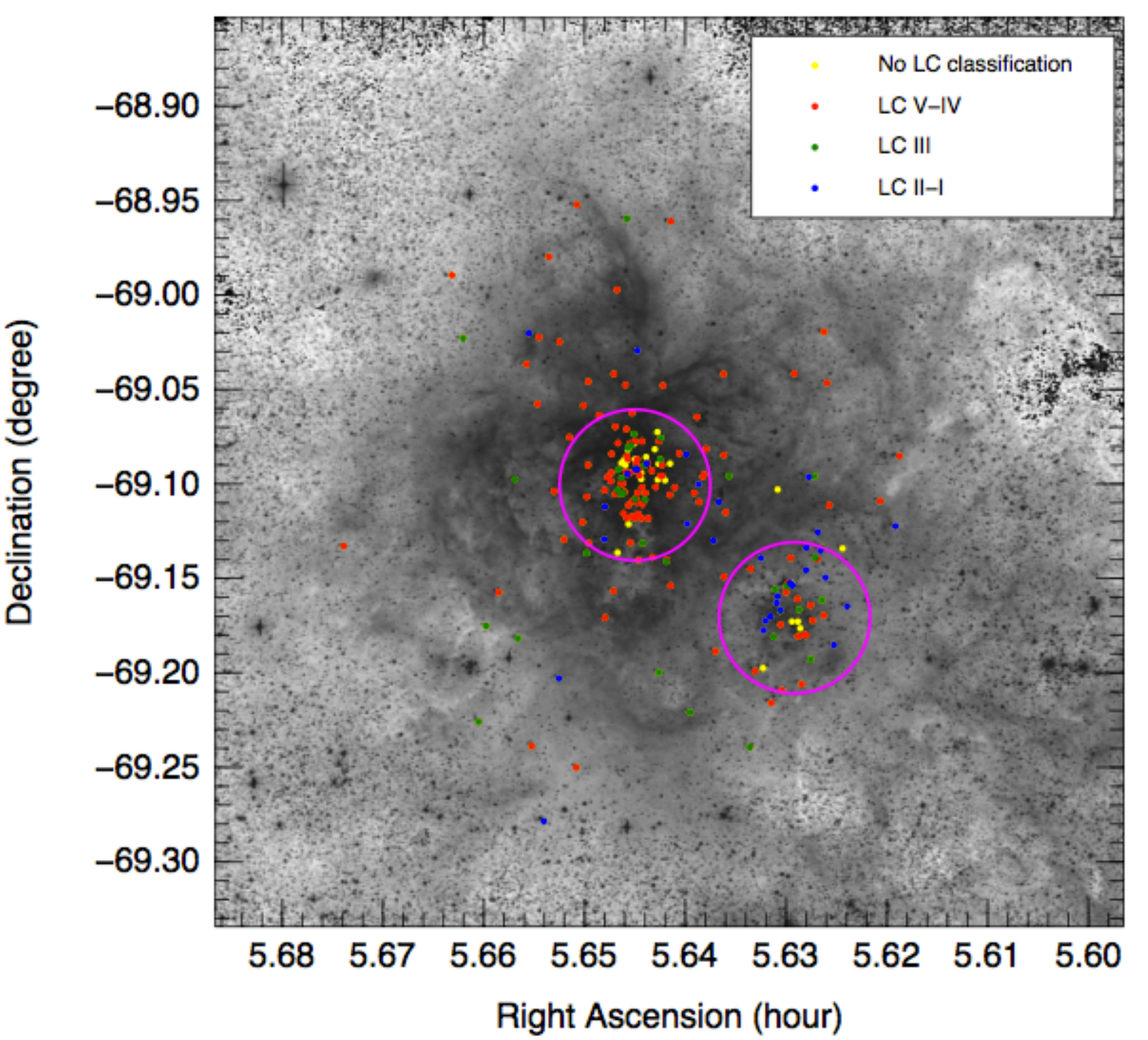}
\caption{SpT of the presumably single O-type stars as a function of the LC. 
The circles define regions within 2.4' of NGC\,2070 (central circle) and NGC\,2060 (SW circle).
Colors have the same meaning as in Fig.\,\ref{fig:SpTLC} (only available in the online version).}
\label{fig:SpTLC_spatial}
\end{figure}
%%%%%%%%%%%%%%%%%%%%%%%%%%%%%%%%%%%%%%%%%%%%%%%%%%%%%%%%%%%%%%%

%%%%%%%%%%%%%%%%%%%%%%%%%%%%%%%%%%%%%%%%%%%%%%%%%%%%%%%%%%%%%%%
\begin{figure}
\centering
\includegraphics[scale=0.4]{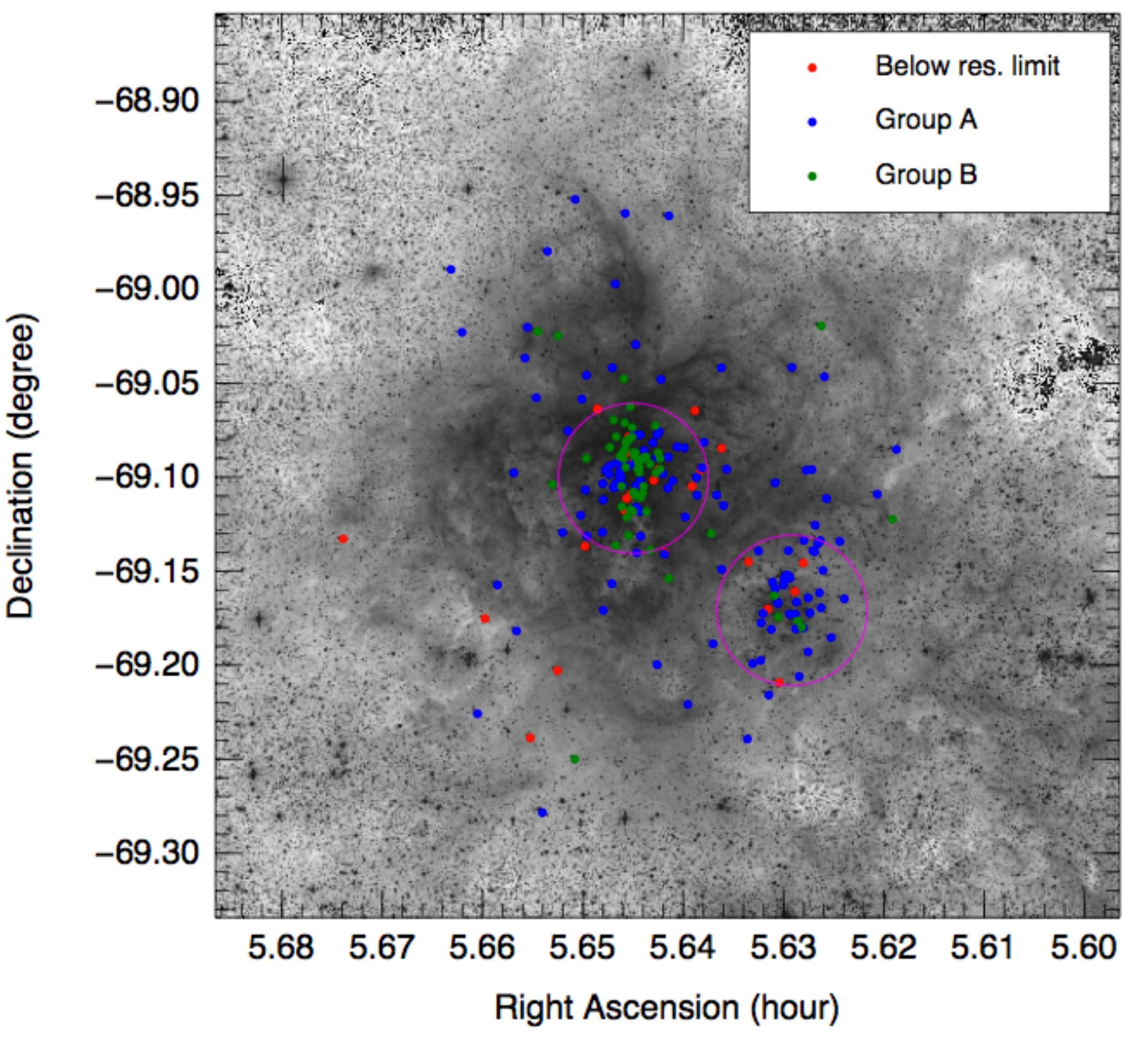}
\caption{SpT of our sample in the 30~Dor field of view. Different colors identify the different groups listed in Table~\ref{table:1} and are identified in the legend.  
%[A color version of the figure is available in the electronic version of the journal.]
} 
\label{fig:Spat_Groups}
\end{figure}
%%%%%%%%%%%%%%%%%%%%%%%%%%%%%%%%%%%%%%%%%%%%%%%%%%%%%%%%%%%%%%%

\section{Sample}\label{sec:sample}

The VFTS project and the data have been described in \citetalias{evans}.  Here, we focus on the presumably single O-type stars that have been observed using the Medusa fibers. The total
Medusa sample contains 332 O-type objects. \citet[][hereafter \citetalias{sana}]{sana} have
identified spectroscopic binaries from multi-epoch radial velocity (RV) measurements: 172 O-type stars show no significant RV
variations and are presumably single; 116 objects
show significant RV variations with a peak-to-peak amplitude ($\rm{\Delta RV}$) larger than 20~\kms\ and are considered
binaries. The remaining 44 objects show low-amplitude significant RV variations ($\rm{\Delta RV}\, \leq$ 20~\kms\,).
The latter variations may be the result of photospheric activity (pulsations and/or wind variability) or indicate a spectroscopic
binary system. In  \citetalias{sana}, we estimated that photospheric variations and genuine binaries contribute to the low-amplitude 
RV variation sample in roughly equal portions. To avoid biasing our analysis against supergiants, which are expected to show 
spectroscopic variability \citep[see e.g.][]{simon1}, we include these 44 low-amplitude RV variable objects in our sample, reaching thus a total sample size of 216 O-type stars. 

Because of the limited number of observing epochs, a subset
%finite number of observations, a 
of stars in our RV-constant sample are expected to be undetected binaries. Given the VFTS binary detection probability among O-type stars \citepalias[of $\approx$ 0.7,][]{sana}, one estimates that 
this may be the case for up to about 25\%\ of our single-star sample. Undetected binaries 
%among the low-amplitude RV variables 
are preferentially wide and/or low-mass ratio systems (see sect. 3 in \citetalias{sana}). Tidal interactions are thus expected
to be negligible for these systems so that the measurement of the rotational velocity
of the main component  should be mostly unaffected by the binary status.

The spectral classifications of the O-type stars in the VFTS will be presented in Walborn et al. (in prep.). Fig.~\ref{fig:SpTLC} shows the spectral type (SpT) and luminosity class (LC)
distribution for our sample, which is dominated by O9-O9.7 stars (52\% of the sample size) and by  dwarfs and sub-giants (LC\,V-IV). As indicated in Fig.~\ref{fig:SpTLC}, a luminosity classification is missing for 22 stars, i.e.\ about 10 \% of our sample. For seven stars the precise spectral type has not been established either due to the poor signal to noise ratio ($\rm{S/N}$) and/or heavy nebular contamination. These seven stars have been excluded from Fig.~\ref{fig:SpTLC} but are still incorporated in our analysis.

The spatial distribution of our sample is shown in Fig.\,\ref{fig:SpTLC_spatial}. The field of view is dominated by the central cluster NGC\,2070. A circle of radius 2.4' (or 37\,pc) around the cluster center contains 105 stars from our sample:
62 are of LC\,V-IV, 15 are LC\,III and 28 are LC\,II-I. A second concentration of 45 O-type stars  is found in a similar sized region around the  NGC\,2060 cluster, located about 6' to the south-west of NGC\,2070. NGC\,2060 is somewhat older than NGC\,2070 \citep[][]{walborn}. 
Accordingly, it contains a larger fraction of LC\,II-I stars (23\,\%) than NGC\,2070 (9\,\%). 
The remaining stars, throughout the paper referred to as the stars outside clusters, are spread throughout the field of view. These may originate from either NGC\,2070 or 2060, but may also have formed in other star-forming events in the 30\,Doradus region at large.

%%%%%%%%%%%%%%%%%%%%%%%%%%%%%%%%%%%%%%%%%%%%%%%%%%%%%%%%%%%%%%%
\begin{figure}[t!]
\centering
%\begin{center}
 \includegraphics[scale=0.5]{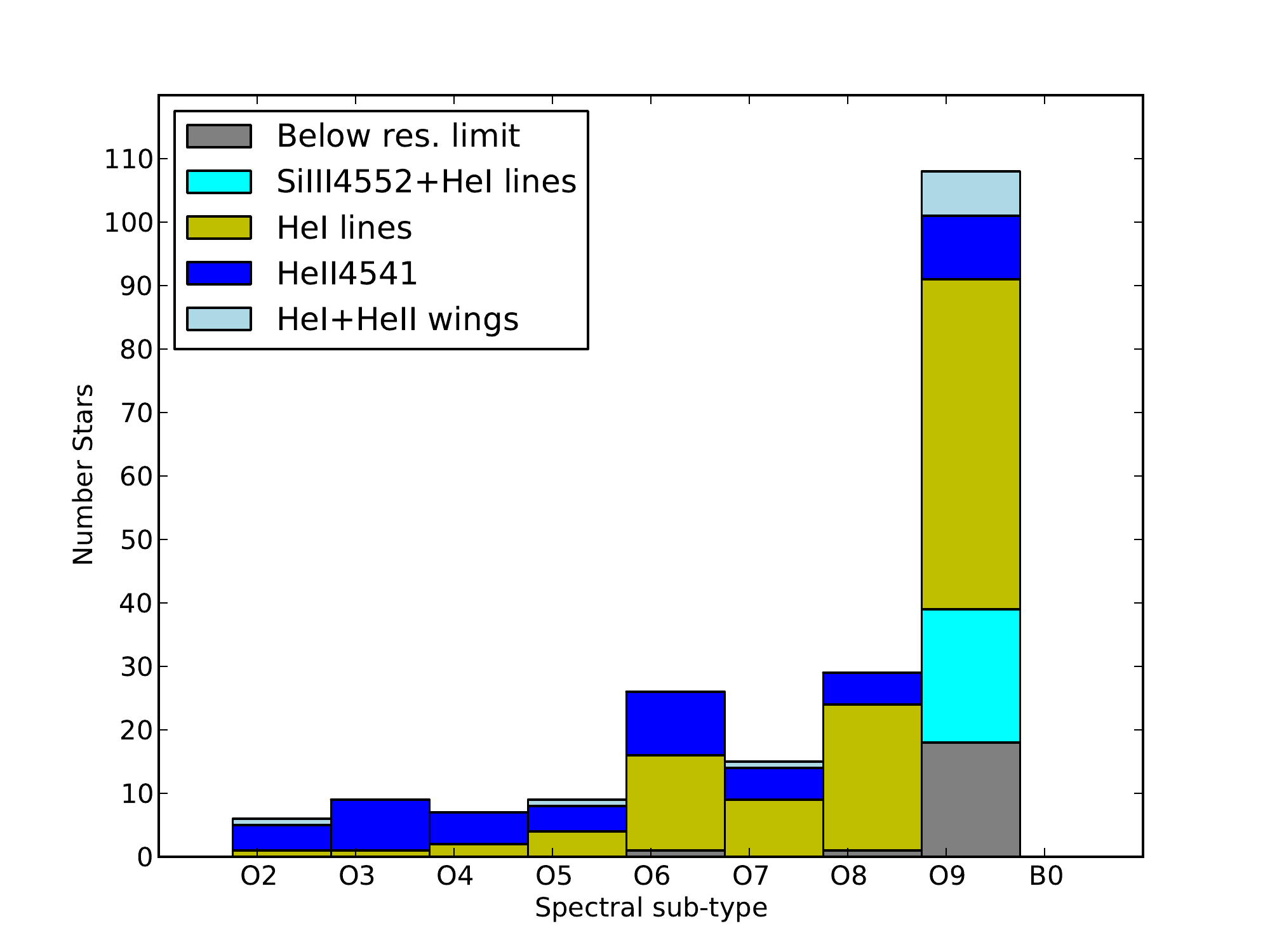}
%\end{center}
\caption{Diagnostic lines used as a function of the SpT of the stars in our sample.
%(see legend). 
%[A color version of the figure is available in the electronic version of the journal.]
}
\label{fig:SpTLC_lines}
\end{figure}
%%%%%%%%%%%%%%%%%%%%%%%%%%%%%%%%%%%%%%%%%%%%%%%%%%%%%%%%%%%%%%%

\section{Measuring the projected rotational velocity} \label{sec:detervsini} 
\subsection{Methodology}\label{subsec:Methodology}
%[Overall introduction on how to measure ve sini]
The projected rotational velocity (\vrot) of stars
can be measured directly from the broadening of their spectral
lines \citep[][]{carroll,Gray}. Commonly used methods for OB-type
stars include direct measurement of the full width at half maximum 
(FWHM; \citealp[e.g.][]{slettebak,herrero,abt}),
cross-correlation of the observed spectrum against a template
spectrum \citep[e.g.][]{penny,howarth}, and comparison with synthetic lines 
calculated from model atmospheres \citep[e.g.][]{mokiem}.
In this study we use a Fourier Transform (FT) method and a line profile fitting method which we refer to as the Goodness of Fit (GOF) method. Both methods are applied to a set of suitable spectral lines present in the VFTS Medusa  LR02 and LR03 spectra (see Section~\ref{subsec:diagnostic_lines}).
A comparison of the measurements obtained from both  methods
allows us to verify the internal consistency of the derived values.

%[Explanation of the FT]
The FT method is explained in \citet{Gray}. It has been systematically applied 
to OB-type stars by \citet{ebbets} and \citet{simon}. 
In summary, the first minimum of the Fourier spectrum
uniquely identifies the value of \vrot. % (see Fig.~\ref{fig:FT} for an example).
One advantage of the FT method is that the derived \vrot\, value is in principle not
strongly affected by the presence of photospheric velocity fields, such as
macro-turbulent motions\footnote{For evolved (B-type) supergiants macro-turbulent 
broadening may be caused by 
non-radial gravity-mode pulsations. For a sub-set of these pulsators the \vrot\ values 
derived using the FT method may be offset by 10-30 \kms\ \citep{aerts2009}.}.
%\footnote{\red{See, however, the warning indicated by \citep{aerts2009} about the 
%reliability of the FT method when macro-turbulent broadening is a result of non-radial pulsations.}}. 
However, the method encounters difficulties in case 
of strong nebular contamination, poor $\rm{S/N}$,
weak lines, and/or when the rotational broadening contribution is of the order 
of the intrinsic broadening of the line.

%[Explanation of GOF]
The GOF method \citep[e.g.][]{ryans,simon1} 
adjusts  a synthetic line profile to the observed  profile
using a least-square fit, taking into account the 
intrinsic, instrumental, rotational, and macro-turbulent broadening contributions
through successive convolutions.
In this study we neglect the intrinsic width of the spectral lines, effectively 
adopting a delta 
function for the  intrinsic profile. The latter is then convolved with an instrumental (Gaussian) function that preserves the equivalent width (EW). For the rotational and macro-turbulent profiles, we follow
the description given by \citet{Gray}.  The rotational profile assumes a linear 
limb darkening law. We finally use a radial-tangential (RT)
model  parametrized with the quantity $\Theta_\mathrm{RT}$ as an appropriate representation of the macro-turbulent profile.
%\footnote{In the case of \hhel{i} and \hhel{ii} lines, this profile turns out to 
%artificially mimic the Stark broadening contribution.} 

%%%%%%%%%%%%%%%%%%%%%%%%%%%%%%%%%%%%%%%%%%%%%%%%%%%%%%%%%%%%%%%
\begin{figure}
\centering
\includegraphics[scale=0.28]{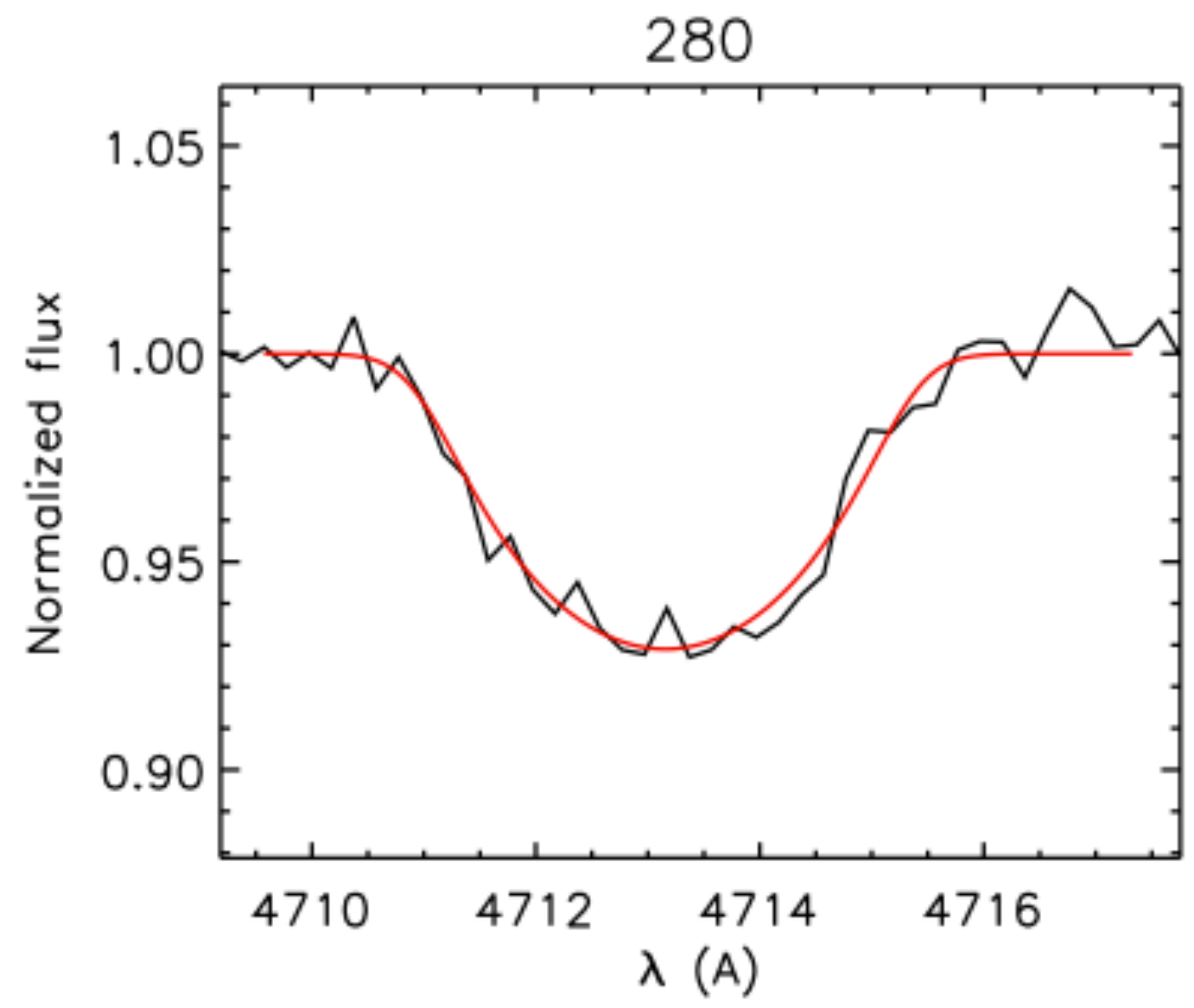}
\includegraphics[scale=0.28]{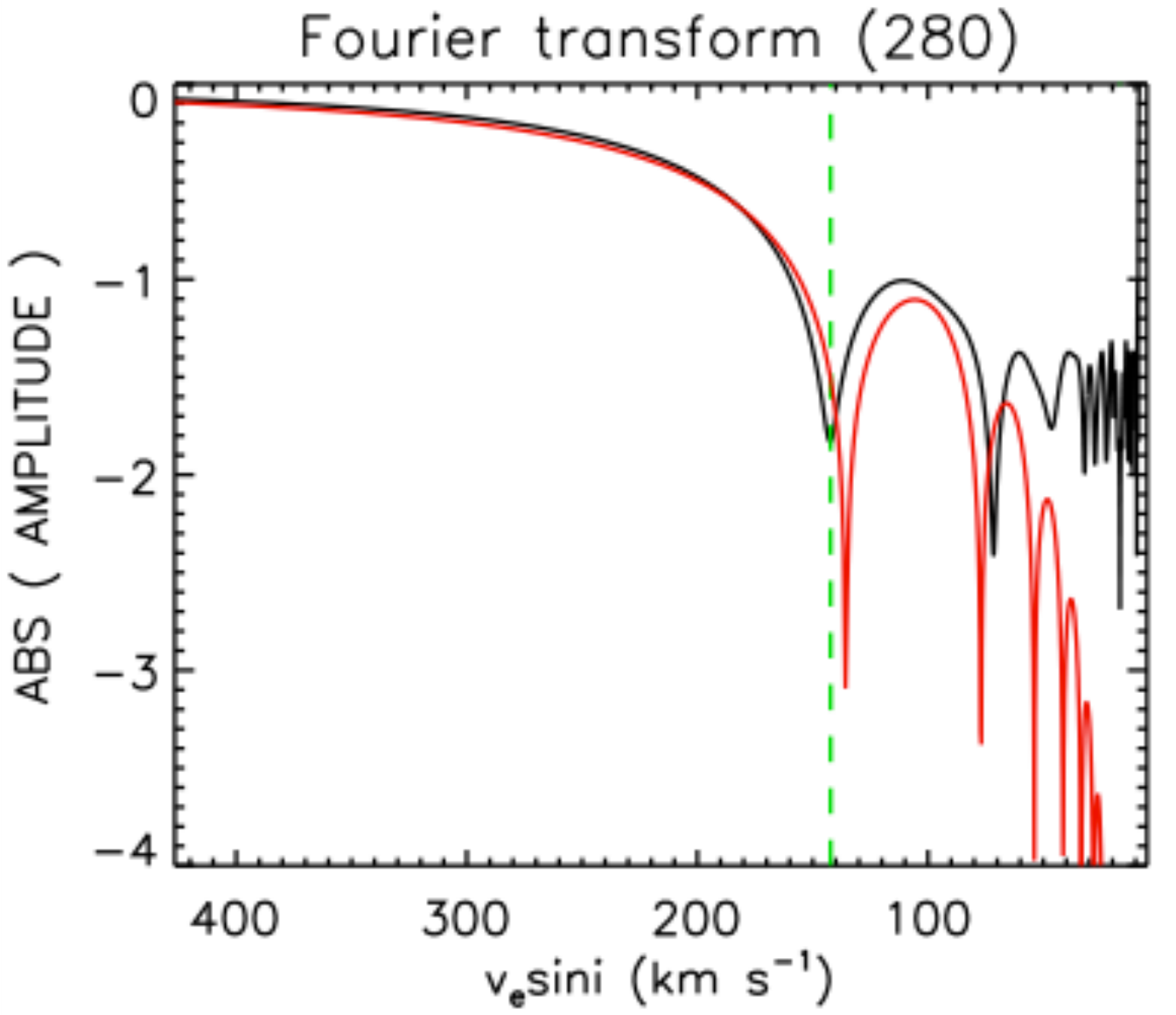} 
\includegraphics[scale=0.28]{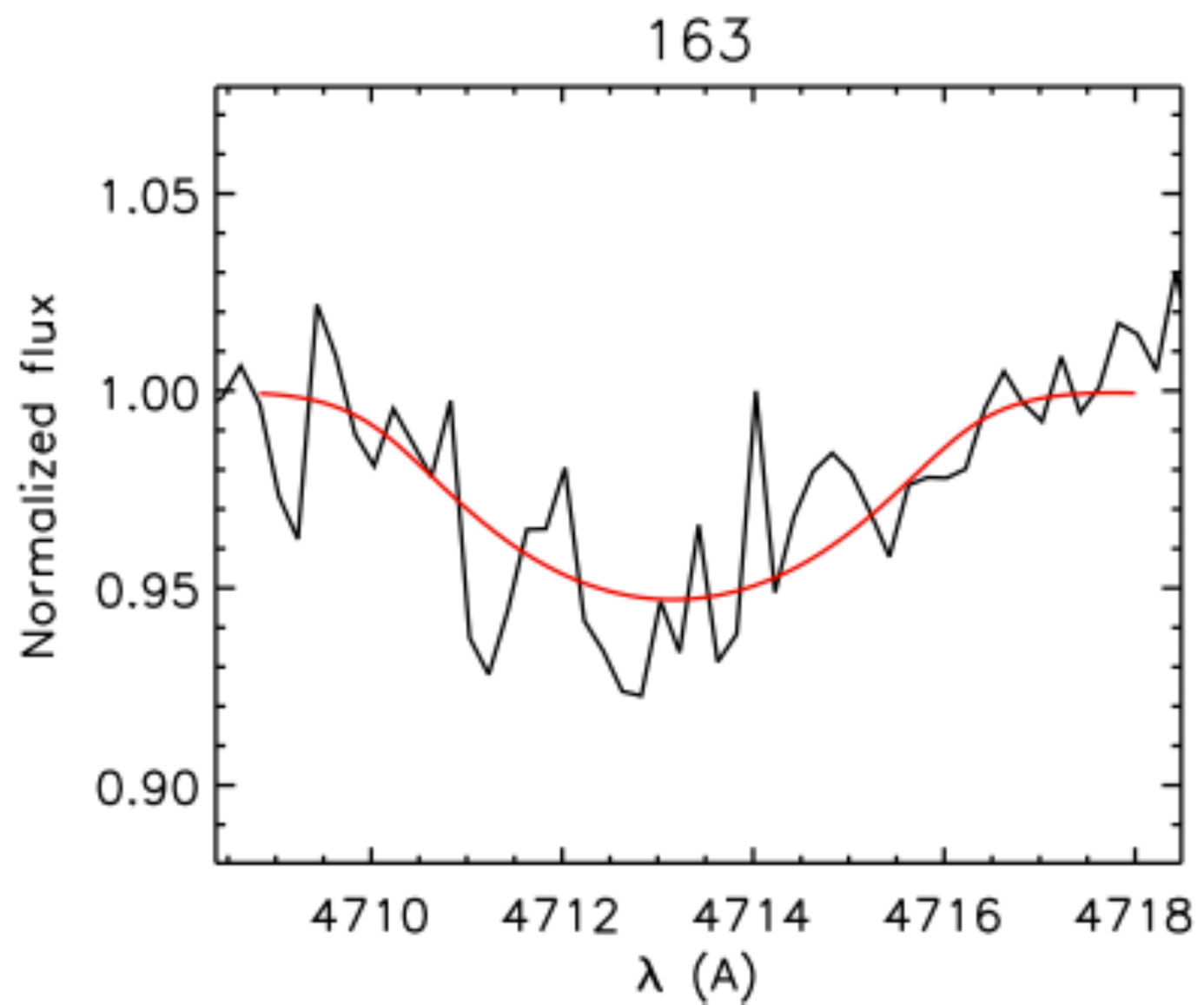}
\includegraphics[scale=0.28]{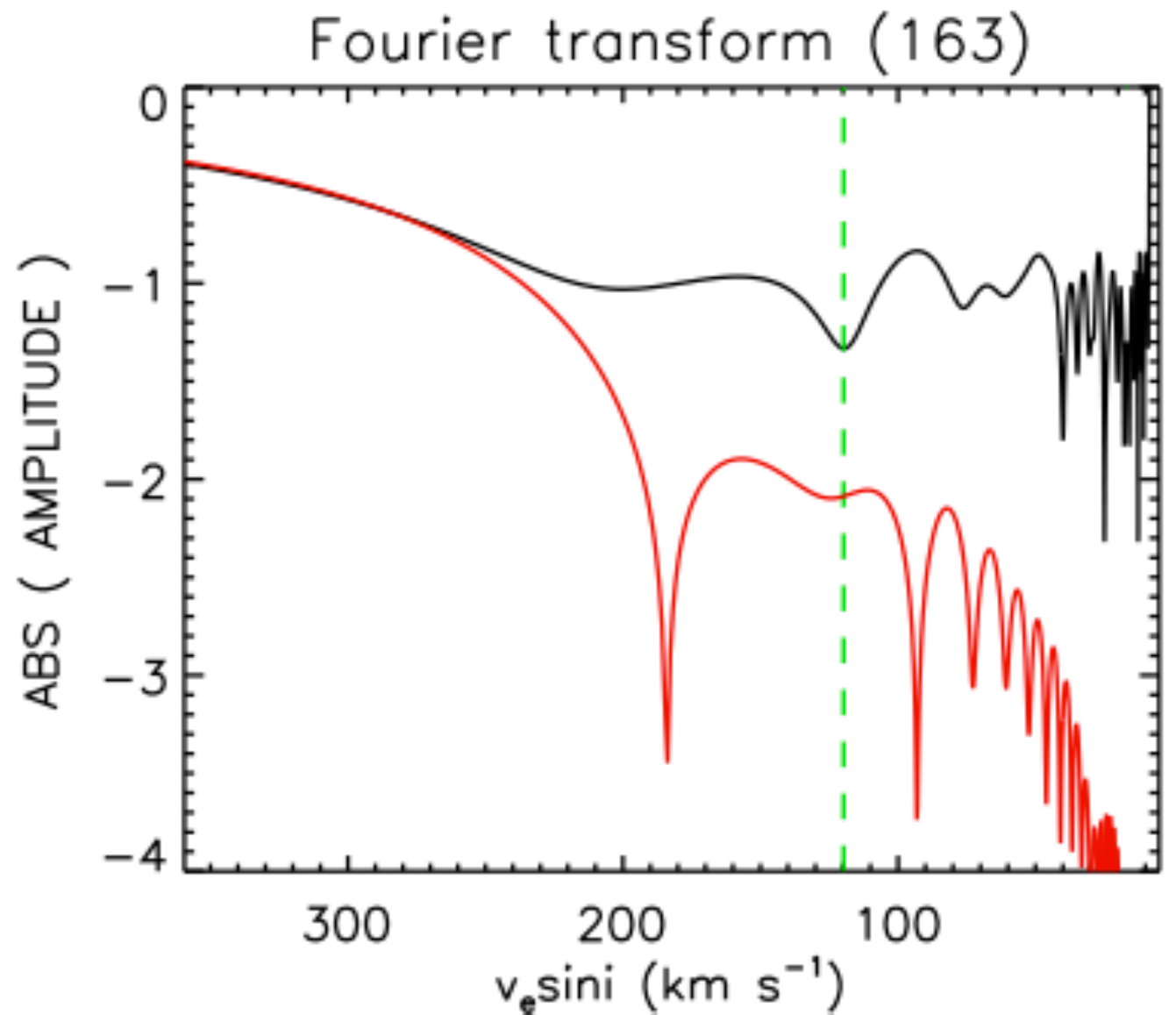} 
\caption{Example of the GOF and FT measurement methods applied to the \hel{i}{4713} line of VFTS 280 (top row, high $\rm{S/N}$) 
and VFTS 163 (bottom row, poor $\rm{S/N}$), to illustrate two different data qualities.
Left-hand column: line profile (black) and its GOF solution (red).  
Right-hand column:  Fourier transform of the observed spectrum (black) and of the best fit GOF profile (red). The dashed lines 
indicate the position of the FT first minimum.  }
\label{fig:FT}
\end{figure}
%%%%%%%%%%%%%%%%%%%%%%%%%%%%%%%%%%%%%%%%%%%%%%%%%%%%%%%%%%%%%%%

\subsection{Diagnostic lines}\label{subsec:diagnostic_lines}

In general, different diagnostic lines 
do not provide the same accuracy in the \vrot\ measurements. 
Metallic lines do not suffer from strong Stark broadening nor from nebular contamination
and hence are best for obtaining \vrot\ estimates.
\Sil{iii}{4552} is the only suitable metal line in our data set for these purposes. Unfortunately, it is only 
present in 6\%\ of the sample and restricted to late spectral sub-types.
Next in line in terms of reliability are
nebular free or weakly contaminated \hhel{i} lines, notably
\hel{i}{4713}, 4922, 4387, and 4471. 
%Although their line profiles suffer from Stark broadening,
%the rotational broadening contribution dominates when \vrot\ is above 80 \kms\ (\\{ref?}).
By using the neutral helium lines we can study an additional 54\% of the sample (see Table~\ref{table:1}).

The remaining sample of stars (31\%) suffers from strong nebular contamination, features only
weak \hhel{i} lines or -- for the earliest spectral sub-types -- does not show these lines at all. In those cases,
amounting to 26\% of our sample, we relied 
%of \hhel{i} lines, or weakness (or absence) of these diagnostic lines in the case of the earliest spectral types
on \hel{ii}{4541}. 
%\hel{ii}{4541} is used in 26\%\ of
%the cases and only used it if no other diagnostic lines are available. 
Finally, 14 
stars\footnote{VFTS 072, 125, 267, 405, 451, 465, 484, 529, 559, 565, 571, 587, 609 and 724.} present strong nebular contamination in \hhel{i}, weak \hhel{ii} lines and no \SSil{iii} line. We have estimated  \vrot\, for these  sources by comparing 
rotationally broadened synthetic line profiles calculated using FASTWIND \citep{puls2005} to the wings of \hhel{i} and \hhel{ii} lines. 
Though FASTWIND does not take into account macro-turbulent broadening it, at least, allows us to derive upper limits on \vrot\ for this small subset of stars.

%For the remaining sample of stars (31\%) due to strong nebular contamination
%of \hhel{i} lines, or weakness (or absence) of these diagnostic lines in the case of the earliest spectral types
%we have then relied on \hel{ii}{4541} as our last-resort diagnostic line. \hel{ii}{4541} is used in 26\%\ of
%the cases and only used it if no other diagnostic lines are available. Finally, 11 
%stars\footnote{VFTS 072, 125, 405, 451, 465, 529, 559, 565, 587, 609 and 724.} present strong nebular contamination in \hhel{i}, weak \hhel{ii} lines and no \SSil{iii} line. We have estimated  \vrot\, for these  sources by comparing spun-up synthetic atmosphere profiles, from FASTWIND \citep{puls2005}, to the wings of \hhel{i} and \hhel{ii} lines.

The assumption of a delta-function for the intrinsic profile ignores a possible contribution
of Stark broadening (more relevant for \hhel{i} and \hhel{ii} lines).
Being conscious of that, we scrutinize the reliability of these lines for \vrot\ determinations in Sect.~\ref{subsec:results}. For practical purposes we refer to the group of stars for which the \Sil{iii}{4552} and/or nebular free \hhel{i} lines can be used as Group A and the remaining sample as Group B. Group A is thus identifying the stars for which
the highest quality diagnostic lines are available, while Group B has diagnostics of a lower quality.

%[There might be a better place for this, but leave it here for now], May 2013: Moved here
The resolving power of the VFTS Medusa LR02 and LR03 is
$\sim$\, 0.6 $\AA$  (\citetalias{evans}), corresponding to 40\,\kms.
We have adopted this value as our resolution limit and hence, 
when  a \vrot\ measurement is below 40~\kms\ 
the corresponding star is systematically assigned to the 0--40~\kms\ bin (see Sect.~\ref{subsec:dist}) without
further indication of the specific \vrot\ value. 

Fig.~\ref{fig:Spat_Groups} shows the spatial distribution of stars below the resolution limit and
of stars in Groups A and B. Group A stars are spread out within 
the two clusters and show a lower fraction in the field of view. 
Group B stars are mainly concentrated in NGC\,2070. There are  two reasons that can explain this difference.
%separation
First, part of the NGC\,2070 cluster lies behind a filament of nebular gas, therefore contamination is stronger.
Second, NGC\,2070 is younger and, pro rata, contains more hot early O-type stars that show neither 
\SSil{iii} nor \hhel{i} lines in their spectra.
%Moreover, stars below the resolution limit (b.r.) show a preferential spatial location but 

Fig.~\ref{fig:SpTLC_lines} shows the spectral lines used for \vrot\, measurements as a
function of SpT. For late SpT, measurements are mostly obtained from \hhel{i}, while \hel{ii}{4541}
is increasingly used for earlier SpT. Stars with \vrot\, below the resolution limit
and stars featuring the \Sil{iii}{4552} line are almost exclusively late O-type stars. Additionally, all
stars showing \SSil{iii}  also display at least one suitable \hhel{i} line.
The accuracy of, and the systematics between, \vrot\ measurements obtained from both methods and for the
different lines are discussed in the next subsections.

%%%%%%%%%%%%%%%%%%%%%%%%%%%%%%%%%%%%%%%%%%%%%%%%%%%%%%%%%%%%%%%
\begin{table}
\caption{Overview of the diagnostic lines used to derive the projected rotational velocity. 
Group A diagnostics provide the most reliable measurements; those in Group B are of a lesser quality. 
The last column lists the fraction of our sample for which each diagnostic can be applied.}             % title of Table
\label{table:1}      % is used to refer this table in the text
\centering                          % used for centering table
\begin{tabular}{c c c}        % centered columns (4 columns)
\hline\hline\\[-7pt]               % inserts double horizontal lines
Group 	& line & Sample (\%) \\[1pt] \hline\\[-8pt]   % table heading
    		& Below the resolution limit    & 9\%             \\[1pt] \hline\\[-7pt]
A  &\Sil{iii}{4552}+\hhel{i} lines    &                 6\%          \\ 
 A  &\hel{i}{4713}      &     \rdelim\}{4}{3mm}[$54\%$]         \\
% A  &\hel{i}{4713}      &     $\left.   \multirow{4}{*}{\right\rbrace $  54\%}           \\
A  &\hel{i}{4922}      &                            \\
A &\hel{i}{4387}      &                             \\
A  &\hel{i}{4471}      &                             \\[2pt]  \hline\\[-9pt]
\multicolumn{2}{c}{\it Total Group A} & 60\%   \\[1pt] \hline\\[-7pt]
B  & \hel{ii}{4541}     &           25\%                \\
B  & \hhel{i}+\hhel{ii} wings     &           6\%                \\[2pt] \hline\\[-9pt]
\multicolumn{2}{c}{\it Total Group B} & 31\%    \\ \hline
                       % inserts single horizontal line
\end{tabular}
\end{table}
%%%%%%%%%%%%%%%%%%%%%%%%%%%%%%%%%%%%%%%%%%%%%%%%%%%%%%%%%%%%%%%

%%%%%%%%%%%%%%%%%%%%%%%%%%%%%%%%%%%%%%%%%%%%%%%%%%%%%%%%%%%%%%%
\begin{figure}
\centering
\includegraphics[scale=0.5]{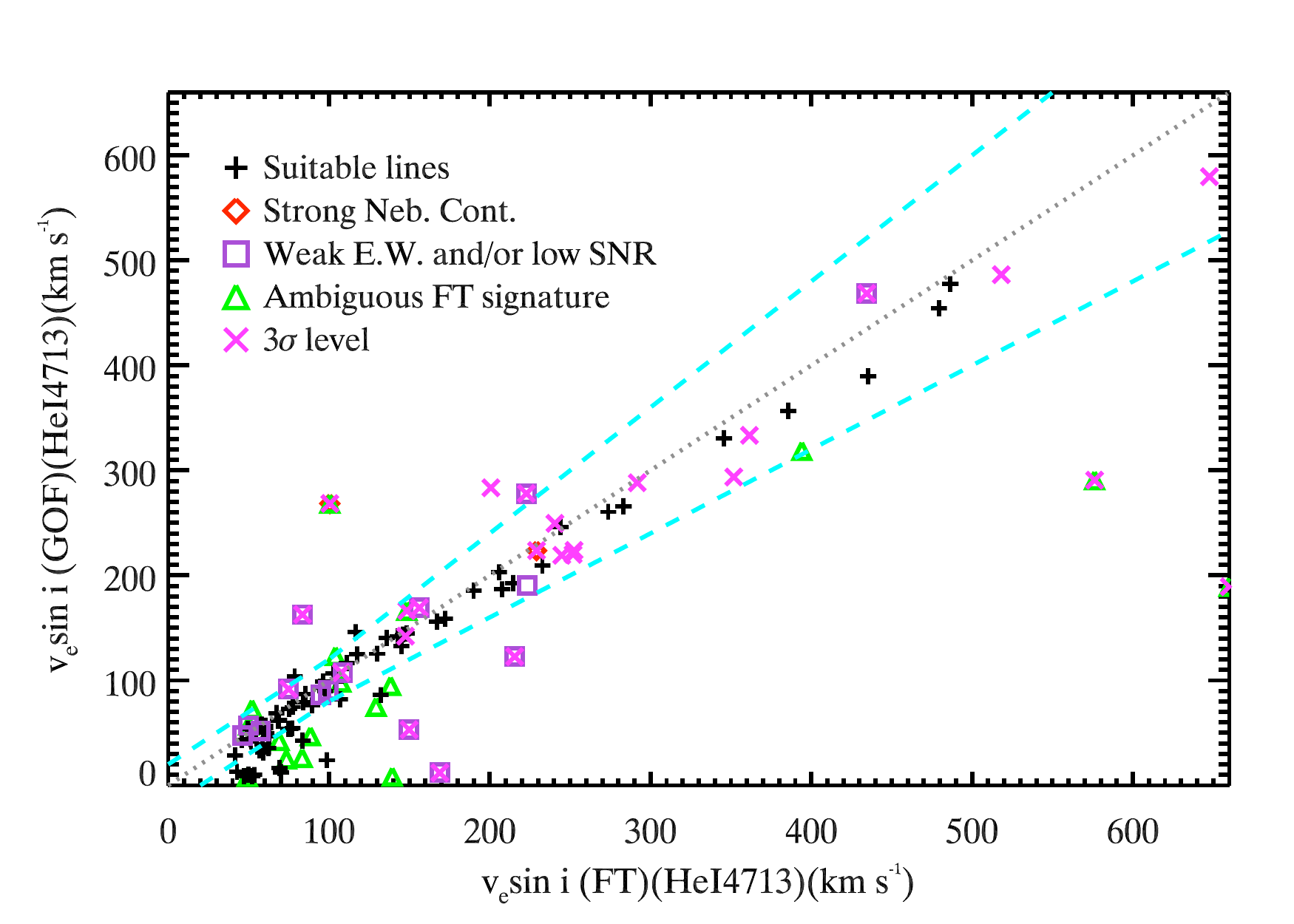}
\includegraphics[scale=0.5]{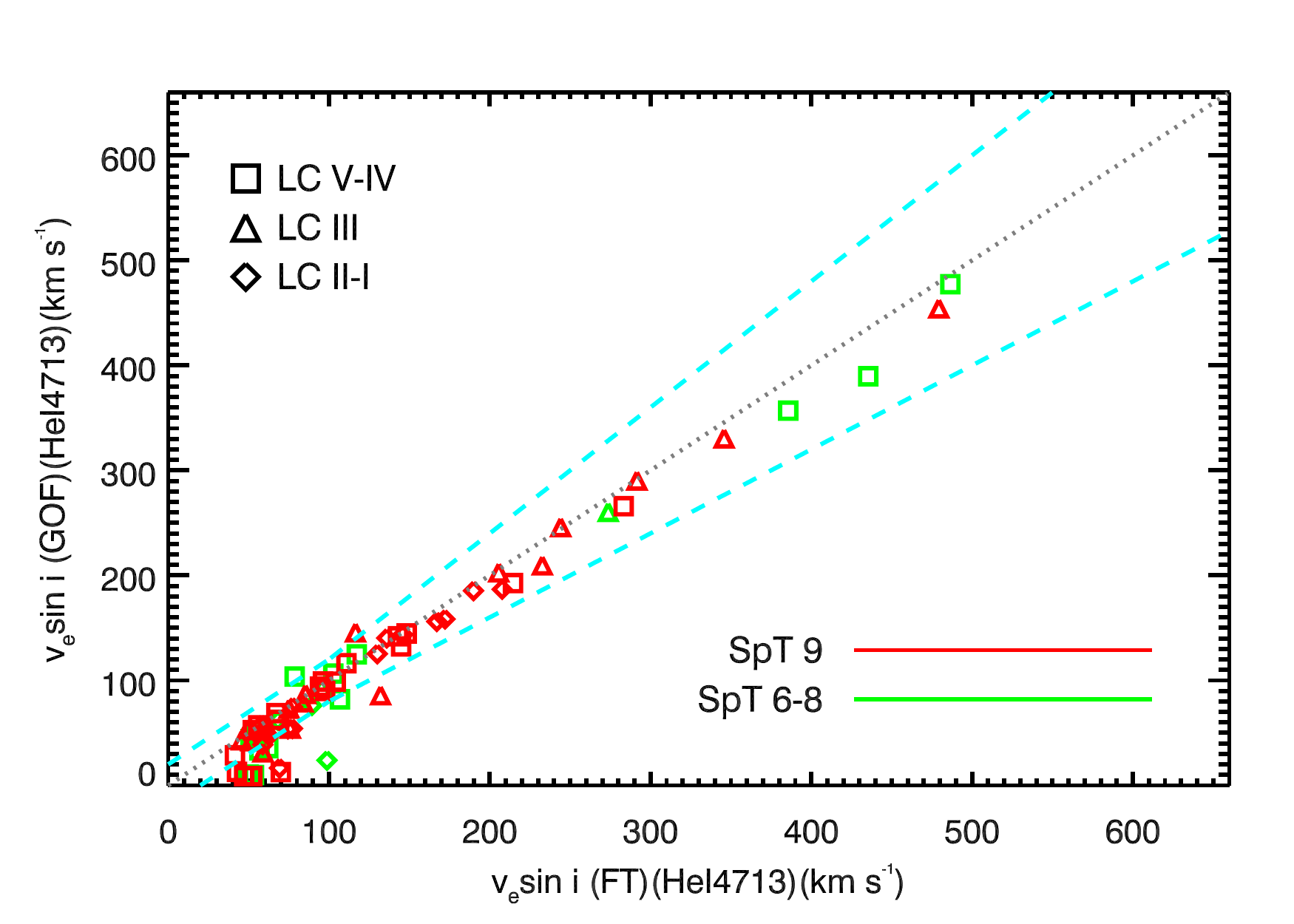}
\caption{Comparison of \vrot\ values obtained from the FT and GOF methods for \hel{i}{4713}. 
{\it Upper panel:} Different symbols indicate different data quality, identified in the legend. 
{\it Lower panel:} Same as above after discarding the lower quality data.
Information on the LC and SpT of the targets is provided by the symbol shapes and colors (see legend).  
In both panels, the dotted lines indicates the 1:1 relation. The dashed lines show the $\pm 20$~\kms\ and/or $\pm\, 20$\%,  
whichever is the largest, around the 1:1 relation. 
%[A color version is available online].
}
\label{fig:Comparison-FT-GOF}
\end{figure}
%%%%%%%%%%%%%%%%%%%%%%%%%%%%%%%%%%%%%%%%%%%%%%%%%%%%%%%%%%%%%%%

\subsection{Comparison of results from FT and GOF methods}\label{subsec:FT_GOF}

An example of FT and GOF \vrot\ measurements for the \hel{i}{4713}  line is presented in Fig.~\ref{fig:FT}
for two different data qualities. In the upper panel the derived \vrot\ from the
FT and the GOF methods are in agreement. In the lower panel, the noise is affecting the FT,
resulting in an erroneous \vrot\ value while the correct \vrot\ value is associated with the shallow FT
minimum at $\sim$ 200~\kms. The combined investigation of results from FT and GOF methods
hence  helps to identify the more ambiguous cases which must be explored in more detail. 

%\[Comparison of the 2 kinds of measurements]
The upper panel in Fig.~\ref{fig:Comparison-FT-GOF} compares \vrot\ values 
obtained from applying the FT and GOF methods to the \hel{i}{4713} 
line. This plot includes all stars in our sample in which the \hel{i}{4713} line
is detected and the \vrot\ (FT) measurements are above the resolution limit (i.e. 116 stars). 
In addition to the full set of good quality measurements, different symbols are used to highlight  
various cases: (i)  strong nebular contamination, 
(ii) low $\rm{S/N}$ ($<60$) and/or a comparatively weak diagnostic line (EW $\leq$ 50 m$\AA$), 
(iii) ambiguous first minimum in Fourier space  (this is, for example, the case 
for VFTS163 in Fig.~\ref{fig:FT}), and (iv) 
a central depth of the absorption-line profile that is smaller than three times the noise level. 
These explain all but 14 deviating points, which correspond to cases in which \vrot\ (GOF) 
is below the resolution limit and \vrot\ (FT)\,$>$\,\vrot\ (GOF). This behaviour  is expected  
when the rotational broadening
contribution to the line profile is similar to the intrinsic broadening 
and the $\rm{S/N}$ is not sufficiently high. In this situation the FT of the line may result in 
a spurious first zero, which is  moved to higher values of \vrot\ compared to the FT's true first zero \citep[see e.g.][]{simon}.
 In these specific cases \vrot\ (FT) measurements must be considered
as upper limits for the actual projected rotational velocity.
Save for one instance, these cases occur when \vrot\, $\leq$ 80~\kms.

Once the poorer quality measurements resulting from cases (i) to (iv) are left aside, the \vrot\ measurements for the 76 stars left in our comparison show a large degree of agreement between FT and GOF estimates over the full range of velocities, save for \vrot\ measurements close to the spectral resolution limit (Fig.~\ref{fig:Comparison-FT-GOF}, bottom panel). Similar conclusions about the large degree of agreement between FT and GOF estimates
can be drawn  for the other \hhel{i}  lines and from \Sil{iii}{4552}.
From now on the analysis will be presented excluding poor quality measurements.

\begin{figure}
\centering
\includegraphics[scale=0.5]{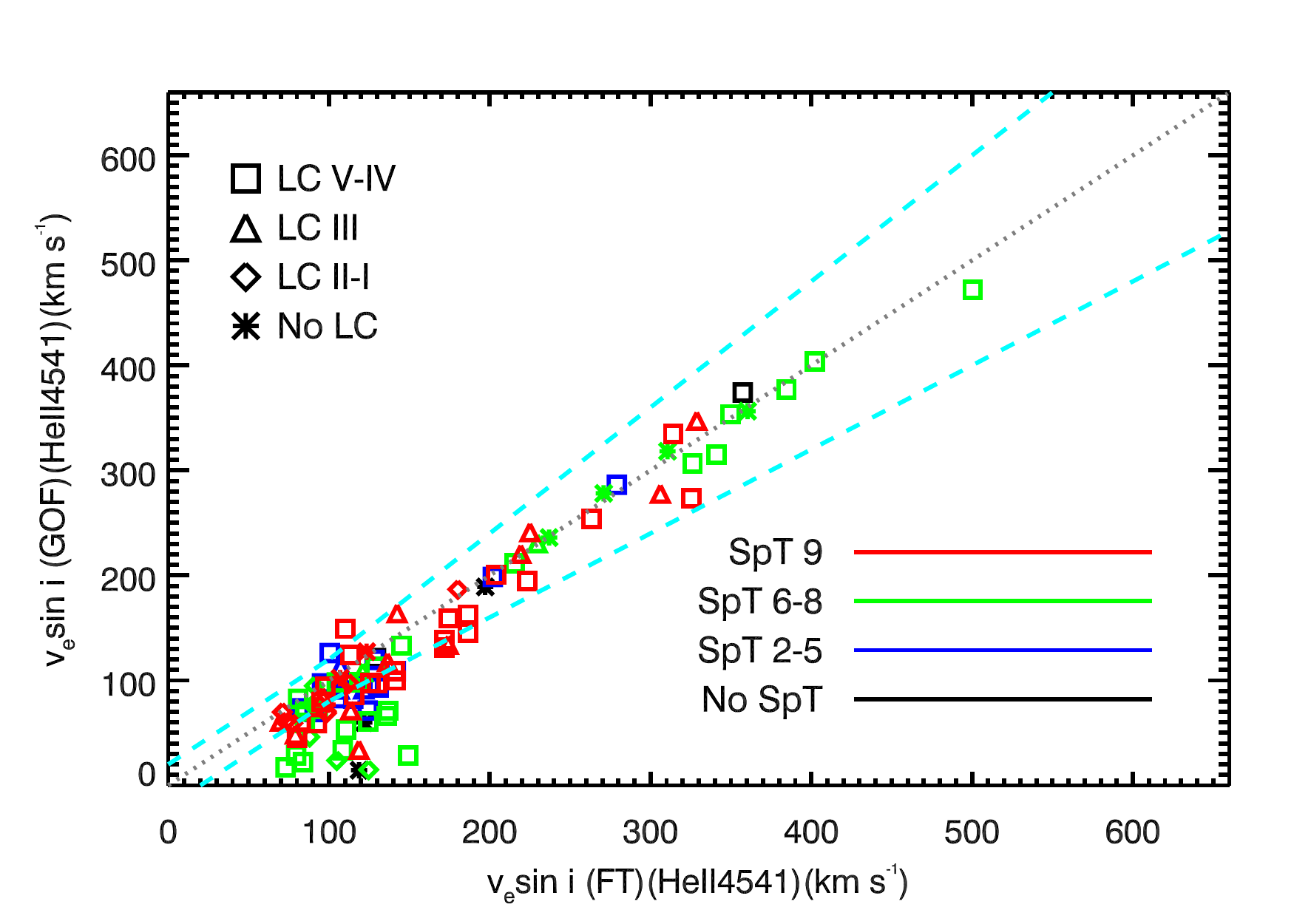}
\caption{Comparison of the \vrot\ measurements from the FT and GOF methods for \hel{ii}{4541}.
Information on the LC and SpT of the targets is provided by the symbol shapes and colors.
Lines have the same meaning as in Fig.~\ref{fig:Comparison-FT-GOF}. %[A color version is available online].
}

\label{fig:HeII_GOF_FT}
%\label{fig:HeliumII}
\end{figure}

The situation is, however, less clear for measurements based on \hel{ii}{4541} (Fig.~\ref{fig:HeII_GOF_FT}, 101 stars).
Above 150\,\kms\ the FT and GOF measurements agree to within 20\%. 
A significant dispersion is, however, observed at lower velocities, 
with FT resulting in systematically higher \vrot\ values than GOF. 
The origin of the observed systematics is discussed further in Section \ref{subsec:test_heii}.

\subsection{Comparison of results from the different diagnostic lines}\label{subsec:results}

In this section we compare the \vrot\, measurements obtained from different diagnostic lines (Table~\ref{table:2}).
Figs.~\ref{fig:Helium_siIII_comp} to \ref{fig:HeliumI_II} 
show different cases of these comparisons and Table~\ref{table:temp} summarizes the degree of agreement between the lines that have been investigated. 
Only \vrot\, measurements that fulfill the quality criteria outlined in Sect.~\ref{subsec:FT_GOF} are considered.

\subsubsection{\Sil{iii}{4552} and \hhel{i} lines}\label{subsec:SiIIIandHeI}

\begin{figure}
\centering
\includegraphics[scale=0.5]{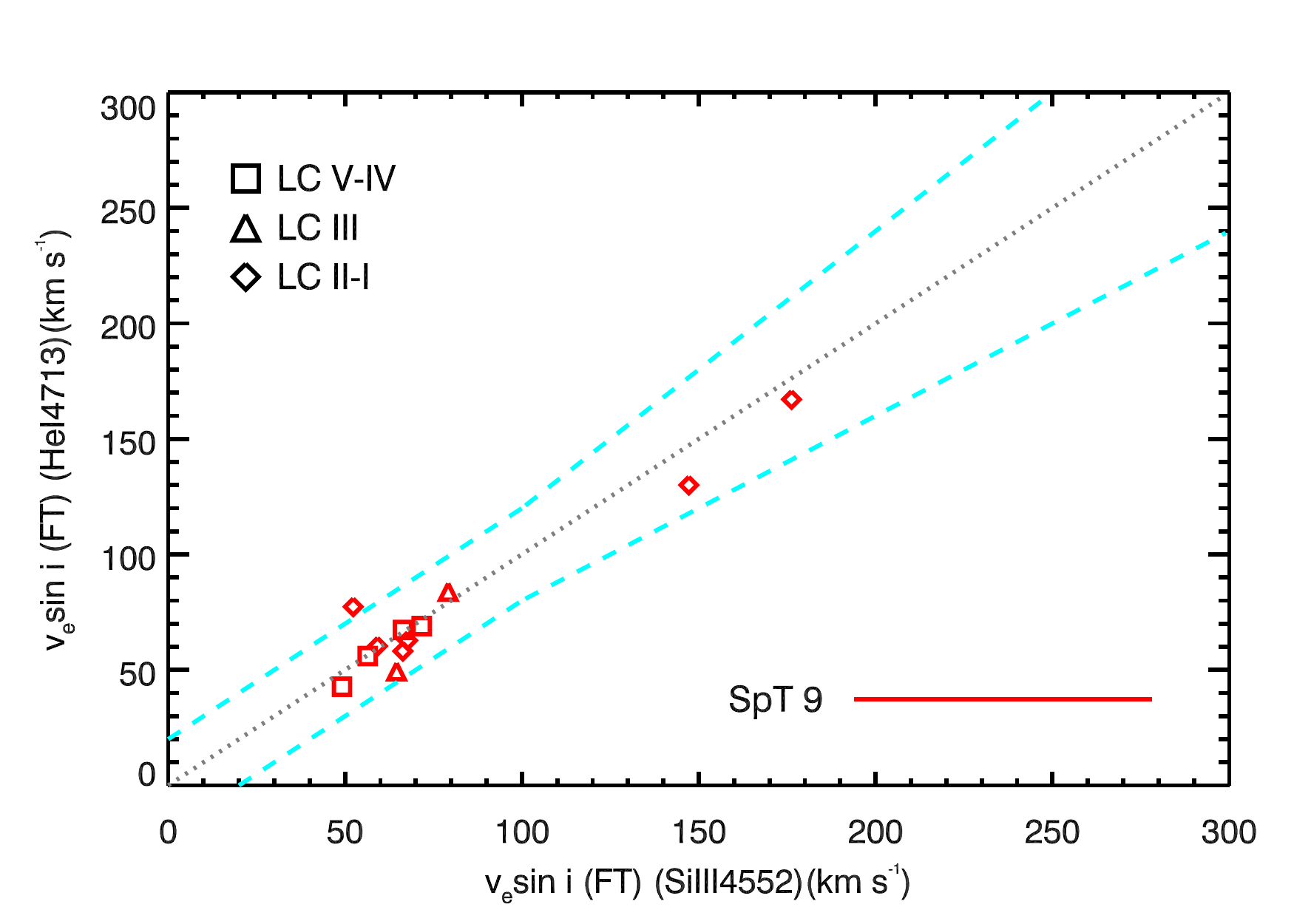}
\caption{
Comparison of \vrot\ values from the FT method for \hel{i}{4713} and \Sil{iii}{4552}.
Information on the LC and SpT of the targets is provided by the symbol shapes and colors.
Lines have the same meaning as in Fig.~\ref{fig:Comparison-FT-GOF}. %[A color version is available online].
}
\label{fig:Helium_siIII_comp}
\end{figure}

\begin{figure}
\centering
\includegraphics[scale=0.5]{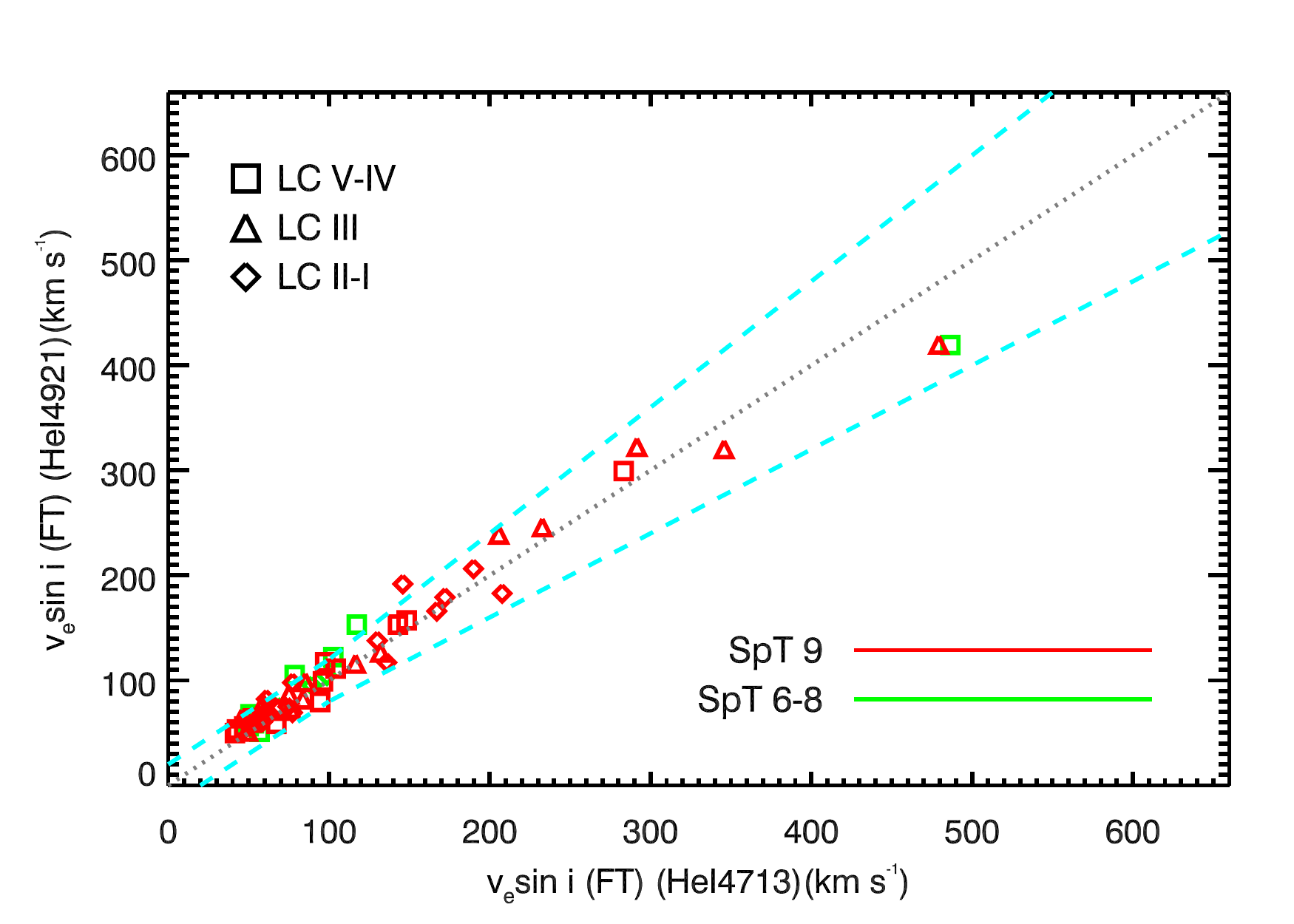}
\caption{
Comparison of \vrot\ obtained from the FT method for \hel{i}{4922} and \hel{i}{4713}.
Information on the LC and SpT of the targets is provided by the symbol shapes and colors.
Lines have the same meaning as in Fig.~\ref{fig:Comparison-FT-GOF}. %[A color version is available online].
}
\label{fig:Heliums_comp}
\end{figure}

Fig.~\ref{fig:Helium_siIII_comp} compares the \vrot\ values obtained from \Sil{iii}{4552} and 
 \hel{i}{4713} for 12 stars. Measurements 
from both lines agree within $\pm$ 20 \%\  and/or $\pm$ 20 \kms, whichever is the largest, in over 90\% of the cases.
A similar degree of agreement is observed between FT measurements of \hel{i}{4713} and \hel{i}{4922} (Fig.~\ref{fig:Heliums_comp}) : 53 of the 58 stars displaying both lines again  agree within $\pm$ 20\% and/or $\pm$ 20~\kms, hence about 91\% of the sample. % where such comparison is possible.
We conclude that the
comparison between \vrot\ measurements obtained from \SSil{iii} and from different \hhel{i} lines (Figs.~\ref{fig:Helium_siIII_comp}, \ref{fig:Heliums_comp} and Table~\ref{table:temp}) reveals no systematic differences. This justifies the grouping of  all the measurements from stars in Group A, irrespectively of the diagnostic line from which they have been obtained.

\subsubsection{\hel{ii}{4541}}\label{subsec:test_heii}

As discussed in Sect.~\ref{subsec:FT_GOF}, the FT and GOF measurements obtained from the \hel{ii}{4541} line 
show a general  agreement well within the 20\% for \vrot\, $\geq$ 150~\kms\ (Fig. \ref{fig:HeII_GOF_FT}).
However, the larger dispersion and systematics observed below 150~\kms\ cast some doubts on the reliability of the 
 \hel{ii}{4541} measurements for slow rotators.
In this section we further explore this issue by comparing the  \hel{ii}{4541} rotational velocities
with those obtained from \hhel{i} lines for those stars in Group A that display both \hhel{i} and \hhel{ii} 
lines. This comparison will allow us to decide which of our two methodologies provides the more reliable 
\vrot\ information for Group B stars.

\begin{figure}
\centering
\includegraphics[scale=0.5]{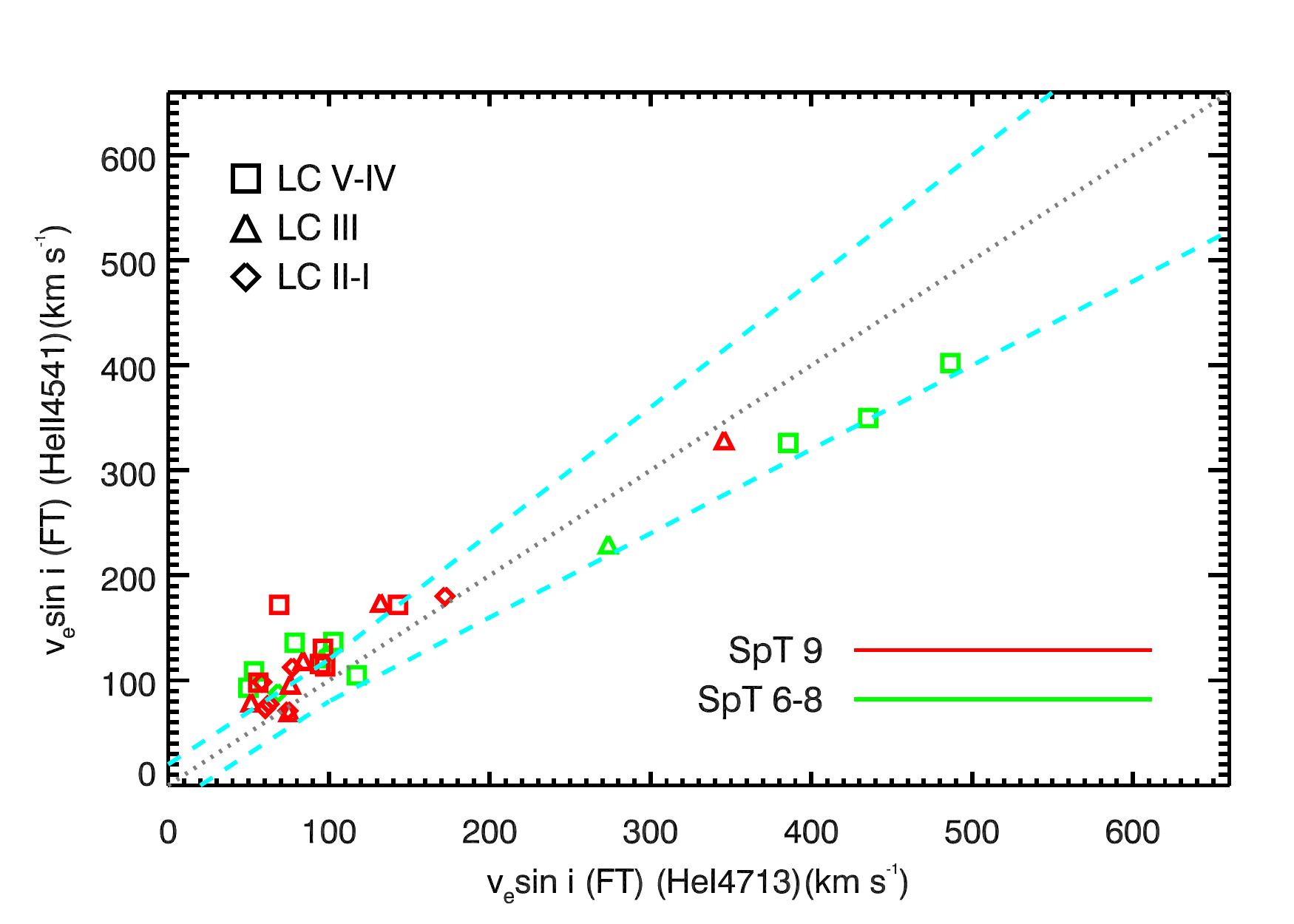}
\includegraphics[scale=0.5]{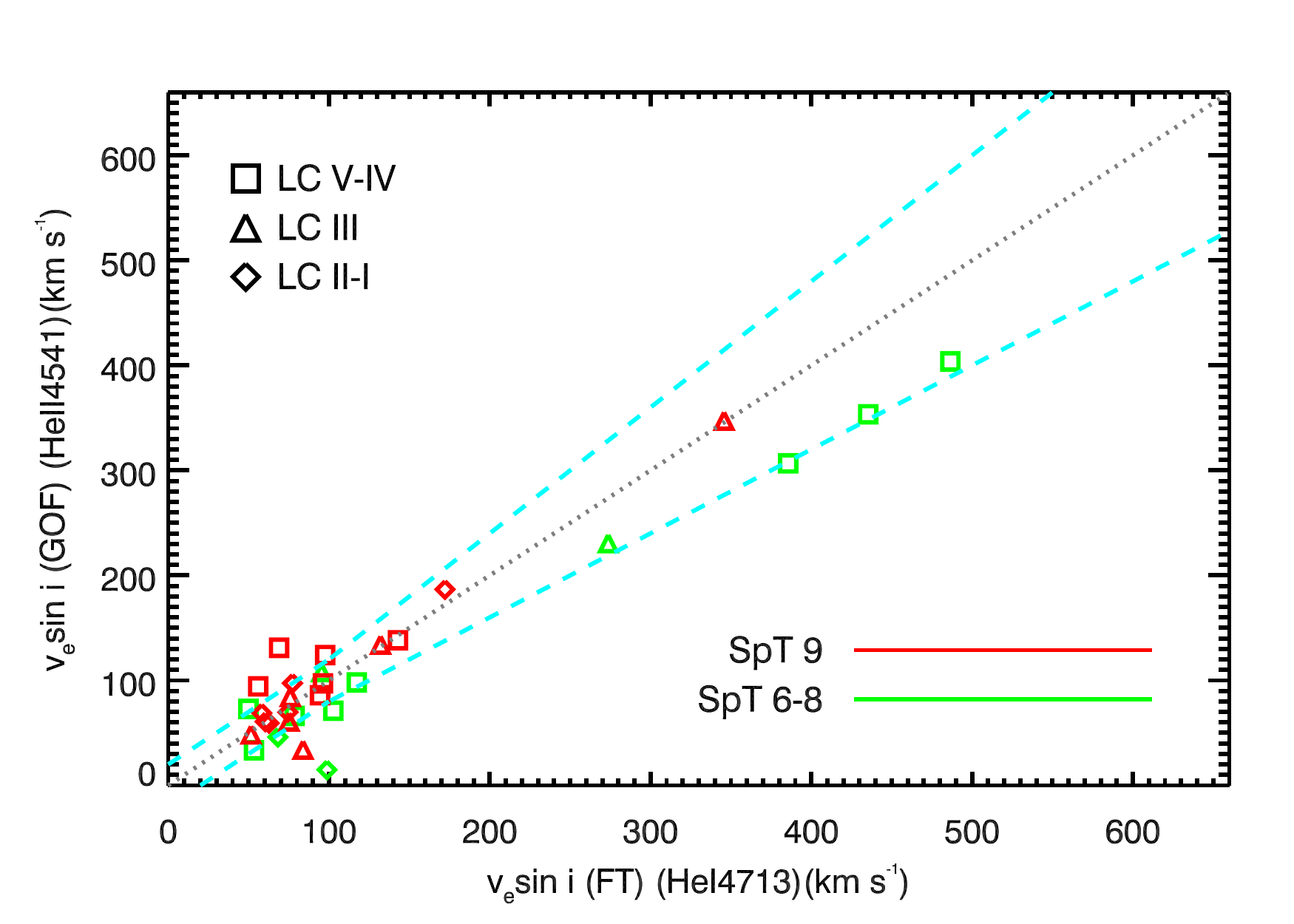}
\caption{Comparison of \vrot\ measurements from  \hel{i}{4713} and \hel{ii}{4541}. 
{\it Upper panel:} both diagnostic lines analysed using the FT method. {\it Lower panel:} \hel{ii}{4541} analyzed using the GOF and \hel{i}{4713} using the FT.
Information on the LC and SpT of the targets is provided by the symbol shapes and colors.
Lines have the same meaning as in Fig.~\ref{fig:Comparison-FT-GOF}. %[A color version is available online].
}
\label{fig:HeliumI_II}
\end{figure}

Fig.~\ref{fig:HeliumI_II} (upper panel) compares \vrot\, measured from \hel{ii}{4541} and from \hel{i}{4713} using the FT method.
In the low-velocity domain ($\vrot\, < 150$~\kms), FT values from \hel{ii}{4541} are systematically larger by about 30 \kms, on average, 
with a dispersion of about 23~\kms\ compared to \hhel{i} FT measurements.
In this domain, the contribution of Stark 
broadening to the \hel{ii}{4541} line causes the power of the first lobe of the FT 
to diminish. As a consequence, the first zero is easily  hidden within the white 
noise if the $\rm{S/N}$ is not high enough. The method then returns \vrot\ values corresponding to the first FT zero that peaks out of the noise. This artificially moves the  measurement to higher \vrot\ values.

As illustrated in the lower panel of Fig.~\ref{fig:HeliumI_II}, the impact 
of the Stark broadening on our \vrot\, measurements is partially mitigated 
by using the GOF rather than the FT method for \hel{ii}{4541}. Although large deviations still
occur in some cases, the average systematic difference in the measurements between 
\hhel{i} and \hhel{ii} drops off to 3 \kms, i.e.\ negligible in comparison with our expected accuracy. Using the GOF instead of the FT for \hhel{ii} measurements thus allows to avoid a systematic bias in the measurements, leaving the derived \vrot\ values only affected by random, though relatively large, uncertainties.

%%%%%%%%%%%%%%%%%%%%%%%%%%%%%%%%%%%%%%%%%%%%%%%%%%%%%%%%%%%%%%%%%%%%%%%%
\begin{table}
\caption{Comparison between different diagnostic lines. Col.~1:
 figure index. Col.~2: total number of stars in the comparison. Cols.~3 and 4: diagnostic
 lines compared. Col.~5: number and fraction of stars in the comparison for which the measurements agree within 
20\%  and/or  20 \kms, whichever is the largest. Measurements from the FT method are used, unless specified otherwise.
}
\label{table:temp}     
\label{table:1a}  
\centering              
\begin{tabular}{c r l l r}        
\hline\hline\\[-8pt]                 % inserts double horizontal lines
Fig. &      Number   & Diagnostic & Diagnostic & $\leq$ 20 \kms \\    % table heading 
       &of stars    & line 1     &   line 2   & or $\leq$ 20 \%\\    % table heading 
\hline\\[-8pt]                        % inserts single horizontal line
%\ref{fig:Comparison-FT-GOF}  b)  &  71  & \hel{i}{4713}  &   \hel{i}{4713} (GOF)  & 51 (72\%)  \\  
%              -               & 112  & \hel{i}{4387}  &   \hel{i}{4387} (GOF)  & 68 (61\%)  \\    
%              -               &  62   & \hel{i}{4471}  &   \hel{i}{4471} (GOF)  & 30 (48\%)  \\    
%              -               &  21  &\Sil{iii}{4552}   &    \Sil{iii}{4552} (GOF) & 18 (86\%)\\   
%              -               &  75   & \hel{i}{4922}  &   \hel{i}{4922} (GOF)  & 44 (59\%)  \\    
%\ref{fig:HeII_GOF_FT}         & 101  & \hel{ii}{4541} &   \hel{ii}{4541} & 59 (58\%)\\ 
%\hline 
\ref{fig:Helium_siIII_comp} &  12  &\Sil{iii}{4552} &   \hel{i}{4713}  & 11 (92\%)\\  
                  -           &  17  &\Sil{iii}{4552} &   \hel{i}{4387}  & 14 (82\%)\\  
                  -           &    8  &\Sil{iii}{4552} &   \hel{i}{4471}  &   7 (88\%)\\  
                  -           &  14  &\Sil{iii}{4552} &   \hel{i}{4922}  & 13 (93\%)\\    
 \hline\\[-8pt]                                  
 \ref{fig:Heliums_comp}        &  58  & \hel{i}{4713}  &   \hel{i}{4922}  & 53 (91\%)\\  
               -              &  66  & \hel{i}{4387}  &   \hel{i}{4713}  & 57 (86\%)\\  
               -              &  44  & \hel{i}{4471}  &   \hel{i}{4713}  & 39 (86\%)\\  
               -              &  52  & \hel{i}{4387}  &   \hel{i}{4471}  & 43 (87\%)\\  
               -              &  72  & \hel{i}{4387}  &   \hel{i}{4922}  & 67 (93\%)\\                 
               -              &  46  & \hel{i}{4387}  &   \hel{i}{4471}  & 41 (89\%)\\  
\hline\\[-8pt]                
\ref{fig:HeliumI_II} a)         &  31  & \hel{i}{4713}  &   \hel{ii}{4541}  & 14 (45\%)\\  
              -                         &  40  & \hel{i}{4387}  &   \hel{ii}{4541}  & 18 (45\%)\\   
              -                         &  31  & \hel{i}{4471}  &   \hel{ii}{4541}  & 20 (65\%)\\   
%              -                         &    7  &\Sil{iii}{4552} &   \hel{ii}{4541}  &  3 (43\%)\\  
              -                         &  31  & \hel{i}{4471}  &   \hel{ii}{4541}  & 15 (48\%)\\     
\hline\\[-8pt]
\ref{fig:HeliumI_II} b)        &  31  & \hel{i}{4713}  &   \hel{ii}{4541}(GOF)  & 21 (68\%)\\ 
             -                         &  40  & \hel{i}{4387}  &   \hel{ii}{4541}(GOF)  & 21 (53\%)\\   
              -                         &  31  & \hel{i}{4471}  &   \hel{ii}{4541}(GOF)  & 23 (74\%)\\   
%              -                         &    7  &\Sil{iii}{4552} &   \hel{ii}{4541}(GOF)  &  4 (57\%)\\  
              -                         &  31  & \hel{i}{4471}  &   \hel{ii}{4541}(GOF)  & 23 (74\%)\\  
\hline                                   %inserts single line
\end{tabular}
\end{table}
%%%%%%%%%%%%%%%%%%%%%%%%%%%%%%%%%%%%%%%%%%%%%%%%%%%%%%%%%%%%%%%%%%%%%%%%

\begin{table*}
\caption{Measured values of \vrot\ (in \kms) for every star and diagnostic line.  The `-' sign indicates that the 
line has not been used.
The full version of the table is available online.}              % title of Table
\label{table:2}      % is used to refer this table in the text	
\centering                          % used for centering table
\begin{tabular}{rrrrrrr}        % centered columns (4 columns)
\hline\hline\\[-8pt]                 % inserts double horizontal lines
VFTS&\hel{i}{4387}&\hel{i}{4471}&\Sil{iii}{4552}&\hel{i}{4713}&\hel{i}{4922}&\hel{ii}{4541}\\    % table heading 
\hline\\[-8pt]                        % inserts single horizontal line
014 &    87 &   104 &   -  &   94 &  79     &    85  \\  
016 &    -   &     -   &   -  &    -  &    -      &     94  \\  
021 &    52 &    67  &   -  &   40 &  57     &     45  \\  
046 &  160 &  166  & 176 &  167 &  166 &    -  \\  
051 &  413 &    -    &   -   &   -    &  -      &      - \\  
064 &    99 &  113  &   -  &   99   &  104 &     15  \\  
065 &  164 &    -    &   -   &   -    &    -   &     200  \\  
067 &    54 &     -   &    -  &   -   &   -    &     -  \\  
070 &  126 &  129  &   -   &   -   &   -    &   -   \\  
%072 &    -   &    -     &   -  &   -   &  -     &     200  \\  
...    & ...   & ...  &  ... &...  &... &    ... \\[2pt]
\hline                                   %inserts single line
\end{tabular}
\end{table*}

In the high-velocity domain (\vrot\, $>$ 250~\kms) a systematically lower \vrot\ is obtained from \hel{ii}{4541} compared to 
\hel{i}{4713}. 
The good agreement found between 
FT and GOF estimates for \hel{ii}{4541} in this \vrot\ domain suggests that 
the difference is real and not related to a methodological bias.  
Fast rotation produces equatorial stretching of stars, which in turn induces a non-uniform 
surface gravity and temperature distribution %; namely the gravity darkening effect 
\citep[][]{zeipel}. In the particular case of fast rotating O-type stars, this % gravity darkening implies 
means that the poles are hotter than the equator. 
As a consequence, photospheric regions closer to the pole (resp.\ equator) will contribute more to the formation of \hhel{ii} (resp.\ \hhel{i}) lines. Projected rotational velocities derived from \hhel{ii} lines are thus expected to be lower than those from HeI lines, as observed in Fig.~\ref{fig:HeliumI_II}.

\subsection{Measurement uncertainties}\label{subsec:errors}

The methods that we have adopted to estimate \vrot\ do unfortunately not provide measurement errors. We have estimated the typical accuracy of our measurements by examining the dispersions observed between measurements based on independent lines.  We outline here the main conclusions. 
At low rotational velocities, absolute uncertainties provide a more sensible estimate of the true measurement error. At high 
rotational velocity, however, uncertainties are better expressed in relative terms. % and we adopt this distinction in the following.

For Group A stars,
%At velocities regime less than $100$~\kms\ 
the root mean square (rms) dispersion between measurements from the different \hhel{i} and \SSil{iii} lines amounts to 
about 10\,\kms\ for projected rotational velocities below 100\,\kms\ (see Figs.~\ref{fig:Helium_siIII_comp}-\ref{fig:Heliums_comp}
and Table~\ref{table:1a}). 
It is of the order of 10\%\ or better above 100~\kms.
For Group B stars, the rms dispersion in the lower panel of the same figure indicates uncertainties not exceeding 30\,\kms\
below $\vrot =  200$\,\kms, and not exceeding 10\% above 200\,\kms\ (see lower panel Fig.~\ref{fig:HeliumI_II}).

% suggests that values better than 30~\kms\ and 10\%, 
%below and above 200~\kms\ respectively, provide an appropriate representation of the statistical uncertainties for stars 
%in Group B.  

As pointed out in the previous section, our \hhel{ii} measurements may underestimate the true rotational velocity of the stars for \vrot\ $\geq$ 250\, \kms. The five stars in Fig.~\ref{fig:HeliumI_II} for which this is relevant indicate that the effect is probably of the order of 15--20\%. 
Similarly, gravity darkening may cause our measurements to underestimate the true \vrot\ at large rotational velocity 
\citep[see also the discussion in][hereafter \citetalias{dufton}]{dufton}.
Though both effects are not negligible, it will not affect our conclusion on the presence of a well-populated high-velocity
tail in the rotational velocity distribution. Correcting these values would even strengthen that outcome.
%correcting for this effect would only make our conclusion on the presence of a well-populated high-velocity tail stronger so that our main results are left unaffected.

As a last point, we note that most of the comparisons performed in this section concern mid- and late-O stars. The \hhel{i} lines in early-O stars (O2-O5) are too weak to provide reliable \vrot\ measurements, depriving us from an anchor point to test the quality of \hhel{ii}-based \vrot\ measurements. In Appendix~\ref{subsec:NV} we compare \vrot\ measurements obtained from \hel{ii}4541 with those obtained from \Nl{v}{4604} for a small set of stars. 
%While metallic \NNl{v} is not widely used to obtain \vrot, it should be little affected by Stark broadening and thus provides an alternative anchor point to test \hhel{ii}-based measurements. 
We find that above 80~\kms, \hhel{ii} and \NNl{v} agree to within 10\%. This adds confidence in the quality of \vrot\ measurements for early-type Group B stars in that regime. The quality of the \hhel{ii}-based  \vrot\ of early-O stars with $\vrot \lesssim 80$~\kms\ cannot be investigated further with the current data set and we thus adopt a formal uncertainty of 30~\kms\ for these 15 stars.

\subsection{Strategy for obtaining \vrot\ estimates}
\label{subsec:strategy}

Based on the analysis presented in Sect.~\ref{subsec:FT_GOF} and
\ref{subsec:results}  we have adopted the following strategy to obtain the final 
 \vrot\ measurements (see Table \ref{table:3}) for the full sample of 216 O-type 
stars: % considered in this paper:

\begin{itemize}
 \item[--] For those stars in Group A with FT measurements below 40\,\kms, we adopt 
an upper limit of 40\,\kms, and include them in the 0\,--\,40 bin of the \vrot\ histograms (see Sect.~\ref{subsec:dist}).

 \item[--] For those stars in Group A with FT measurements above 40\,\kms, we compute
a non-weighted average of the individual values obtained from the available \hhel{i} and \SSil{iii}
lines (see Table \ref{table:2}).

 \item[--] For all stars in Group B where the \hel{ii}{4541} diagnostic line is available, we  adopt the GOF measurement obtained from that line. If the obtained value is below 40\,\kms, we consider the object below our resolution limit and assign it to the lowest velocity bin in our histograms.

 \item[--] For the remaining 14 stars for which the above diagnostics can not be applied, we use the wings of the lines and compare them to synthetic spectra as described in Sect.~\ref{subsec:diagnostic_lines}.
\end{itemize}

While the GOF measurements of \vrot\ based on \hhel{ii} lines are good enough for the purposes of this study, 
individual values must be handled with care, as the measurements have a large dispersion. In particular, some of the Group B stars included in the first three bins of the \vrot\
distributions presented in next sections may move between adjacent bins due to sizeable
uncertainties. Because we verified earlier that there are no systematic effects, the larger uncertainties for some stars in Group B do not affect the overall distributions presented in this paper.
 
%%%%%%%%%%%%%%%%%%%%%%%%%%%%%%%%%%%%%%%%%%%%%%%%%
\begin{table}
\caption{Averaged \vrot\, measurements and their 1$\sigma$ dispersion. 
The number of lines used and the category are also indicated.
The full version of the table is available online.}             % title of Table
\label{table:3}      % is used to refer this table in the text
\centering                          % used for centering table
\begin{tabular}{c c c c c }        % centered columns (4 columns)
\hline\hline\\[-8pt]                  % inserts double horizontal lines
 VFTS & Group & \vrot\, (\kms) & $\sigma$ (\kms) & \# of lines \\[1pt] \hline\\[-8pt]
014  &  A   &  91    &   10    &    4\\     
016  &  B   &  94    &   30    &    1\\      
021  &  A   &  54    &   11    &    4\\      
046  &  A   & 167    &    5    &    5\\      
051  &  A   & 413    &   20    &    1\\      
064  &  A   & 104    &    6    &    4\\      
065  &  A   & 164    &   20    &    1\\      
067  &  A   &  54    &   20    &    1\\     
070  &  A   & 128    &   20    &    2\\
%072 	& B 	& 200    &	  30 	 &   1\\
...     & ...   & ...         &  ...      &  ... \\[2pt]  
\hline                        % inserts single horizontal line
\end{tabular}
\end{table}
%%%%%%%%%%%%%%%%%%%%%%%%%%%%%%%%%%%%%%%%%%%%%%%%%

\section{The \vrot\ distribution}\label{subsec:dist}

\begin{figure}
\centering
\includegraphics[scale=0.5]{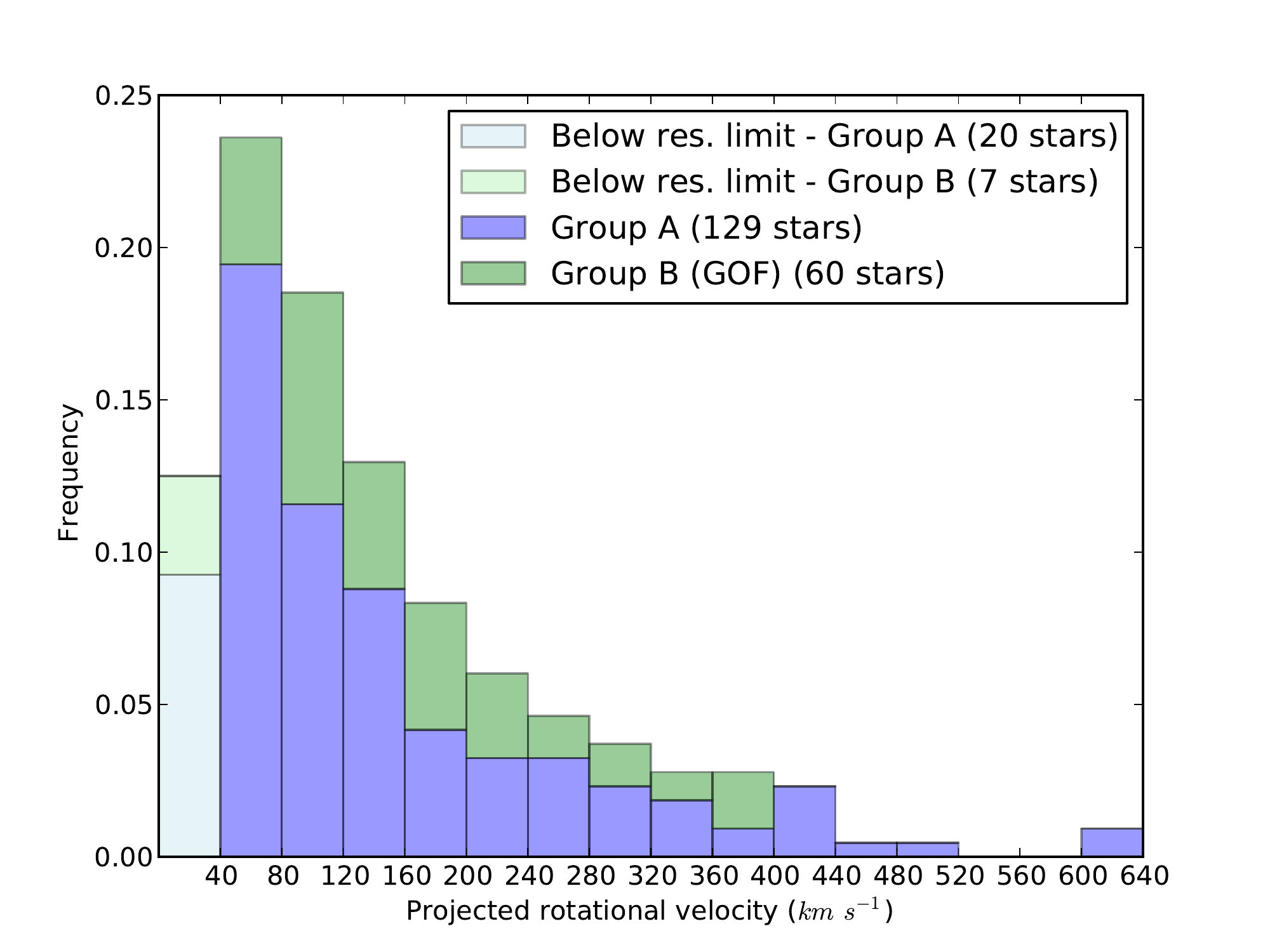}
\caption{Histogram of the projected rotational velocities of our sample of 216 O-type stars in 30~Dor.
%[A color version of the plot is available in the electronic version of the journal.]
}
\label{fig:vrotsini_dist}
\end{figure}

In this section, we %use the \vrot\ measurements obtained following the strategy outline in Sect.~\ref{subsec:strategy} to
construct and investigate the  \vrot\ distributions of the overall O-type star population within 30 Dor as well as within sub-populations of stars in our sample. We approximate the probability density function (pdf) using histograms with a bin size of 40~\kms. Such a bin size is consistent with our resolution limit (see Sect.~\ref{sec:sample}) and is large enough to mitigate the effect of measurement uncertainties on the appearence of the histograms.
%probability density function (pdf)
Fig.~\ref{fig:vrotsini_dist} shows the overall \vrot\ distribution. It is dominated by a clear peak at fairly low \vrot\  and a well-populated high-velocity tail which extends continuously to 500~\kms. With a measured \vrot\ of 609 and 610~\kms, VFTS~285 and VFTS~102 \citep{dufton1}, complete the projected rotational distributions at extreme rotation rates.

In the following sub-sections, we analyse the \vrot\ distribution as a function of spatial location,
LC and SpT. We use  Kuiper tests \citep[KP,][]{kuiper} to search for significant differences between the considered
sub-populations. Results are reported in Table~\ref{table:4}.
Specifically, KP tests allow us to test the null hypothesis that two observed distributions are randomly drawn
from the same parent population. Compared to the more widely used Kolmogorov-Smirnov test \citep[KS,][]{smirnov}, the KP test has the advantage to be similarly sensitivity to differences in the tails and around the median of the distributions while the KS test is known to be less sensitive if differences are located in the tails of the distribution. To support our discussion of the outcomes of the KP tests, we also show the cumulative distribution functions (cdf) because they provide a more direct view of the location of the differences between the distributions.   

%%%%%%%%%%%%%%%%%%%%%%%%%%%%%%%%%%%%%%%%%%%%%%%%%
%%%%%%%%%%%%%%%%%%%%%%%%%%%%%%%%%%%%%%%%%%%%%%%%%

\subsection{Spatial variations}
\label{subsec:Spatialvar}

\begin{figure}
\centering
\includegraphics[scale=0.5]{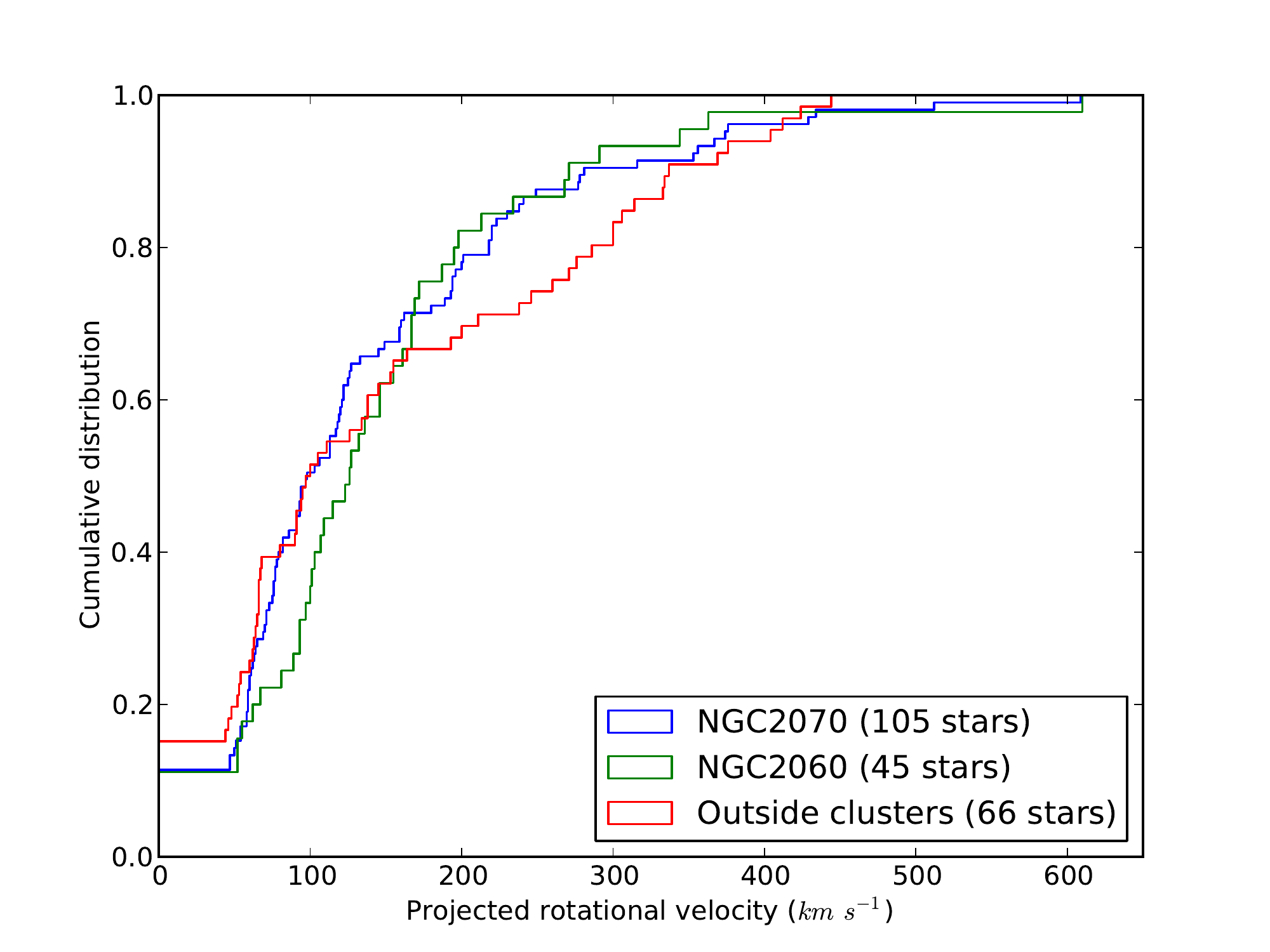}
\includegraphics[scale=0.5]{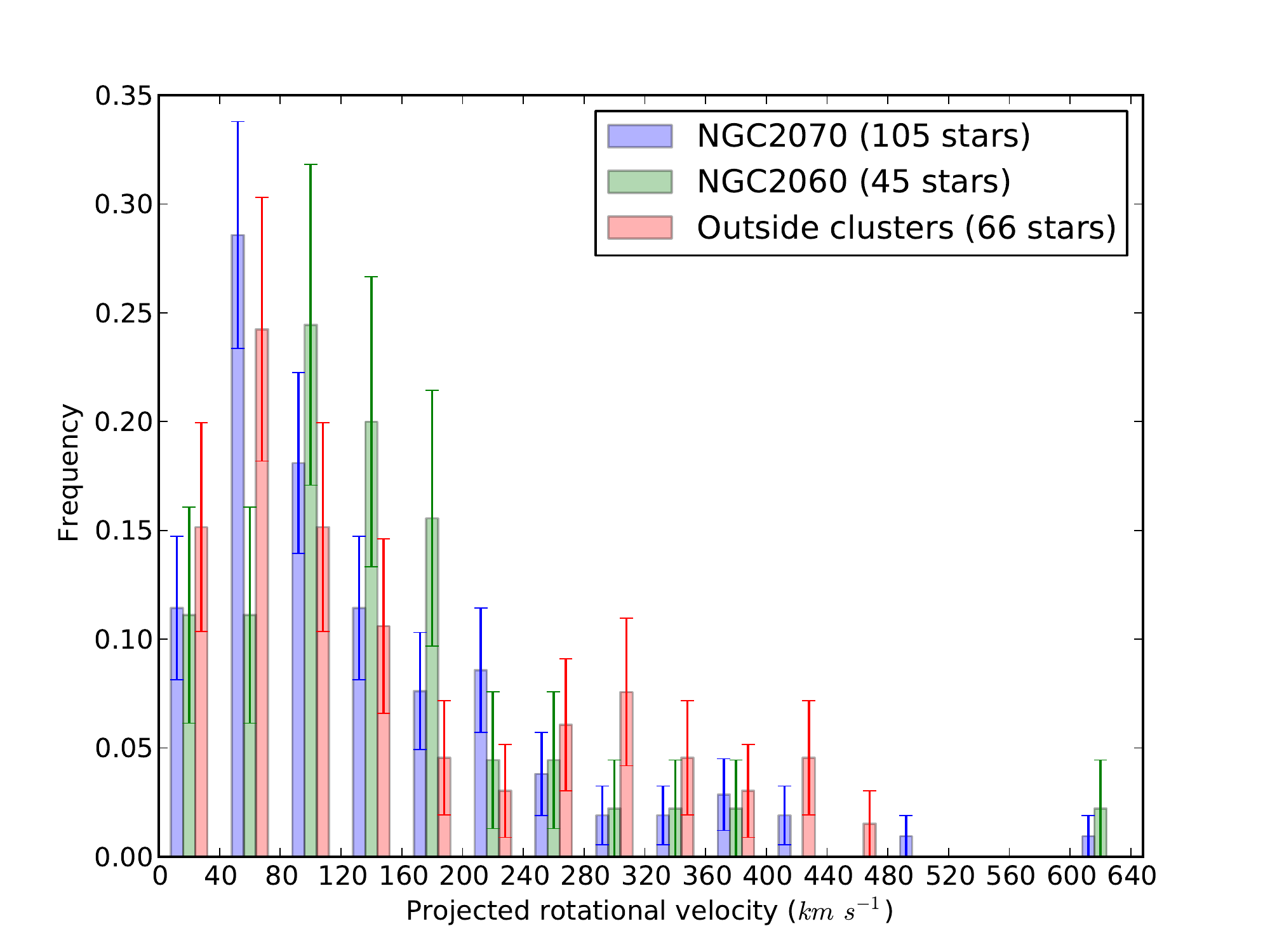}
\caption{
Cumulative (upper panel) and frequency (lower panel, with poisson error bars) distributions of the projected rotational velocities of the O-type stars
for three spatially selected groups.}
\label{fig:freq_clusters}
\end{figure}

Figure~\ref{fig:freq_clusters} shows the \vrot\ distributions  for stars in the NGC\,2070 and NGC\,2060 
clusters and for stars outside these two clusters. 
Qualitatively all three distributions are similar with
a  peak in the low-velocity region and a high-velocity tail. 
The distributions for stars in the two clusters are statistically compatible.
A KP test, however, indicates that the \vrot\, distribution of stars outside the clusters shows differences with the NGC\,2070 and NGC\,2060 distributions, at the 9 and 3\% significance level, respectively. 
For NGC\,2070 the difference manifests itself in the high-velocity regime,
the fraction of rapid rotators being larger outside of the clusters.
For NGC\,2060 both the low- and high-velocity region contribute
to the significance of the differences detected  by the KP test.

The difference in age between the two clusters, hence in the evolutionary stage of their O-type star population, does not seem to have strongly affected the \vrot\ distribution of NGC\,2060 compared to the younger NGC\,2070. This suggests that standard evolutionary effects, such as spin down through wind mass-loss and/or envelope expansion are modest.
The presence of a larger fraction of fast rotators outside clusters may, however, be linked to binary evolution if the field is relatively overpopulated with post-interaction objects. 
Binary interaction is indeed expected to produce rapidly rotating stars either through mass and angular momentum transfer to the secondary star during Roche lobe overflow or through the merging of the two components 
\citep{selma}. A relative overpopulation of the field can be obtained in two ways. Some of these binary products may have 
 %originally been born in previous massive star formation events and have 
 outlived their co-eval single star counterparts thanks to the rejuvenation effects resulting from the interaction process. Alternatively,  post-interaction systems may have been  ejected from their natal cluster when the primary experienced its supernova explosion. The correlation between rapid spin and large radial velocity of the field stars identified in the VFTS O-type
star sample (Sana et al. in prep.) is an argument in favor of the latter scenario. This will be the topic of a separate investigation within the VFTS series of papers.

\begin{figure}
\centering
\includegraphics[scale=0.5]{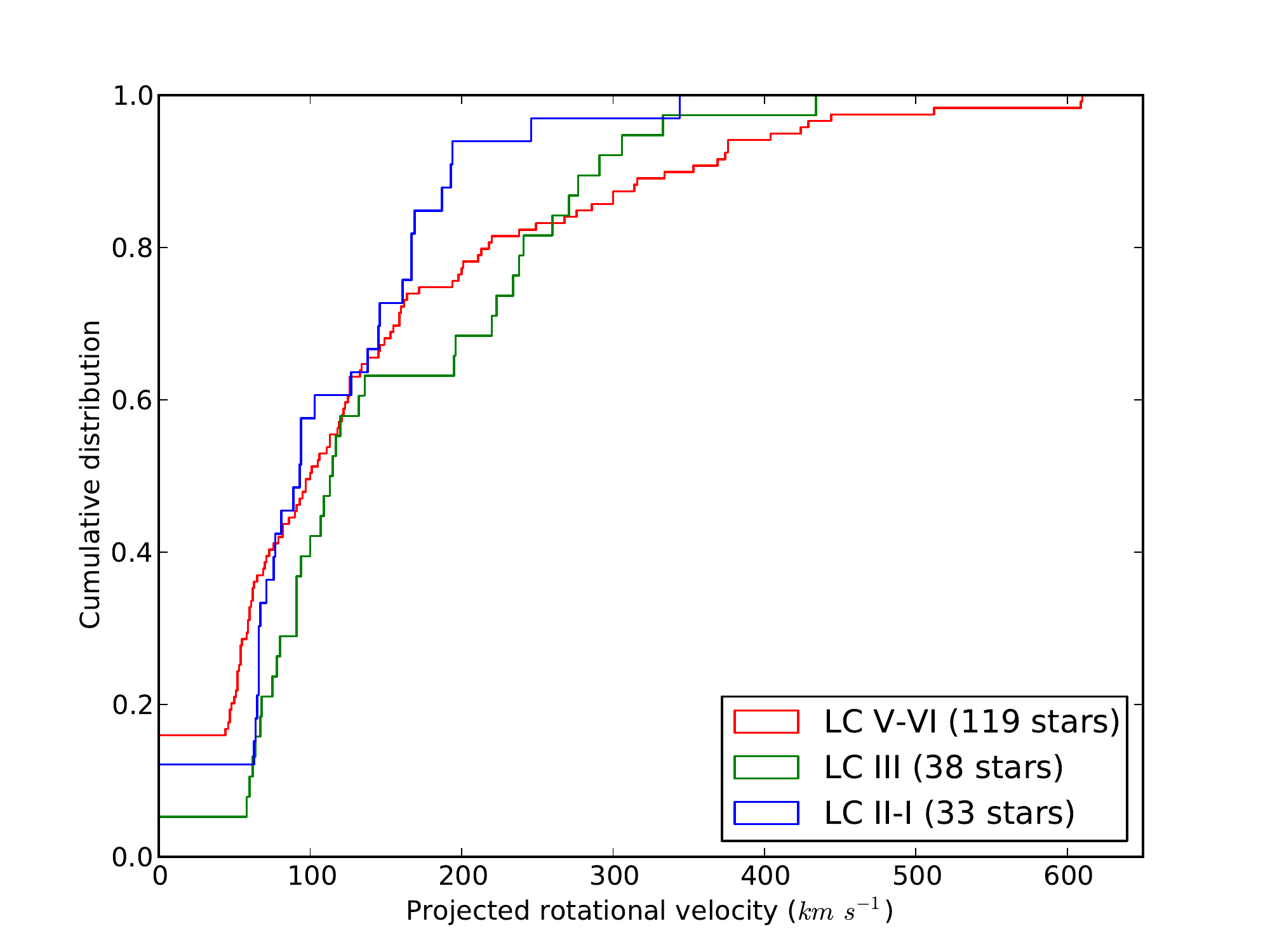}
\includegraphics[scale=0.5]{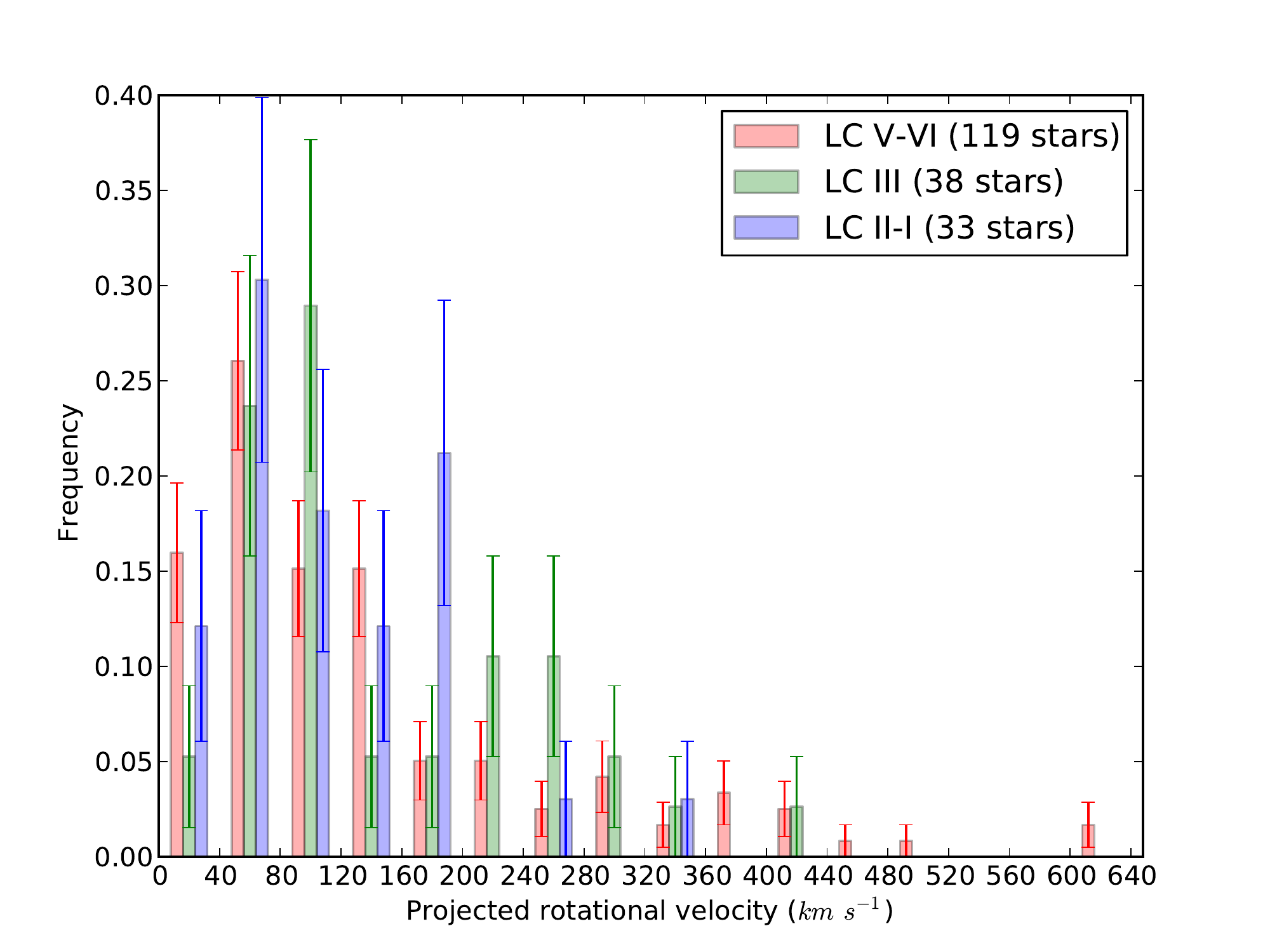}
\caption{Cumulative (upper panel) and frequency (lower panel, with poisson error bars)  distributions of the projected rotational velocities of the O-type stars 
for three different luminosity class categories.}
\label{fig:freq_LC}
\end{figure}

Qualitatively the origin of the lower number of slow rotators within NGC\,2060 remains unclear. 
It may be related to the specific evolutionary stage of most of the stars in the clusters or reveal differences in the initial \vrot\ distribution. The fact
that  dwarfs and sub-giants (LC\,V-IV) and bright giants and supergiants (LC\,II-I) have very similar distributions at  low velocities (Fig.~\ref{fig:freq_LC}) tends to rule out differences in the evolutionary stage of the two clusters as a straightforward explanation.

%%%%%%%%%%%%%%%%%%%%%%%%%%%%%%%%%%%%%%%%%%%%%%%%%
%%%%%%%%%%%%%%%%%%%%%%%%%%%%%%%%%%%%%%%%%%%%%%%%%

%%%%%%%%%%%%%%%%%%%%%%%%%%%%%%%%%%%%%%%%%%%%%%%%%
%%%%%%%%%%%%%%%%%%%%%%%%%%%%%%%%%%%%%%%%%%%%%%%%%

%%%%%%%%%%%%%%%%%%%%%%%%%%%%%%%%%%%%%%%%%%%%%%%%%%%%%%%%%%%%
\begin{figure}
\centering
\includegraphics[scale=0.5]{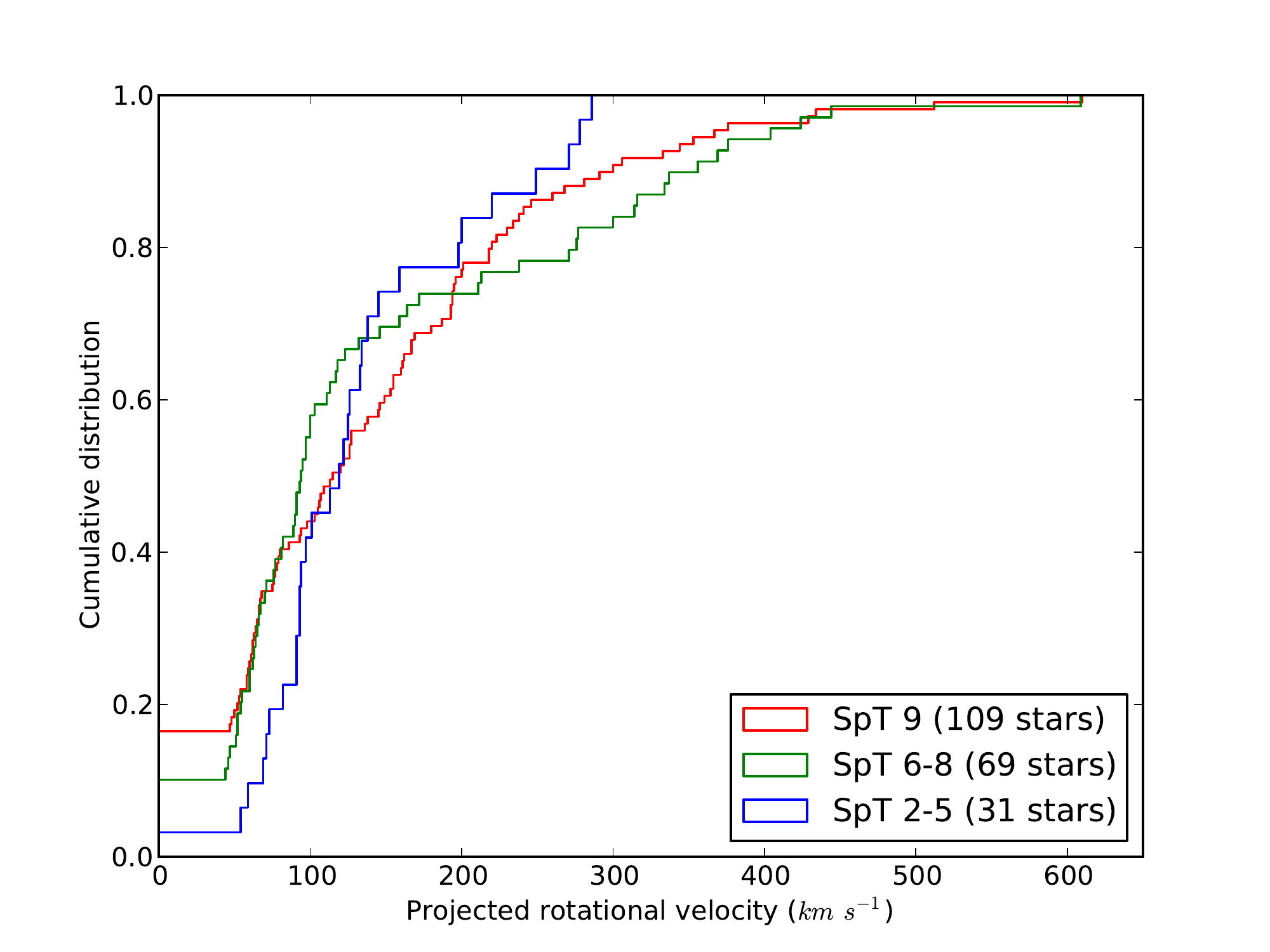}
\includegraphics[scale=0.5]{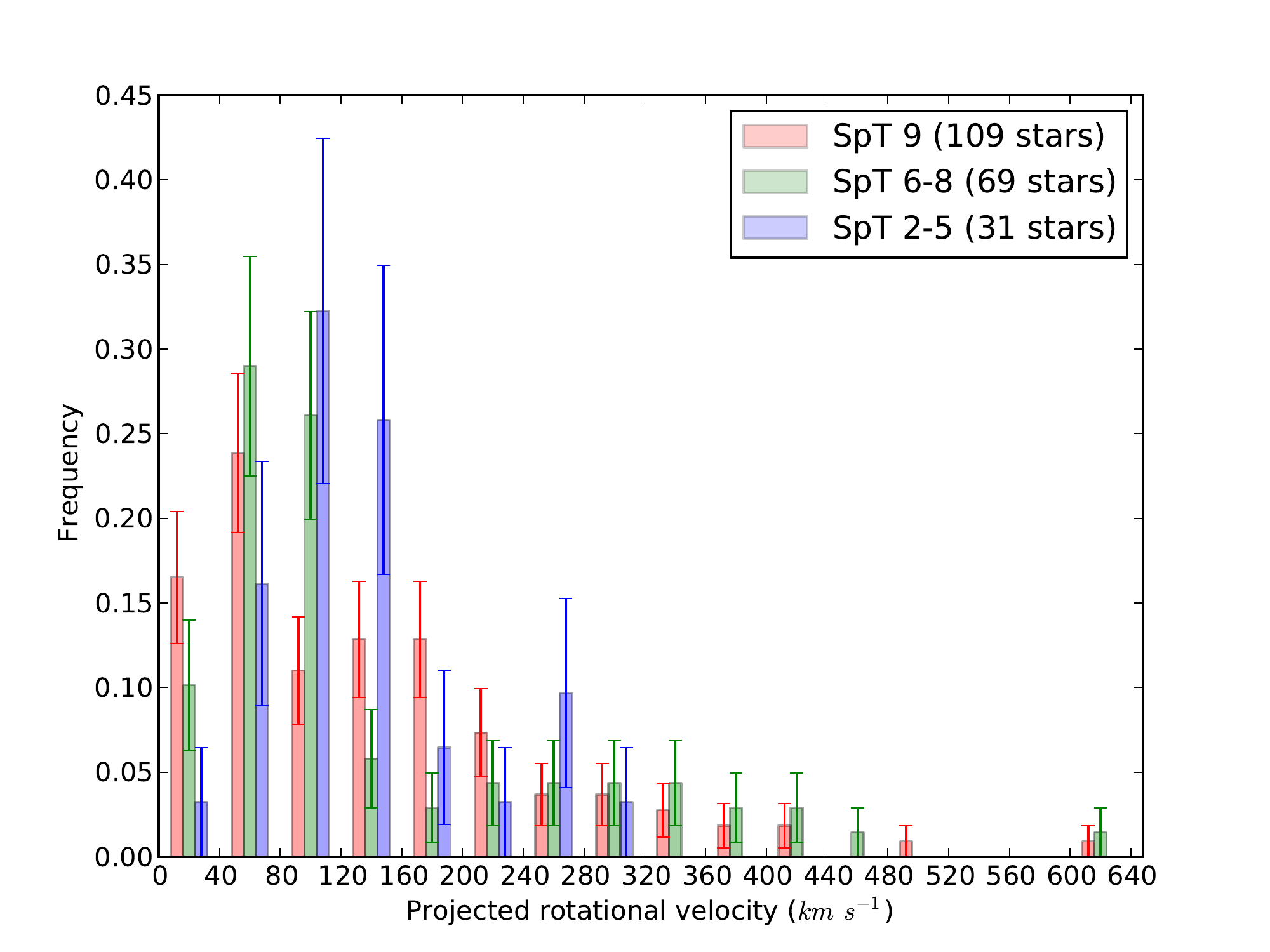}
\caption{
Cumulative (upper panel) and frequency (lower panel, with poisson error bars) distributions of the projected rotational velocities of the O-type stars 
for three different spectral type categories.}
\label{fig:freq_SpT}
\end{figure}
%%%%%%%%%%%%%%%%%%%%%%%%%%%%%%%%%%%%%%%%%%%%%%%%%%%%%%%%%%%%

\subsection{Luminosity Class}
\label{subsec:LC}

Figure~\ref{fig:freq_LC} shows the \vrot\ distributions for three luminosity groups: 
V-IV (119 stars),  III (38 stars), and II-I (33 stars).
There are 26 stars in our sample with no LC classification available and therefore we only used 190 stars. 
KP tests do not reveal  statistically significant differences between the distributions of giants (LC\,III) and
bright giants and supergiants (LC\,II-I).
Both distributions are, however, statistically different
from that of the dwarfs and sub-giants (LC\,V-IV), this latter group dominating the extreme of the high-velocity tail (\vrot\ $\geq$ 450~\kms).

The difference in the high-velocity regime between the V-IV class and the other classes can be explained from
evolutionary considerations. % to result from evolutionary effects. 
Following \citet{weiderandvink2010} some 18\% of our total sample presents masses above 40~\msun\ (see Sect.~\ref{sec:lowtail}).
The bulk of our O-type star sample consists thus of main sequence stars initially less massive than about 40 \msun. 
For such stars the loss of angular momentum as a result of mass loss in a stellar wind is quite limited, therefore 
this is not expected to be an efficient mechanism for spin down \citep[see][]{vink2010,brott}.  
As the O-type stars evolve away from the zero-age main sequence their radii expand and one may naively expect 
to observe a spin down of the surface layers for more evolved stars. The spin down with increasing radius is however prevented
by a simultaneous contraction of the stellar core and the efficient transport of angular momentum from the
core to the envelope \citep[][]{ekstrom,brott}.

The critical rotation rate of a star, i.e.\ the maximum surface velocity above which the centrifugal force 
exceeds gravity, is, however, mostly defined by its size. Dwarfs are more compact than supergiants 
by about a factor two to three \citep{martins},
which corresponds to break-up velocities a factor up to $\sqrt{3}$ larger. Break-up velocities for dwarfs and supergiants typical 
of our sample are around 700 and 400~\kms, respectively \citep[][]{selma}. These estimates match well the maximum rotational 
velocities observed in Fig.~\ref{fig:freq_LC}. It also  explains the absence of very rapid rotators among supergiants: such stars 
simply can not rotate faster than about 400~\kms.

The low-velocity region shows a lower frequency of giants compared to the other luminosity classes.
%to dwarfs (V-IV) and supergiants (II-I). 
Although the cause of this is unclear,
%We do not have a clear scenario to explain this feature. 
possible explanations include different parent populations, different  effects of the macroturbulent velocity field and/or low number statistics. Further investigations with higher-resolution data are needed to search for the origin of the observed paucity of giants with \vrot\ $<$ 120~\kms.

\begin{table}
\caption{Kuiper tests (KP) statistics for different sets of the O-type stars (Cols. 1 and 2). The number
of stars in every sub-sample are given in Cols. 3 and 4. Col. 5 specifies the deviation between both distributions. 
Col. 6 gives the probability $p_{\rm K}$ of the KP statistics, stating the confidence level that the two populations
are randomly drawn from the same parent distribution.}             % title of Table
\label{table:4}      % is used to refer this table in the text
\centering                          % used for centering table
\begin{tabular}{llrrrr}        % centered columns (4 columns)
\hline\hline\\[-8pt]                 % inserts double horizontal lines
Sample 1 & Sample 2 &  n1 & n2&  D   & $p_{\rm K}$\,(\%)  \\[1pt]  \hline\\[-8pt]    % table heading
%         &          &  n1 & n2&  D   & p(\%)  \\  \hline    % table heading 
%\vspace*{-3mm}\\
\multicolumn{6}{l}{Spatial distribution}   \\ \hline\\[-8pt]
NGC\,2060  &   NGC\,2070          &  45 & 105& 0.24 & 23    \\     
NGC\,2070  &   Outside clusters   & 105 &  66& 0.24 &  9    \\     
NGC\,2060  &   Outside clusters   &  45 &   66& 0.34 &  3   \\     
\hline\\[-8pt]                        % inserts single horizontal line
\multicolumn{6}{l}{Luminosity class}      \\ \hline\\[-8pt]
LC II-I    &    LC III            &  33 &   38& 0.31 & 26    \\     
LC III     &     LC V             &  38 &  119& 0.31 &  4    \\     
LC II-I    &     LC V             &  33 &  119& 0.41 &  0.2  \\     
\hline\\[-8pt]                        % inserts single horizontal line
\multicolumn{6}{l}{Spectral type}      \\ \hline\\[-8pt]
SpT 2-5    &    SpT 6-8           &  31 &   69& 0.37  & 3     \\     
SpT 6-8    &    SpT 9             &  69 &  109& 0.24 & 8      \\     
SpT 2-5    &    SpT 9             &  31 &  109& 0.40 &  0.5   \\     
\hline                        % inserts single horizontal line
\end{tabular}
\end{table}

\subsection{Spectral type}\label{subsec:SpT}

Figure~\ref{fig:freq_SpT} shows the \vrot\ distributions for the different spectral type categories. 
There are seven stars with no SpT available and thus we only consider 209 stars here. 
Though once again the overall appearance of the distributions seems similar, the
early-type group (O2-5) does not contain stars that rotate faster than
300 \kms\ while the later SpT groups show a more extended high-velocity tail. 
%\red{Certainly such stars present masses larger than 40~\msun\ and therefore mass loss may be important.}
Indeed a KP test indicates that the O2-5 rotation rates are statistically different from
the \vrot\ distribution of the later types.
The true maximum rotation rate of O2-5 stars could be slightly higher than the observed 300~\kms, since \hel{ii}{4541} is the only rotational velocity diagnostic for the early-O stars and systematically  underestimates
\vrot\ by 15 to 20\% compared to measurements based on  HeI lines (see Sect. \ref{subsec:test_heii}). 
Even taking this into consideration does not make the early-O stars spin as fast as the later type stars.

The peak of the velocity distribution is also at somewhat higher \vrot\ for the earlier
type stars compared to the later types. The fraction of earlier type
stars in the first bin (below our resolution limit) increases
%shows a correlated increase 
when progressing to later groups of spectral sub-types. 
This trend continues in the B-star domain as shown in \citetalias{dufton}. Because of the limitation of the current data set,  it is
difficult  to decide whether this trend  traces a real effect or  results from a measurement artifact. We discuss both options below. 
% or indicate a physical effect.  

In Sect.~\ref{subsec:test_heii} we showed that 
there is a significant dispersion for \vrot\ (\hhel{ii}) with respect to \vrot\ (\hhel{i})
measurements for \vrot\, $<$ 150~\kms. This comparison 
is done for mid-O stars only, as a similar comparison cannot be performed for stars earlier than O6. 
If broadening due to the linear Stark effect and as a
result of macro-turbulent velocity fields is stronger among early-O-type stars 
\citep[see e.g. fig.~1 in][for the case of Galactic O-type stars]{simon2013} 
and if our method cannot discriminate between rotation and extra broadening for early-O stars 
as well as for mid- and late-O stars, the rotation rates of the earliest types may be somewhat overestimated. It is, however, unclear why the trend should continue for late O- and early B-type stars. 

If the signal is real, it suggests either that more massive stars cannot be spun down as efficiently as lower mass stars or that there exists an additional line broadening mechanism which strength correlates with spectral type. Such a mechanism remains to be identified. A similar conclusion is reached by Markova et al. (in prep.) and Sim\'on-D\'iaz et al. (in prep.) using high resolution data of a Galactic samples of O-type stars.

The absence of a high-velocity tail among early O-type stars may also be related to binary evolution effects (see Sect.~\ref{sec:hightail}). 
During a mass-transfer event, the secondary less massive star accretes mass
and angular momentum from the primary star.  As a result secondaries are efficiently spun up to break-up velocities.
As the masses of primary stars are larger than those of secondaries, and as the distribution of the mass ratio ($q=M_2/M_1$)
favors relatively lower mass companions ($f_q \propto q^{-1  }$ in the VFTS field; \citetalias{sana}), 
the secondaries in the O+O binaries in 30\,Dor will much more frequently  be mid- and late-O stars than early-O stars. 
A population of unidentified post-interacting binary products in our sample
could explain the strong preference for mid- or late-type stars in the high-velocity tail of Fig.~\ref{fig:freq_SpT}. Alternatively, the earliest O-type stars, having the strongest winds, may already have experienced a larger spin down due to mass loss compared to later spectral sub-type stars. 
However, current evolutionary models do not predict this effect to be sufficient, unless of course the adopted mass-loss recipes 
have underestimated the true mass-loss rates of these more extreme objects.

%\textbf{Our sample stars are dominated by LC V-IV (see Fig.~\ref{fig:SpTLC}), then it would mean
%that more massive dwarfs and bright giant stars.}

\subsection{Comparison with earlier studies}\label{subsec:erlierres}

%{\it Preliminary}

%Fig.~\ref{fig:vrotsini_dist_comp} compares our cumulative \vrot\ distribution to the results of recent surveys
%by \cite{penny2009} and \citet{huanggies1,huanggies2}. 
%The most relevant observational studies in early O- late B-type stars to our work are:
%\citet{penny2009}, \citet{huanggies1,huanggies2} and \citet{dufton}.

\citet{penny2009} used FUSE UV observations with a spectral resolving power of about 20\,000
and a cross-correlation method to measure \vrot\ in a sample of 258 stars in the Galaxy (97 stars in total), 
LMC (106) and SMC (55). 
The cross-correlation method measures the overall broadening of the line and 
does not allow to distinguish between rotation and macro-turbulence. The method is, however, versatile in that it allows to identify double-lined spectroscopic binaries\footnote{In those 
cases \citet{penny2009} provided \vrot\ values for both the primary and the secondary star.}. 
%Stars in their sample consisted of 
%were divided in the three different metallicity environments as follows:
%97 Galactic, 55 SMC, and 106 \textbf{LMC} OB-type stars. 
For all three metallicity environments, the authors reported a lack 
of very slowly  rotating stars (i.e.\ \vrot\ $\leq$ 50~\kms),
which they interpreted as the signature of additional-broadening due to macro-turbulent motions. 

%With the aim to investigate the evolution of angular momentum as a function of evolutionary status and metallicity 
\citet{penny2009} also divided their total sample into relatively unevolved (V-IV) and evolved (II-I) stars, omitting LC\,III stars. 
For their Galactic sample, the unevolved and evolved samples were statistically different for $\vrot < 200$ \kms. 
The authors interpret this as the result of a stronger photospheric macro-turbulence
of the evolved stars. For the LMC and SMC stars such a behavior was not observed, suggesting 
that macro-turbulence is metallicity dependent.
%a metallicity dependance of the macro-turbulence effects.  
Similar comparisons at high projected rotational velocities (i.e.\ $>$ 200 \kms) revealed no significant statistical 
difference between the unevolved and evolved objects. 

The slowly rotating dwarfs in the three metallicity environments showed statistical differences, but
no clear trend could be identified, while the more rapidly rotating dwarfs ($>$ 200 \kms) showed compatible distributions.
For the evolved low \vrot\ stars a trend could be identified: the higher the metallicity the smaller the fraction of slow rotators.  
This too is consistent with a metallicity dependence of macro-turbulent effects, in that such effects seem more prominent for higher metal content.
The rapidly spinning evolved objects did not show a clear trend, though the SMC environment appears to be
somewhat richer in these objects.

\citet{huanggies1}  used moderate resolution spectra to study projected rotational velocities of 496 presumably single OB
stars (%selected based on the absence of large RV variations, i.e.\ 
$\rm{\Delta RV}\, \leq$ 30~\kms).
Their sample stars belong to 19 different open clusters in the Galaxy and span an age of 7-73 Myr.
By using different diagnostic lines, such as \hel{i}{4026}, 4387 and 4471, and \Mgl{ii}{4481}, they  fitted \vrot\ 
using TLUSTY-SYNSPEC stellar atmospheres \citep{hubenyandlanz1995}.
%by means of 
%the stellar atmosphere code TLUSTY and the radiative transfer code SYNSPEC \citep{hubenyandlanz1995}. 
%In the cases of double-lined spectroscopic binaries (SB2), the \vrot\ value of the star was determined by fitting the dominant component.
They found that the \vrot\ distribution of B-type field stars contains a larger fraction of slow rotators 
than the B-type stars in clusters. Their high-velocity tail extends up to 350~\kms\ and 400~\kms,
respectively. They finally suggested that some of these rapid rotators may have been spun up through 
mass transfer in close binary systems. 
%Finally, the \vrot\ and intrinsic (\veq\) distribution show .

As a part of the VFTS project, \citetalias{dufton}
studied a sample of around 300 stars spanning SpT from O9.5 to B3, excluding supergiants. 
In addition to the set of diagnostic lines used here, \citetalias{dufton} also made use of
\hel{i}{4026}, \Mgl{ii}{4481}, \Cl{ii}{4267}, and \Ol{ii}{4661}. 
As in the present work, the FT method was used to estimate the \vrot\ 
for stars that do not show significant radial velocity variations.
Because of uncertainties in the initial crude VFTS classification, 47 stars, predominantly O9.5 and O9.7 stars, were analyzed both in \citetalias{dufton} and in the present paper. The agreement in \vrot\ in the overlapping set of stars between our and their analysis is excellent (see Appendix ~\ref{subsec:Common_OB}).  
The  \vrot\ distribution and the deconvolved \veq\ distribution of the B-type stars in VFTS show a distinct bi-modal structure with
25\% of the sample having \veq\, $<$ 100~\kms\ (see \citetalias{dufton}). 
The components of the bi-modal structure do not correlate with different episodes of star formation
nor with different locations in the field of view.

%The bi-modal structure does not show
%a correlation with respect to different episodes of star formation, and/or field and cluster stars.

%%%%%%%

\begin{figure}
\centering
\includegraphics[scale=0.5]{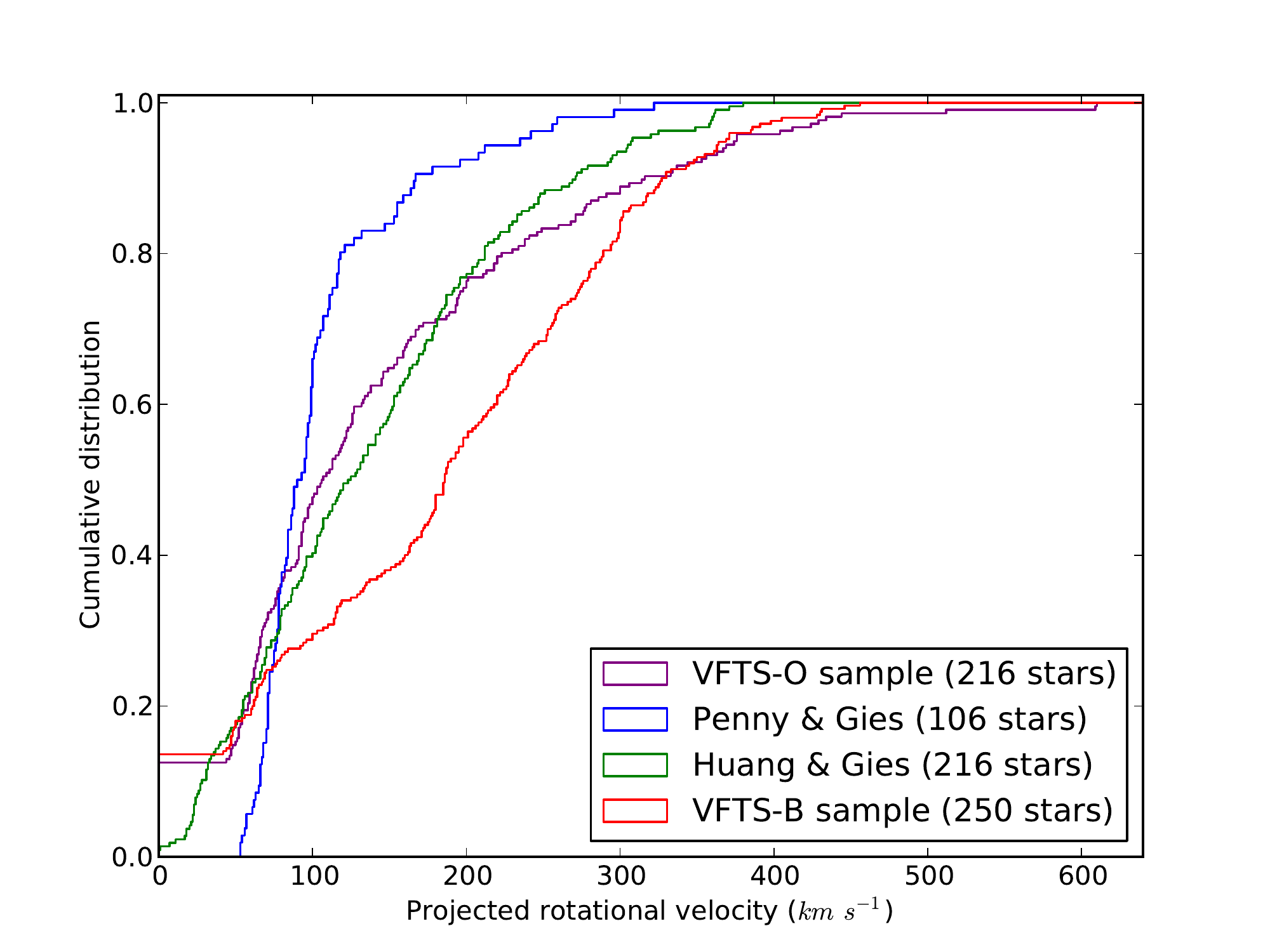}
\caption{Comparison of the cumulative distribution of projected rotational velocities of our work 
(VFTS-O sample - purple), \citeauthor{penny2009} (blue), \citeauthor{huanggies1} (green), and the VFTS-B sample of
\cite{dufton} (red).}
\label{fig:vrotsini_dist_comp}
\end{figure}

%%%%%%%

Figure~\ref{fig:vrotsini_dist_comp} compares our \vrot\ cumulative
distribution of the O-type stars in 30~Dor with those from the studies summarized above. %different works previously discussed.
%1
For \citeauthor{penny2009}, we only show the LMC sample (106 stars).  
%Qualitatively over 85 percent of their stars rotate at a rate less 
%than 200 \kms and the high velocity tail extends up to 350 \kms. 
%\citet{mokiem} studied \vrot\ properties for 17 of these stars 
%and \citet[see Fig.~4]{penny2009} showed that no systematic differences are found. 
%2
From the total sample of \citeauthor{huanggies1}, Fig.~\ref{fig:vrotsini_dist_comp} presents a sub-sample of 216 stars spanning SpT from O9.5 to B1.5
(see figure~4 of their work). 
%Their cumulative distribution at \vrot\ = 40\,\kms\ is naively similar to ours, though 
%the method and resolution in their spectra allowed them to also provide \vrot\ values in the very slow rotating region. 
%Qualitatively their and our distribution are similar, being the probability of their sample slightly higher 
%for projected rotational velocities below 200 \kms\ and the other way around for larger velocities.
Their higher spectral resolution allowed them to 
extract lower \vrot\ values than is possible for our data set (i.e. \vrot\, $\leq$ 40\, \kms, see Sect. \ref{subsec:diagnostic_lines}).
%3
Finally, the VFTS B-type star distribution excludes the 47 late-O stars in common with the present work, to preserve the independency of the two samples.

%Conlcusion
KP tests indicate that the \citeauthor{penny2009} and VFTS B-star distributions are statistically different, with a confidence level better than 1\%, while the \citeauthor{huanggies1} distribution marginally agrees with our O-star distribution ($ p \sim 11$\%). The fact that 
\citeauthor{penny2009} % Penny \& Gies 
do not correct for macro-turbulence is a straightforward explanation for the absence of slow rotators in their sample. The other three distributions agree well with respect to the fraction of extremely slow rotators. 
The fraction of VFTS O- and B-stars below our \vrot\ resolution limit (see Sect. \ref{subsec:strategy}), for instance, is roughly similar.

The distribution of \vrot\ of the \citeauthor{penny2009} sample peaks at the same projected rotational velocity as in our distribution. The lack of stars spinning faster than 300 \kms\ in their sample
%The lack of fast rotators among the  Penny \& Gies sample 
is intriguing, but may result from a selection effect. Indeed the FUSE archives may not be representative of the population of fast rotators in the LMC, as  individual observing programs may have focused on stars most suitable for their respective science aims, possibly excluding fast rotators as these are notoriously difficult to analyze. % In that case their sample could suffer from biases.}
%rather than attempting to build an unbiased sample. 

The similarities between the  O-star distribution in 30~Dor and the distribution of late-O and early-B Galactic stars of 
\citeauthor{huanggies1} 
suggests a limited influence of metallicity. This is %in line \cancel{with the fact that we do not}
consistent with our expectation that stellar winds do not play a significant role in shaping the rotational velocity distributions in both samples, being dominated by stars less massive than 40~\msun. 

The differences with the VFTS B-star sample are striking, and lack a straightforward explanation. 
The B-type stars show a bi-modal population of very slow rotators and fast rotators, with few stars rotating at rates that are 
typical for the low-velocity peak seen in the VFTS O-type stars. We return to this issue in Sect.~\ref{sec:discuss}.

%\red{Finally, the differences with the VFTS B-star sample is striking, the latter showing a larger populations of extremely 
%slow rotators as well as fast rotators. CALL FOR IDEAS: what does this comparison tells us more?}

\begin{figure}
\centering
\includegraphics[scale=0.5]{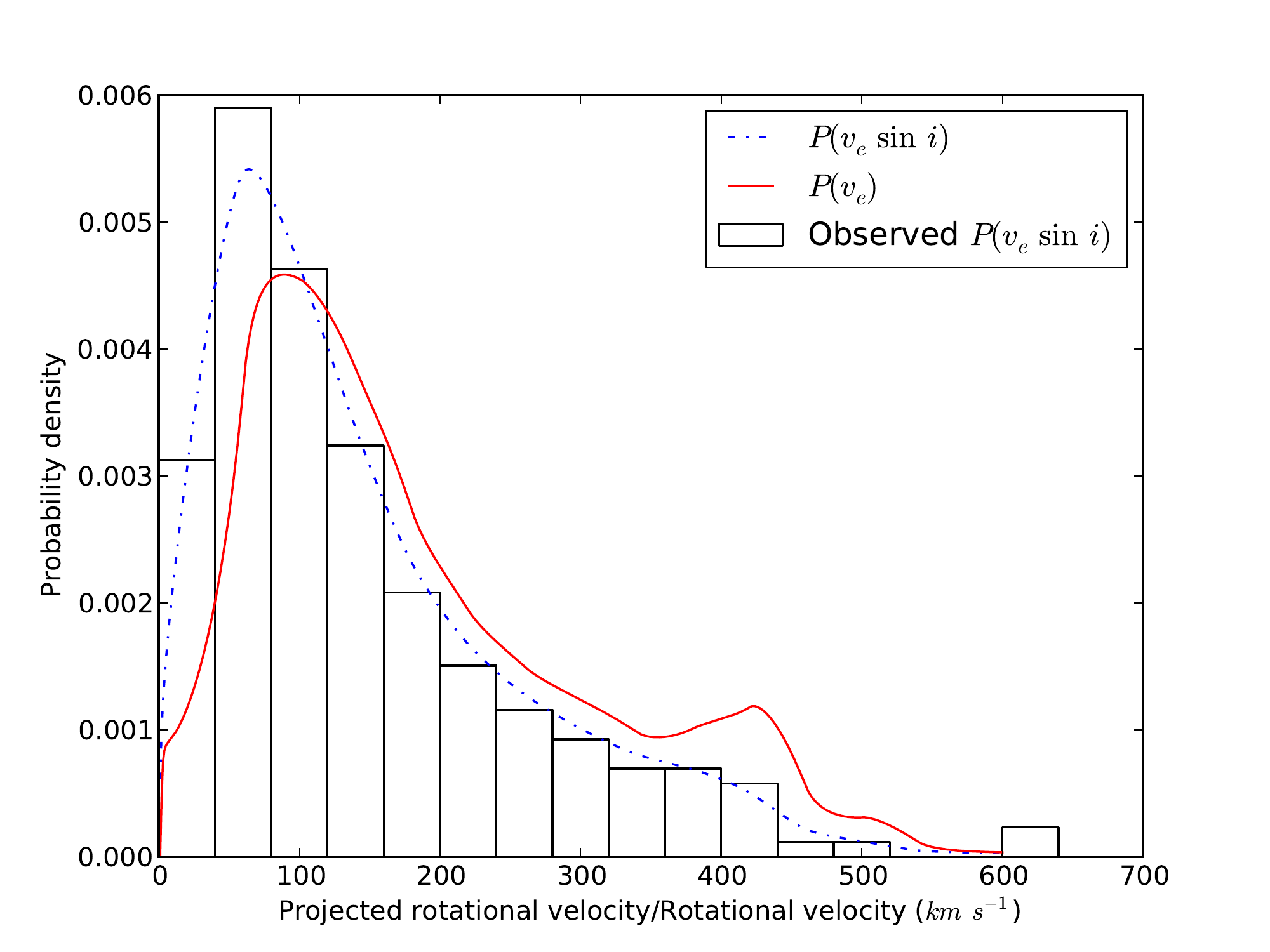}
\caption{Observed \vrot\ and Lucy-deconvolved \veq\ distributions. The dot-dashed line shows the estimates,
after 4 iterations in the Lucy-deconvolution, of the probability density function for the projected rotational 
velocity distribution.
The solid line shows the probability density function of the actual rotational velocities.
%P($\rm{v_\mathrm{e}}$).
}
\label{fig:deconvol}
\end{figure}

\subsection{Analytical representation of the \veq\ distribution} %Intrinsic  \vrot\ distribution}
\label{subsec:deconv}
The size of our sample is large enough to investigate the distribution of
intrinsic rotational velocities. By assuming that the rotation axes are randomly 
distributed, we infer the probability density function of the rotational velocity distribution $P(\varv_{e})$ from 
that of \vrot. We adopt the iterative procedure of \cite{lucy}, as applied in \citetalias{dufton} for the B-type
stars in the VFTS, to estimate the pdf of the projected rotational 
velocity $\rm{P(\vrot)}$ % ; $\tilde{\phi}^4$ in Lucy's notation)
and of the corresponding deprojected pdf velocity  $\rm{P(\veq)}$. %; $\psi^4$ in Lucy's notation). 
As expected, $\rm{P(\veq)}$ moves to larger
velocities compared to $\rm{P(\vrot)}$ due to the effect of inclination.
At \veq\ $\geq$ 300 \kms, $\rm{P(\veq)}$ presents small scale fluctuations
that probably result from small numbers in the observed distribution. 
The two extremely fast rotators at \vrot\ $\gtrsim$ 600 \kms\ are excluded from the 
deconvolution for numerical stability reasons.

\begin{figure}
\includegraphics[scale=0.5]{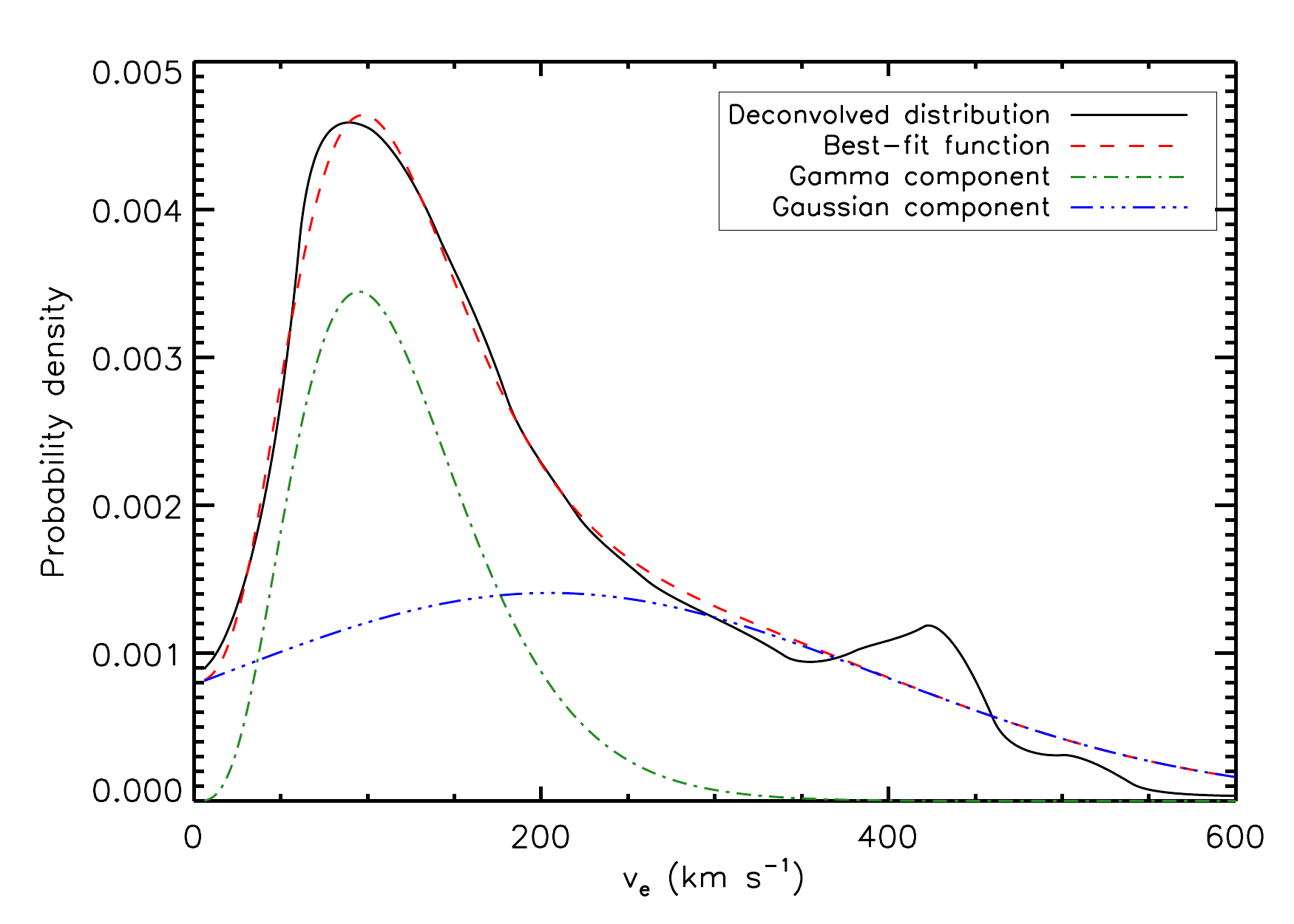}
\caption{Analytical representation of the deconvolved rotational velocity distribution (Eq. \ref{eq: analytic}).}
\label{fig:model}
\end{figure}

We can well approximate the deconvolved rotational velocity distribution by an analytical function with two components. We use a gamma distribution for the low-velocity peak and a Normal distribution to model the high-velocity contribution:

\begin{equation}
P(\mathrm{v_e}) \approx I_\gamma\ g(\mathrm{v_e}; \alpha,\beta) + I_N\ N(\mathrm{v_e}; \mu, \sigma^2)
\end{equation}
where
\begin{eqnarray}
%\begin{equation}
g(x; \alpha,\beta) &=& \frac{\beta^\alpha}{\Gamma(\alpha)} x^{\alpha-1} e^{- \beta x},\\
%\end{equation}
%
%\begin{equation}
N(x; \mu, \sigma^2) &=& \frac{1}{\sqrt{2 \pi} \sigma } e^{-(x-\mu)^2 / 2 \sigma^2 },
%\end{equation}
\end{eqnarray}
and $I_\gamma$ and $I_N$ are the relative contributions of both distributions to $P(v_\mathrm{e})$.
The best representation, shown in Fig.~\ref{fig:model}, is obtained for:
\begin{eqnarray}
P(v_\mathrm{e}) & \approx & 0.43\, g \left(\alpha=4.82,\beta=1/25 \right)  \nonumber \\
             & + &  0.67\, N \left(\mu=205~\kms, \sigma^2= \left(190~\kms \right)^2 \right)
             \label{eq: analytic}
\end{eqnarray}
The function is normalized to 0.99 to allow for the inclusion of an additional 1\%\ component to represent the two extremely
fast rotators is our sample. % with \vrot $>$ 600~\kms. 
One should note that the reliability of the fit function is limited by the sample size at extreme rotational velocities.
This analytical representation of the intrinsic rotational velocity distribution may be valuable in stellar population synthesis models 
that account for rotational velocity distributions.

\section{Discussion}
\label{sec:discuss}

The most distinctive feature of the \vrot\ and \veq\ distributions of the 
O-type stars in 30~Dor is its two-component structure: 
a low-velocity peak and an extended high-velocity tail. 
%This behavior could correspond to different star forming conditions and/or be a consequence of differences in the evolution of stars.
%Standard evolutionary theory of single stars is not able to fully explain it. We address the
%binary interaction as a likely scenario to produce the high velocity tail (see Sect.~\ref{subsec:SpT}). 
In this section, we consider possible physical mechanisms that may be responsible for 
the global shape of our \veq\ distribution. We start, however, by assessing the projected
spin rates \vrot\ relative to the critical spin rates.

\subsection{\vrot\ relative to the critical rotation rate}\label{sec:vcrit}
%\subsection{The low velocity peak}\label{sec:lowtail}

%Fig.~\ref{fig:vcrit} shows the projected rotational velocities in units of the
%critical velocity as obtained from the spectral sub-type (SpT)
%calibration for rotating LMC stars by \citet{weiderandvink2010}.

\begin{figure}
\includegraphics[scale=0.45]{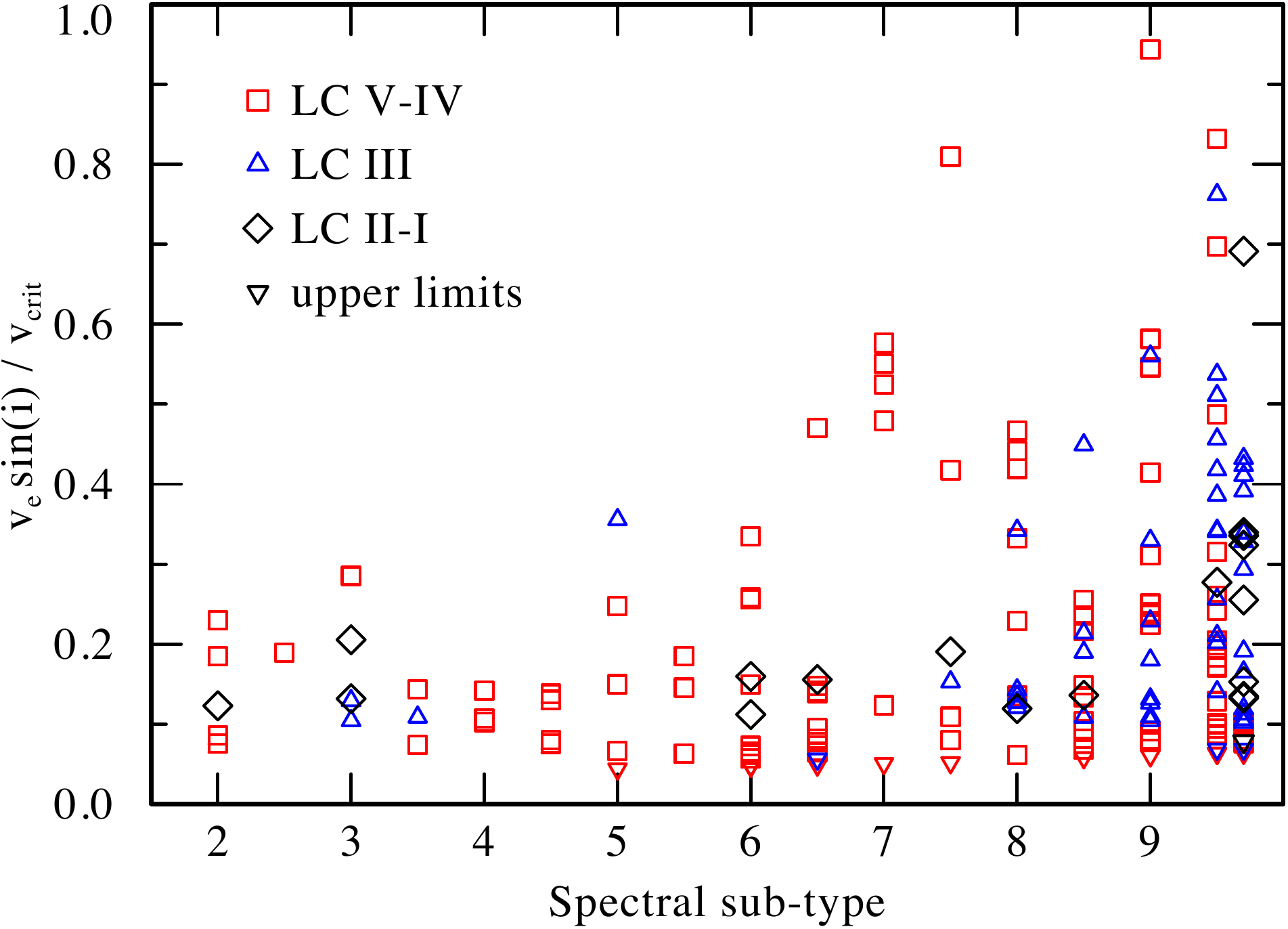}
\caption{Projected rotational velocity normalized to critical velocity, \vrot$/\rm{v_{crit}}$, vs. spectral sub-type 
for our sample stars with known spectral types. Critical velocities are estimated using the SpT calibration by \citeauthor{weiderandvink2010}. 
See legend for the information on the luminosity class. 
Stars indicated in the legend as ``upper limits" refer to stars with \vrot\ $\leq$ 40\, \kms.
}
\label{fig:vcrit}
\end{figure}

To estimate the critical rotation rate, \vcrit, the sample is divided into three groups
with LC\,V-IV, III, and II-I. Stellar masses ($M$),
luminosities ($L$), and radii ($R$) are then obtained from the SpT
calibration for the representative LC (V, III, or I) from the spectral sub-type
calibration for rotating LMC stars by \citet{weiderandvink2010}.
We approximate $\vcrit^2 = \mathrm{G} M (1-\Gamma) / R $
by correcting the mass for the effect of radiation pressure on
free electrons with an Eddington factor $\Gamma$.
%The masses are corrected for the effect of radiation pressure on
%free electrons with an Eddington factor $\Gamma$, so that 
%$\vcrit^2 = \mathrm{G} M (1-\Gamma) / R $. 
Around 50\% of our sample stars are found in the low-velocity peak with \vrot\ between 50 and 150 \kms, 
implying rotation rates less than 20\% of the critical rotation rate  \vcrit. %(Fig.~\ref{fig:vcrit})
Fig.~\ref{fig:vcrit} confirms the general behaviour found in Sect.~\ref{subsec:SpT} that earlier O-type stars in our sample lack extremely fast rotators relative to
%rotate slower than 
later spectral sub-types, either in absolute \vrot\ or as a fraction of  \vcrit. 
%\red{High values of $\frac{\vrot}{\vcrit}\, \rm{>}$ 0.4 are only found for spectral sub-types later
%than O6 (LC V-IV),  O8 (LC III), and O 9.5 (LC II-I).
%According to \citeauthor{weiderandvink2010}
%SpT calibration these limits correspond to initial masses of 37(-5/+3) \msun\ (LC V-IV), 
%30(-4/+5) \msun\ (LC III), and 37(-10/+13) \msun\ (LC II-I).}

\subsection{The low-velocity peak}\label{sec:lowtail}

The low-velocity peak contains the large majority of the stars in our sample. 
%It represents thus a fundamental aspect, 
The origin of this peak may be 
related to the formation or early evolution of these stars. Massive stars inherit their angular momentum from 
their parental cloud that contains more than sufficient angular
momentum to spin up the proto-star to critical rotation \citep[see][]{larson}. 
Interestingly, \citet{lin} found that gravitational torques prohibit a star
from rotating above $\sim$50\%\ of its break-up speed during formation. Magnetic coupling between the
massive proto-star and its accretion disk is expected to be insufficient in spinning down the star
further \citep{rosen}. As pointed out earlier, stellar winds and (single-star) evolution are also not
effective in reducing the rotation rate during most of the main sequence phase, save for objects initially more massive than 40 \msun. An additional braking mechanism is thus needed.

\citet{meynet2011} and \citet{potter} have explored stellar evolution models
for magnetic main sequence stars. Based on a model for magnetic braking of \citet{uddoula2002},
they both predict that a massive star rotates at only a modest fraction of its break-up velocity if it has a surface magnetic field strength
of the order of 2 kG. Most O-type stars in the Milky Way have no measured magnetic field   and
the majority of the few known magnetic O-type star have a magnetic field strength of the order of several hundreds of G to a couple of kG \citep{Donati2009,2012AIPC.1429...67G}. If magnetic braking is indeed the mechanism that slows down the stars after their birth, most of the spin down has to occur within the first Myr, after which the strength of the stellar magnetic field has to decrease below the detection limit of the current surveys.

Within the context of a main-sequence magnetic braking scenario, the absence of a large population of very slow rotators 
(\vrot\ $<$ 40~\kms) in our sample may indicate that either magnetic fields disappear before being able to fully spin down 
the star or that the generation of the magnetic field itself is related to the high rotational velocity. The analysis of spin rates for
the early-B stars in the VFTS field \citepalias{dufton} shows a bi-modal distribution with a low-velocity component that peaks at lower
velocities relative to that for the O-stars. If this difference is connected to magnetic braking it may indicate that magnetic
fields are initially stronger and/or more efficient in spinning down early-B stars. %, as the incidence fraction of magnetic fields in 
%O and early-B stars seem comparable \citep{2012AIPC.1429...67G}.}

\subsection{The high-velocity tail}\label{sec:hightail}

Although most of the stars must spin down quickly after their formation in order to produce the low-velocity peak 
(see Sect.~\ref{sec:lowtail}), it is possible that some very young stars -- potentially 
rotating faster than average -- are still present in our sample. 
Fig.~\ref{fig:vcrit}  shows that the later spectral sub-types contain a pro rata
larger percentage of fast rotating stars, almost reaching up to critical speed. 
The compatible rotational distributions of the stars in NGC\,2060 and the younger NGC\,2070 
(see Sect.~\ref{subsec:Spatialvar}) and the fact that star formation has likely stopped 
in NGC\,2060, however, argue against newborn stars being a suitable explanation for the high-velocity tail.

The presence of a high-velocity tail is, however, predicted by recent population synthesis
computations that study  the influence of binary evolution 
on the projected rotation rate of massive stars \citep{selma}. Those simulations can create
a population of stars with high rotational rates through  
binary interaction. Such a population is composed of mergers and of secondary stars that have been spun up by mass transfer. In a second paper, de Mink et al. (in prep.) also argue that most post-interaction binary products cannot be identified by RV investigations and will thus contaminate our ``single-star'' sample.
Interestingly, we find that the fraction of fast rotators (\veq\ $>$ 300~\kms)  observed in the 30\,Dor O-type star population 
(19\%\ of our `single-star' sample, hence $\sim$11\%\ of the whole VFTS O-type star population) is of the 
same order of magnitude as the numerical predictions by \citeauthor{selma} Dedicated simulations 
that take into account the star formation history of 30\,Dor are, however, desirable to further investigate this scenario.
\citeauthor{sana} (\citeyear{sana}; \citetalias{sana}) report
that the measured O-type star binary fraction in 30 Dor (51\%) is apparently smaller than this fraction measured in young nearby Milky Way clusters \citep[69\%;][]{sanaa}. The assumption that most of the stars in our high 
rotational-velocity tail are post-binary interaction products could potentially conciliate these two measurements.
The pronounced high-velocity component in the \vrot\ distribution of the early-B stars in the VFTS \citepalias{dufton} may 
also be in line with a post-binary nature, as secondaries more often are of spectral type B than O.

%(\textbf{13}\%\ of our ``single star'' sample, hence $\approx 8$\%\

If our interpretation of the high-velocity tail as resulting from binary interaction is correct, it suggests that 
the low-velocity part of our distribution is a cleaner original single-star sample. As discussed by \citet{selma}
the contamination of the high-velocity end
of the distribution by unresolved binary products complicates, and may even invalidate, 
%It also complicates or even invalidates
surface abundance analysis aiming to test or calibrate rotational mixing theories, as such surface enrichment may 
also be the result of mass transfer or mixing in merger products.

\subsection{The single and binary channel for long-duration gamma-ray bursts}

The nature of the high-velocity tail of the distribution of rotation rates as discussed in Sect.~\ref{sec:hightail} 
has important implications for the evolutionary origin of systems that produce long-duration gamma-ray bursts (LGRBs).
In both the collapsar model \citep{woosley1993} and the millisecond-magnetar model \citep{lyutikov} the Wolf-Rayet progenitor system 
of a stellar explosion producing a long (at least two second) burst of gamma rays is required to have a rapidly spinning core 
\citep[see e.g.][]{langer2012}.  Most, perhaps all \citep{niino2011}, of such gamma-ray bursts occur in regions of their host galaxies 
that have a low-metal content \citep{fruchter2006,modjaz2008,grafener2008}.  Two channels leading up to a GRB have been proposed.  First, a close
interacting binary system may lead to the late production and spin-up of a Wolf-Rayet star \citep{izzard2004,fryer2005,podsiadlowski2010,tout2011}.
With only limited time left before the supernova explosion, the stellar wind of the Wolf-Rayet is not able to remove sufficient angular momentum to prevent a gamma-ray burst, especially not at low-metallicity where the outflow is less dense \citep{vink2005}. Second, a LGRB may occur for a single star that rotates so fast that mixing processes cause the interior to become quasi-chemically homogeneous and the star as a whole to remain compact \citep{yoon2005,woosley}. As such a system develops Wolf-Rayet characteristics early on, a low-metallicity environment is required to avoid wind-induced spin down.

The single-star LGRB progenitors need O-type star descendants that at birth spin faster than $\sim$ 300-400 \kms\ 
and are more massive than about 20 \msun\ \citep{brott}. \citet{mokiem} identified presumably-single candidate objects for such evolution, a result that actually spurred the above mentioned groups to put forward the possibility of a single-star channel.  If indeed the high-end tail of the velocity distribution is dominated by, or is exclusively due to, post-interaction binaries and mergers, the relative importance of the single-star channel is reduced. Indeed, if the high-end tail is exclusively composed of binary products then the existence of the single-star GRB channel may be challenged altogether.

\section{Conclusions}\label{sec:conclusions}

We have estimated projected rotational velocities for the presumably single O-type stars in the VFTS sample (216 stars). 
We find that the most distinctive feature of the \vrot\ distribution of O stars in 30\,Dor is a two-component structure: 
a low-velocity peak at $\sim$80 \kms\ and a high-velocity tail extending up to 
$\sim$600 \kms. 
The presence of the low-velocity peak is consistent with previous LMC surveys, but we conclusively 
find a considerable population of rapidly spinning stars (\veq\ $>$ 300 \kms\ for 20\%\ of the sample). 
The homogeneity and size of the sample also allows us to study the \vrot\ distribution as a function of spatial 
distribution, luminosity class and spectral type. %LC and SpT. 
%Among these categories we find that there is not any significant difference in the low velocity region. 
%The high velocity tail frequency distribution is dominated  by 
%stars outside the clusters with LC V-IV and mid-late spectral sub-type.

Based on expectations of star-formation and single-star evolution, most of the stars seem to have to spin down 
shortly after their formation, 
from critical or half-critical rotation rates to a much smaller fraction of their break-up velocity. 
For the bulk of O stars angular momentum loss in a stellar wind is insufficient and another
mechanism should act
%We claim that stellar winds are not sufficient 
%in this process and that 
%there should exist another mechanism 
to efficiently spin down the stars;  magnetic fields being prime candidates.

The presence of a well populated high-velocity tail is compatible with expectations from binary evolution and 
qualitatively agrees with recent population synthesis predictions \citep{selma}.
The nature of the high-velocity tail of the distribution has an important implication for the evolutionary origin 
of systems that produce long-duration gamma-ray bursts. 
If the objects in the high-velocity tail are predominantly products of binary interaction and mergers this would imply a scenario for long-duration GRB production without a preferred metallicity (range) for the progenitor systems, unless binarity itself presents a metallicity dependence in LGRB progenitor production.  If the high-velocity tail is dominated by single stars after all, a low metallicity environment seems required for LGRBs.

\begin{acknowledgements}
S.d.M. acknowledges support by NASA through an Einstein Fellowship grant, PF3-140105, and a Hubble Fellowship grant, HST-HF-51270.01-A, awarded by the STScI operated by AURA under contract NAS5-26555.
%S.d.M. acknowledges support through a Hubble Fellowship grant HST-HF-51270.01-A awarded by the STScI, operated by AURA, Inc., under contract NAS 5-26555 and an %Einstein Fellowship grant PF3-140105 awarded by the Chandra X-ray Center, operated SAO under contract NAS8-03060. 
JMA acknowledges support from the Spanish Government Ministerio
de Educaci\'on y Ciencia through grants AYA2010-15081 and AYA2010-17631
and the Consejer\'{i}a de Educaci\'on of the Junta de Andaluc\'{i}a through grant P08-
TIC-4075. SS-D and AH acknowledge financial support from the Spanish Ministry of Economy and Competitiveness
(MINECO) under the grants AYA2010-21697-C05-04, Consolider-Ingenio 2010 CSD2006-00070, and 
Severo Ochoa SEV-2011-0187, and by the Canary Islands Government under grant PID2010119.
FN ackowledges support by the Spanish MINECO under grants
  AYA2010-21697-C05-01 and FIS2012-39162-C06-01.
\end{acknowledgements}

\bibliographystyle{aa}	% (uses file "plain.bst")
%\bibliography{bibliography}

\begin{thebibliography}{71}
\expandafter\ifx\csname natexlab\endcsname\relax\def\natexlab#1{#1}\fi

\bibitem[{{Abt} {et~al.}(2002){Abt}, {Levato}, \& {Grosso}}]{abt}
{Abt}, H.~A., {Levato}, H., \& {Grosso}, M. 2002, \apj, 573, 359

\bibitem[{{Aerts} {et~al.}(2009){Aerts}, {Puls}, {Godart}, \&
  {Dupret}}]{aerts2009}
{Aerts}, C., {Puls}, J., {Godart}, M., \& {Dupret}, M.-A. 2009, \aap, 508, 409

\bibitem[{{Brott} {et~al.}(2011{\natexlab{a}}){Brott}, {de Mink}, {Cantiello},
  {Langer}, {de Koter}, {Evans}, {Hunter}, {Trundle}, \& {Vink}}]{brott}
{Brott}, I., {de Mink}, S.~E., {Cantiello}, M., {et~al.} 2011{\natexlab{a}},
  \aap, 530, A115

\bibitem[{{Brott} {et~al.}(2011{\natexlab{b}}){Brott}, {Evans}, {Hunter}, {de
  Koter}, {Langer}, {Dufton}, {Cantiello}, {Trundle}, {Lennon}, {de Mink},
  {Yoon}, \& {Anders}}]{brotta}
{Brott}, I., {Evans}, C.~J., {Hunter}, I., {et~al.} 2011{\natexlab{b}}, \aap,
  530, A116

\bibitem[{{Carroll}(1933)}]{carroll}
{Carroll}, J.~A. 1933, \mnras, 93, 478

\bibitem[{{Collins}(1963)}]{collins1963}
{Collins}, II, G.~W. 1963, \apj, 138, 1134

\bibitem[{{Collins} \& {Harrington}(1966)}]{collins1966}
{Collins}, II, G.~W. \& {Harrington}, J.~P. 1966, \apj, 146, 152

\bibitem[{{de Koter} {et~al.}(1998){de Koter}, {Heap}, \&
  {Hubeny}}]{dekoter1998}
{de Koter}, A., {Heap}, S.~R., \& {Hubeny}, I. 1998, \apj, 509, 879

\bibitem[{{de Mink} {et~al.}(2013){de Mink}, {Langer}, {Izzard}, {Sana}, \& {de
  Koter}}]{selma}
{de Mink}, S.~E., {Langer}, N., {Izzard}, R.~G., {Sana}, H., \& {de Koter}, A.
  2013, \apj, 764, 166

\bibitem[{{Donati} \& {Landstreet}(2009)}]{Donati2009}
{Donati}, J.-F. \& {Landstreet}, J.~D. 2009, \araa, 47, 333

\bibitem[{{Dufton} {et~al.}(2011){Dufton}, {Dunstall}, {Evans}, {Brott},
  {Cantiello}, {de Koter}, {de Mink}, {Fraser}, {H{\'e}nault-Brunet},
  {Howarth}, {Langer}, {Lennon}, {Markova}, {Sana}, \& {Taylor}}]{dufton1}
{Dufton}, P.~L., {Dunstall}, P.~R., {Evans}, C.~J., {et~al.} 2011, \apjl, 743,
  L22

\bibitem[{{Dufton} {et~al.}(2013){Dufton}, {Langer}, {Dunstall}, {Evans},
  {Brott}, {de Mink}, {Howarth}, {Kennedy}, {McEvoy}, {Potter},
  {Ram{\'{\i}}rez-Agudelo}, {Sana}, {Sim{\'o}n-D{\'{\i}}az}, {Taylor}, \&
  {Vink}}]{dufton}
{Dufton}, P.~L., {Langer}, N., {Dunstall}, P.~R., {et~al.} 2013, \aap, 550,
  A109 (\citetalias{dufton})

\bibitem[{{Ebbets}(1979)}]{ebbets}
{Ebbets}, D. 1979, \apj, 227, 510

\bibitem[{{Ekstr{\"o}m} {et~al.}(2012){Ekstr{\"o}m}, {Georgy}, {Eggenberger},
  {Meynet}, {Mowlavi}, {Wyttenbach}, {Granada}, {Decressin}, {Hirschi},
  {Frischknecht}, {Charbonnel}, \& {Maeder}}]{ekstrom2012}
{Ekstr{\"o}m}, S., {Georgy}, C., {Eggenberger}, P., {et~al.} 2012, \aap, 537,
  A146

\bibitem[{{Ekstr{\"o}m} {et~al.}(2008){Ekstr{\"o}m}, {Meynet}, {Maeder}, \&
  {Barblan}}]{ekstrom}
{Ekstr{\"o}m}, S., {Meynet}, G., {Maeder}, A., \& {Barblan}, F. 2008, \aap,
  478, 467

\bibitem[{{Evans} {et~al.}(2011){Evans}, {Taylor}, {H{\'e}nault-Brunet},
  {Sana}, {de Koter}, {Sim{\'o}n-D{\'{\i}}az}, {Carraro}, {Bagnoli}, {Bastian},
  {Bestenlehner}, {Bonanos}, {Bressert}, {Brott}, {Campbell}, {Cantiello},
  {Clark}, {Costa}, {Crowther}, {de Mink}, {Doran}, {Dufton}, {Dunstall},
  {Friedrich}, {Garcia}, {Gieles}, {Gr{\"a}fener}, {Herrero}, {Howarth},
  {Izzard}, {Langer}, {Lennon}, {Ma{\'{\i}}z Apell{\'a}niz}, {Markova},
  {Najarro}, {Puls}, {Ramirez}, {Sab{\'{\i}}n-Sanjuli{\'a}n}, {Smartt},
  {Stroud}, {van Loon}, {Vink}, \& {Walborn}}]{evans}
{Evans}, C.~J., {Taylor}, W.~D., {H{\'e}nault-Brunet}, V., {et~al.} 2011, \aap,
  530, A108 (\citetalias{evans})

\bibitem[{{Fruchter} {et~al.}(2006){Fruchter}, {Levan}, {Strolger},
  {Vreeswijk}, {Thorsett}, {Bersier}, {Burud}, {Castro Cer{\'o}n},
  {Castro-Tirado}, {Conselice}, {Dahlen}, {Ferguson}, {Fynbo}, {Garnavich},
  {Gibbons}, {Gorosabel}, {Gull}, {Hjorth}, {Holland}, {Kouveliotou}, {Levay},
  {Livio}, {Metzger}, {Nugent}, {Petro}, {Pian}, {Rhoads}, {Riess}, {Sahu},
  {Smette}, {Tanvir}, {Wijers}, \& {Woosley}}]{fruchter2006}
{Fruchter}, A.~S., {Levan}, A.~J., {Strolger}, L., {et~al.} 2006, \nat, 441,
  463

\bibitem[{{Fryer} \& {Heger}(2005)}]{fryer2005}
{Fryer}, C.~L. \& {Heger}, A. 2005, \apj, 623, 302

\bibitem[{{Gibson}(2000)}]{gibson2000}
{Gibson}, B.~K. 2000, \memsai, 71, 693

\bibitem[{{Gr{\"a}fener} \& {Hamann}(2008)}]{grafener2008}
{Gr{\"a}fener}, G. \& {Hamann}, W.-R. 2008, \aap, 482, 945

\bibitem[{Gray(1976)}]{Gray}
Gray, D. 1976, The Observation and Analysis of Stellar Photospheres, third
  edition edn. (Cambridge University Press)

\bibitem[{{Grunhut} {et~al.}(2012){Grunhut}, {Wade}, \& {MiMeS
  Collaboration}}]{2012AIPC.1429...67G}
{Grunhut}, J.~H., {Wade}, G.~A., \& {MiMeS Collaboration}. 2012, in American
  Institute of Physics Conference Series, Vol. 1429, American Institute of
  Physics Conference Series, ed. J.~L. {Hoffman}, J.~{Bjorkman}, \&
  B.~{Whitney}, 67--74

\bibitem[{{Herrero} {et~al.}(1992){Herrero}, {Kudritzki}, {Vilchez}, {Kunze},
  {Butler}, \& {Haser}}]{herrero}
{Herrero}, A., {Kudritzki}, R.~P., {Vilchez}, J.~M., {et~al.} 1992, \aap, 261,
  209

\bibitem[{{Howarth} {et~al.}(1997){Howarth}, {Siebert}, {Hussain}, \&
  {Prinja}}]{howarth}
{Howarth}, I.~D., {Siebert}, K.~W., {Hussain}, G.~A.~J., \& {Prinja}, R.~K.
  1997, \mnras, 284, 265

\bibitem[{{Huang} \& {Gies}(2006)}]{huanggies1}
{Huang}, W. \& {Gies}, D.~R. 2006, \apj, 648, 580

\bibitem[{{Huang} {et~al.}(2010){Huang}, {Gies}, \& {McSwain}}]{huang2010}
{Huang}, W., {Gies}, D.~R., \& {McSwain}, M.~V. 2010, \apj, 722, 605

\bibitem[{{Hubeny} \& {Lanz}(1995)}]{hubenyandlanz1995}
{Hubeny}, I. \& {Lanz}, T. 1995, \apj, 439, 875

\bibitem[{{Hunter} {et~al.}(2008){Hunter}, {Lennon}, {Dufton}, {Trundle},
  {Sim{\'o}n-D{\'{\i}}az}, {Smartt}, {Ryans}, \& {Evans}}]{hunter}
{Hunter}, I., {Lennon}, D.~J., {Dufton}, P.~L., {et~al.} 2008, \aap, 479, 541

\bibitem[{{Izzard} {et~al.}(2004){Izzard}, {Ramirez-Ruiz}, \&
  {Tout}}]{izzard2004}
{Izzard}, R.~G., {Ramirez-Ruiz}, E., \& {Tout}, C.~A. 2004, \mnras, 348, 1215

\bibitem[{{Kolmogorov}(1933)}]{smirnov}
{Kolmogorov}, A. 1933, Sulla determinazione empirica di una legge di
  distribuzione (Inst. Ital. Attuari)

\bibitem[{{Kuiper}(1960)}]{kuiper}
{Kuiper}, N.~H. 1960, Proceedings of the Koninklijke Nederlandse Akademie Van
  Wetenshappen, 63, 38

\bibitem[{{Langer}(2012)}]{langer2012}
{Langer}, N. 2012, \araa, 50, 107

\bibitem[{{Larson}(2010)}]{larson}
{Larson}, R.~B. 2010, Reports on Progress in Physics, 73, 014901

\bibitem[{{Lin} {et~al.}(2011){Lin}, {Krumholz}, \& {Kratter}}]{lin}
{Lin}, M.-K., {Krumholz}, M.~R., \& {Kratter}, K.~M. 2011, \mnras, 416, 580

\bibitem[{{Lucy}(1974)}]{lucy}
{Lucy}, L.~B. 1974, \aj, 79, 745

\bibitem[{{Lyutikov} \& {Blackman}(2001)}]{lyutikov}
{Lyutikov}, M. \& {Blackman}, E.~G. 2001, \mnras, 321, 177

\bibitem[{{Maeder}(1980)}]{maeder1980}
{Maeder}, A. 1980, \aap, 92, 101

\bibitem[{{Maeder} \& {Meynet}(2000)}]{maederymeynet2000}
{Maeder}, A. \& {Meynet}, G. 2000, \aap, 361, 159

\bibitem[{{Martins} {et~al.}(2005){Martins}, {Schaerer}, \&
  {Hillier}}]{martins}
{Martins}, F., {Schaerer}, D., \& {Hillier}, D.~J. 2005, \aap, 436, 1049

\bibitem[{{Massey} \& {Hunter}(1998)}]{masseyhunter1998}
{Massey}, P. \& {Hunter}, D.~A. 1998, \apj, 493, 180

\bibitem[{{Meynet} {et~al.}(2011){Meynet}, {Eggenberger}, \&
  {Maeder}}]{meynet2011}
{Meynet}, G., {Eggenberger}, P., \& {Maeder}, A. 2011, \aap, 525, L11

\bibitem[{{Modjaz} {et~al.}(2008){Modjaz}, {Kewley}, {Kirshner}, {Stanek},
  {Challis}, {Garnavich}, {Greene}, {Kelly}, \& {Prieto}}]{modjaz2008}
{Modjaz}, M., {Kewley}, L., {Kirshner}, R.~P., {et~al.} 2008, \aj, 135, 1136

\bibitem[{{Mokiem} {et~al.}(2006){Mokiem}, {de Koter}, {Evans}, {Puls},
  {Smartt}, {Crowther}, {Herrero}, {Langer}, {Lennon}, {Najarro}, {Villamariz},
  \& {Yoon}}]{mokiem}
{Mokiem}, M.~R., {de Koter}, A., {Evans}, C.~J., {et~al.} 2006, \aap, 456, 1131

\bibitem[{{Niino}(2011)}]{niino2011}
{Niino}, Y. 2011, \mnras, 417, 567

\bibitem[{{Penny}(1996)}]{penny}
{Penny}, L.~R. 1996, \apj, 463, 737

\bibitem[{{Penny} \& {Gies}(2009)}]{penny2009}
{Penny}, L.~R. \& {Gies}, D.~R. 2009, \apj, 700, 844

\bibitem[{{Podsiadlowski} {et~al.}(2010){Podsiadlowski}, {Ivanova}, {Justham},
  \& {Rappaport}}]{podsiadlowski2010}
{Podsiadlowski}, P., {Ivanova}, N., {Justham}, S., \& {Rappaport}, S. 2010,
  \mnras, 406, 840

\bibitem[{{Potter} {et~al.}(2012){Potter}, {Chitre}, \& {Tout}}]{potter}
{Potter}, A.~T., {Chitre}, S.~M., \& {Tout}, C.~A. 2012, \mnras, 424, 2358

\bibitem[{{Puls} {et~al.}(2005){Puls}, {Urbaneja}, {Venero}, {Repolust},
  {Springmann}, {Jokuthy}, \& {Mokiem}}]{puls2005}
{Puls}, J., {Urbaneja}, M.~A., {Venero}, R., {et~al.} 2005, \aap, 435, 669

\bibitem[{{Rosen} {et~al.}(2012){Rosen}, {Krumholz}, \& {Ramirez-Ruiz}}]{rosen}
{Rosen}, A.~L., {Krumholz}, M.~R., \& {Ramirez-Ruiz}, E. 2012, \apj, 748, 97

\bibitem[{{Ryans} {et~al.}(2002){Ryans}, {Dufton}, {Rolleston}, {Lennon},
  {Keenan}, {Smoker}, \& {Lambert}}]{ryans}
{Ryans}, R.~S.~I., {Dufton}, P.~L., {Rolleston}, W.~R.~J., {et~al.} 2002,
  \mnras, 336, 577

\bibitem[{{Sabbi} {et~al.}(2012){Sabbi}, {Lennon}, {Gieles}, {de Mink},
  {Walborn}, {Anderson}, {Bellini}, {Panagia}, {van der Marel}, \& {Ma{\'{\i}}z
  Apell{\'a}niz}}]{sabbi}
{Sabbi}, E., {Lennon}, D.~J., {Gieles}, M., {et~al.} 2012, \apjl, 754, L37

\bibitem[{{Sana} {et~al.}(2013){Sana}, {de Koter}, {de Mink}, {Dunstall},
  {Evans}, {H{\'e}nault-Brunet}, {Ma{\'{\i}}z Apell{\'a}niz},
  {Ram{\'{\i}}rez-Agudelo}, {Taylor}, {Walborn}, {Clark}, {Crowther},
  {Herrero}, {Gieles}, {Langer}, {Lennon}, \& {Vink}}]{sana}
{Sana}, H., {de Koter}, A., {de Mink}, S.~E., {et~al.} 2013, \aap, 550, A107
  (\citetalias{sana})

\bibitem[{{Sana} {et~al.}(2012){Sana}, {de Mink}, {de Koter}, {Langer},
  {Evans}, {Gieles}, {Gosset}, {Izzard}, {Le Bouquin}, \& {Schneider}}]{sanaa}
{Sana}, H., {de Mink}, S.~E., {de Koter}, A., {et~al.} 2012, Science, 337, 444

\bibitem[{{Selman} {et~al.}(1999){Selman}, {Melnick}, {Bosch}, \&
  {Terlevich}}]{selman}
{Selman}, F., {Melnick}, J., {Bosch}, G., \& {Terlevich}, R. 1999, \aap, 347,
  532

\bibitem[{{Sim{\'o}n-D{\'{\i}}az} {et~al.}(2013){Sim{\'o}n-D{\'{\i}}az},
  {Castro}, {Herrero}, {Aerts}, {Puls}, \& {Markova}}]{simon2013}
{Sim{\'o}n-D{\'{\i}}az}, S., {Castro}, N., {Herrero}, A., {et~al.} 2013, in ASP
  Conf.\ Series, Vol. 465, Four decades of research on massive stars, ed.
  L.~{Drissen}, C.~{Rubert}, N.~{St-Louis}, \& A.~F.~J. {Moffat}, 19

\bibitem[{{Sim{\'o}n-D{\'{\i}}az} \& {Herrero}(2007)}]{simon}
{Sim{\'o}n-D{\'{\i}}az}, S. \& {Herrero}, A. 2007, \aap, 468, 1063

\bibitem[{{Sim{\'o}n-D{\'{\i}}az} {et~al.}(2010){Sim{\'o}n-D{\'{\i}}az},
  {Herrero}, {Uytterhoeven}, {Castro}, {Aerts}, \& {Puls}}]{simon1}
{Sim{\'o}n-D{\'{\i}}az}, S., {Herrero}, A., {Uytterhoeven}, K., {et~al.} 2010,
  \apjl, 720, L174

\bibitem[{{Slettebak} {et~al.}(1975){Slettebak}, {Collins}, {Parkinson},
  {Boyce}, \& {White}}]{slettebak}
{Slettebak}, A., {Collins}, II, G.~W., {Parkinson}, T.~D., {Boyce}, P.~B., \&
  {White}, N.~M. 1975, \apjs, 29, 137

\bibitem[{{Tout} {et~al.}(2011){Tout}, {Wickramasinghe}, {Lau}, {Pringle}, \&
  {Ferrario}}]{tout2011}
{Tout}, C.~A., {Wickramasinghe}, D.~T., {Lau}, H.~H.-B., {Pringle}, J.~E., \&
  {Ferrario}, L. 2011, \mnras, 410, 2458

\bibitem[{{ud-Doula} \& {Owocki}(2002)}]{uddoula2002}
{ud-Doula}, A. \& {Owocki}, S.~P. 2002, \apj, 576, 413

\bibitem[{{Vink} {et~al.}(2010){Vink}, {Brott}, {Gr{\"a}fener}, {Langer}, {de
  Koter}, \& {Lennon}}]{vink2010}
{Vink}, J.~S., {Brott}, I., {Gr{\"a}fener}, G., {et~al.} 2010, \aap, 512, L7

\bibitem[{{Vink} \& {de Koter}(2005)}]{vink2005}
{Vink}, J.~S. \& {de Koter}, A. 2005, \aap, 442, 587

\bibitem[{{Vink} {et~al.}(2011){Vink}, {Muijres}, {Anthonisse}, {de Koter},
  {Gr{\"a}fener}, \& {Langer}}]{vink2011}
{Vink}, J.~S., {Muijres}, L.~E., {Anthonisse}, B., {et~al.} 2011, \aap, 531,
  A132

\bibitem[{{von Zeipel}(1924)}]{zeipel}
{von Zeipel}, H. 1924, \mnras, 84, 665

\bibitem[{{Walborn} \& {Blades}(1997)}]{walborn}
{Walborn}, N.~R. \& {Blades}, J.~C. 1997, \apjs, 112, 457

\bibitem[{{Weidner} \& {Vink}(2010)}]{weiderandvink2010}
{Weidner}, C. \& {Vink}, J.~S. 2010, \aap, 524, A98

\bibitem[{{Woosley}(1993)}]{woosley1993}
{Woosley}, S.~E. 1993, in AIP Conf.\ Proc., Vol. 280, Compton gamma‐ray
  observatory, ed. M.~{Friedlander}, N.~{Gehrels}, \& D.~J. {Macomb}, 995

\bibitem[{{Woosley} \& {Heger}(2006)}]{woosley}
{Woosley}, S.~E. \& {Heger}, A. 2006, \apj, 637, 914

\bibitem[{{Yoon} \& {Langer}(2005)}]{yoon2005}
{Yoon}, S.-C. \& {Langer}, N. 2005, \aap, 443, 643

\bibitem[{{Zinnecker} \& {Yorke}(2007)}]{zinnecker2007}
{Zinnecker}, H. \& {Yorke}, H.~W. 2007, \araa, 45, 481

\end{thebibliography}
%\begin{thebibliography}{}

%\end{thebibliography}

\begin{appendix} %First online appendix
%\section{Comparison of more diagnostics lines}
%Because the optical images used in this analysis...

%\begin{figure*}
%\centering
%\includegraphics[scale=0.48]{Annex1_1}
%\includegraphics[scale=0.48]{Annex1_2}
%\includegraphics[scale=1.0]{Annex1_31}
%\caption{Differences of \vrot\, for the \hhel{i} lines as a function of \hel{i}{4713}.
%}
%\label{fig:deconvol}
%\end{figure*}

\section{ \Nl{v}{4604} line}\label{subsec:NV}

Though the \NNl{v} line is not widely used to obtain \vrot,
the line is little affected by Stark broadening and may thus provide an alternative
anchor point for \hhel{ii}-based measurements. As \Nl{v}{4604} is intrinsically
strong it may be partly formed in the stellar outflow. It can thus only be used for
\vrot\ determination if the wind is weak and the line is of photospheric origin, i.e.
it should be symmetric.

\begin{figure}
\centering
\includegraphics[scale=0.52]{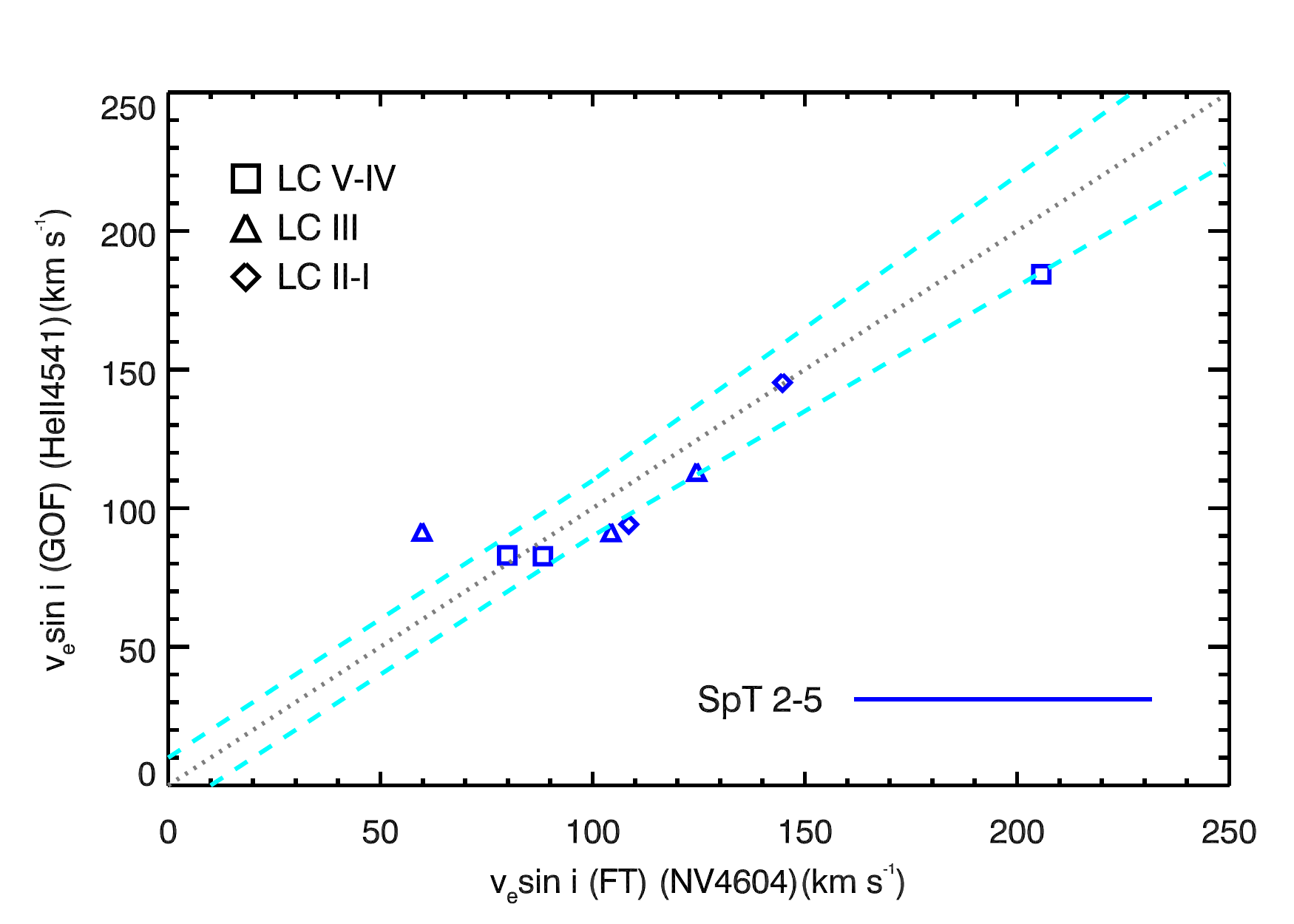}
\caption{Comparison of the \vrot\ measurements for \hel{ii}{4541} and \Nl{v}{4604}.
Information on the LC of the targets is provided by the symbol shapes. % and colors.
%The dashed and dotted lines have the same meaning as in Fig.~\ref{fig:Comparison-FT-GOF}.
The dashed lines show the $\pm 10$~\kms\ and/or $\pm\, 10$\%,  
whichever is the largest, around the 1:1 relation. 
}
\label{fig:NV}
\end{figure}

Eight stars in our sample  (VFTS~016, 072, 180, 267, 506, 518, 566, 599, 621) show both \Nl{v}{4604} and  \hel{ii}{4541}.VFTS~180 was discarded due to its asymmetric profile.
Fig.~\ref{fig:NV} compares the \vrot\ measurements obtained, for the remainder sample, from both diagnostic lines. 
Save for one target, all measurements agree within $\pm 10$~\kms\ and/or $\pm\, 10$\% of the 1:1 relation. 
The comparison lacks stars with low projected spin rates. The sole case for which the comparison is poor has the
lowest spin rate when using the N\,{\sc v} diagnostic. On the basis of one case we cannot conclude whether this indicates
a systematic effect at low \vrot\ or whether it is the result of the relatively large dispersion of spin rate measurements based on
\hhel{ii} at modest rotational velocities (see Sect.~\ref{subsec:test_heii}).
%\blue{Measurements for stars with projected rotational velocities agree within  10\%, 
% the use of the \hhel{ii}, at least for \vrot > 80\,\kms.} % velocity range.
% The only one star with \vrot $<$ 80~\kms\ show a deviation of 30~\kms\ with respect to the 1:1 relation. It is unfortunately not possible to decide whether this indicate a systematic effect or whether this just reflect the large statistical dispersion of \vrot\ measurements based on \hhel{ii} at modest rotational velocity. 

\section{Comparison with overlapping sample of the O and B-type stars in the VFTS}\label{subsec:Common_OB}

Rotational velocities  for the single B-type stars in the Tarantula Survey 
have been presented in \citetalias{dufton}. There 
the FT technique was also used, but a different set of lines was applied reflecting the different temperature range. 
Due to uncertainties in the preliminary spectral classification, 47 late-O and early-B 
stars were included in both data sets. 
\begin{figure}
\centering
\includegraphics[scale=0.52]{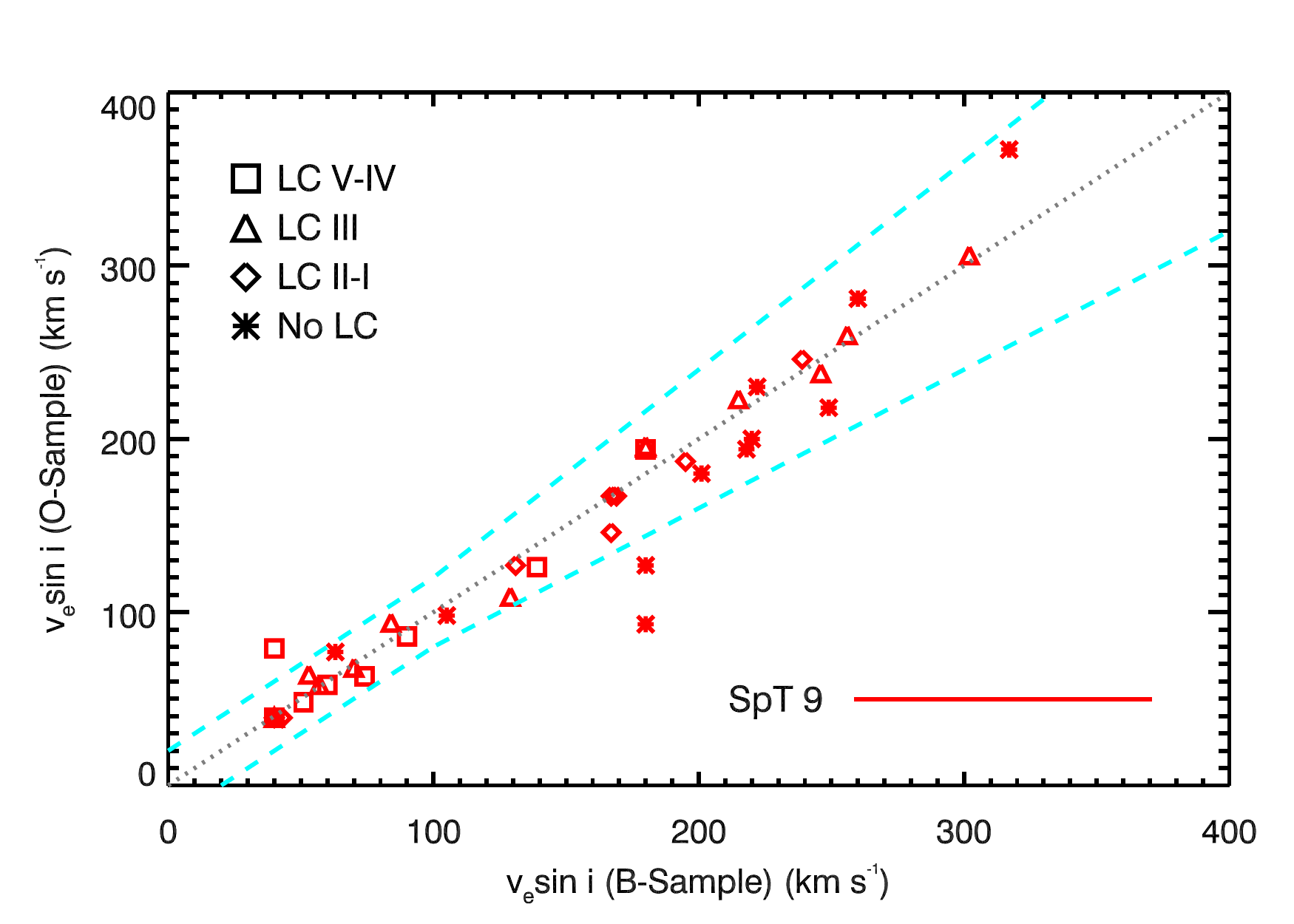}
\caption{Comparison of the \vrot\ measurements of the overlapping late O-type sample in \citet{dufton} and
this study.
%from the VFTS O and B common sample.
Information on the LC of the targets is provided by the symbol shapes.
%The dashed and dotted lines have the same meaning as in Fig.~\ref{fig:NV}.
The dashed and dotted lines have the same meaning as in Fig.~\ref{fig:Comparison-FT-GOF}.
}
\label{fig:Common_OB}
\end{figure}

Fig.~\ref{fig:Common_OB} compares both \vrot\ measurements. 
We find that there are no significant systematic differences between the two sets of measurements.
Despite two outliers (VFTS~412 and 594 at about $\sim$180~\kms), the overall  comparison presents an excellent 
agreement, with a rms dispersion of about  10\%\ of the 1:1 relation.

\end{appendix}

\end{document}